\documentclass[letterpaper,twocolumn,english,aps,prx,citeautoscript,superscriptaddress,longbibliography]{revtex4-2}
\usepackage[T1]{fontenc}
\usepackage[latin9]{inputenc}
\usepackage{color}
\usepackage{amsmath}
\usepackage{amssymb}
\usepackage{graphicx}

\makeatletter

\pdfpageheight\paperheight
\pdfpagewidth\paperwidth

\providecommand{\tabularnewline}{\\}

\usepackage{hyperref}
\hypersetup{colorlinks=true, citecolor=black, hidelinks = true}
\usepackage{amsfonts}
\usepackage{bbm}
\usepackage{graphics}

\def\ket#1{\left| #1\right>}
\def\bra#1{\left< #1\right|}

\def\<{\langle}
\def\>{\rangle}

\usepackage{leftidx}

\usepackage{etoolbox}
\patchcmd{\subsection}{\centering}{\raggedright}{}{}

\makeatother

\usepackage{babel}
\begin{document}
\title{Cavity-mediated entanglement of parametrically driven spin qubits
via sidebands}
\author{V. Srinivasa}
\email{vsriniv@uri.edu}

\affiliation{Department of Physics, University of Rhode Island, Kingston, RI 02881,
USA}
\author{J. M. Taylor}
\affiliation{Joint Quantum Institute, University of Maryland, College Park, Maryland
20742, USA}
\affiliation{Joint Center for Quantum Information and Computer Science, University
of Maryland, College Park, Maryland 20742, USA}
\affiliation{National Institute of Standards and Technology, Gaithersburg, Maryland,
20899, USA}
\author{J. R. Petta}
\affiliation{Department of Physics and Astronomy, University of California--Los
Angeles, Los Angeles, California 90095, USA}
\affiliation{Center for Quantum Science and Engineering, University of California--Los
Angeles, Los Angeles, California 90095, USA}
\date{\today}
\begin{abstract}
We consider a pair of quantum dot-based spin qubits that interact
via microwave photons in a superconducting cavity, and that are also
parametrically driven by separate external electric fields. For this
system, we formulate a model for spin qubit entanglement in the presence
of mutually off-resonant qubit and cavity frequencies. We show that
the sidebands generated via the driving fields enable highly tunable
qubit-qubit entanglement using only ac control and without requiring
the qubit and cavity frequencies to be tuned into simultaneous resonance.
The model we derive can be mapped to a variety of qubit types, including
detuning-driven one-electron spin qubits in double quantum dots and
three-electron resonant exchange qubits in triple quantum dots. The
high degree of nonlinearity inherent in spin qubits renders these
systems particularly favorable for parametric drive-activated entanglement.
We determine multiple common resonance conditions for the two driven
qubits and the cavity and identify experimentally relevant parameter
regimes that enable the implementation of entangling gates with suppressed
sensitivity to cavity photon occupation and decay. The parametrically
driven sideband resonance approach we describe provides a promising
route toward scalability and modularity in spin-based quantum information
processing through drive-enabled tunability that can also be implemented
in micromagnet-free electron and hole systems for spin-photon coupling.
\end{abstract}
\maketitle

\section{Introduction}

Scaling to many-qubit systems represents a current challenge in the
implementation of quantum information processing \cite{DiVincenzo2000FortschrPhys,Ladd2010}
due to the highly complex electronics required to control even a few
qubits in most realizations, combined with the need to minimize dissipation
of quantum information into the environment. One approach to addressing
this challenge is provided by modularity \cite{Taylor2005,Jiang2007,Monroe2014,Vandersypen2017,Tosi2017,Jnane2022},
which enables scalability by linking existing, relatively well-controlled,
and locally optimized few-qubit modules via robust long-range interactions.
For semiconductor spin qubits, which represent a promising quantum
information processing platform \cite{Loss1998,Kane1998,Hanson2007RMP,Zwanenburg2013,Petit2020,Chatterjee2021,Noiri2022,Xue2022,Mills2022,Weinstein2023,Burkard2023},
such long-range interactions can be achieved by coupling spins to
photons in a microwave cavity using the approach of circuit quantum
electrodynamics (cQED) \cite{Childress2004,Blais2004,Wallraff2004,Blais2007,Majer2007,Sillanpaa2007,Blais2021,Clerk2020,Burkard2020,Burkard2023}. 

Building on the promise of long coherence times for spins in silicon
\cite{Veldhorst2014,Veldhorst2015,Yoneda2018,Noiri2022,Xue2022,Mills2022},
strong spin-photon coupling \cite{Mi2018,Samkharadze2018,Landig2018,Landig2019,Yu2023}
as well as coherent photon-mediated interaction of two single-electron
silicon spin qubits \cite{Borjans2020,Harvey-Collard2022} have now
been achieved. While these results provide a path to scalability for
spin-based quantum information processing, tuning and scaling challenges
remain for applying this approach to more than two qubits. For resonant
cavity-mediated qubit-qubit coupling \cite{Borjans2020}, all qubit
frequencies must be tuned into simultaneous resonance with the cavity
frequency, and the micromagnets required in silicon for achieving
sufficient spin-charge coupling must be precisely positioned for each
qubit. In the standard dispersive approach to cavity-mediated coupling
\cite{Blais2004,Blais2007,Landig2019,Harvey-Collard2022,Benito2019b,Warren2019},
these constraints are partially relaxed as the cavity virtually mediates
the interaction and its frequency is distinct from those of the qubits.
However, an entangling interaction between the two qubits still requires
their (cavity-shifted) frequencies to be tuned into mutual resonance.
To allow for increased flexibility in achieving qubit-qubit entanglement,
off-resonant coupling approaches have been developed for cQED systems
with superconducting qubits \cite{Blais2007,Rigetti2005,Wallraff2007,Leek2009,Rigetti2010,Chow2011,Beaudoin2012,McKay2016,Blais2021}
and related two-qubit gates have also recently been explored in the
context of semiconductor qubits \cite{Tosi2018,Sigillito2019npjQI,Ruskov2021,Warren2021,Hansen2021,McMillan2022,Mielke2023}.
These approaches enable qubits to be fixed at optimal operation points
where decoherence is minimized, while interactions are activated via
external microwave driving of either the coupling or one or more qubits. 

\begin{figure}
\begin{centering}
\includegraphics[viewport=0bp 0bp 600bp 400bp,width=1\columnwidth]{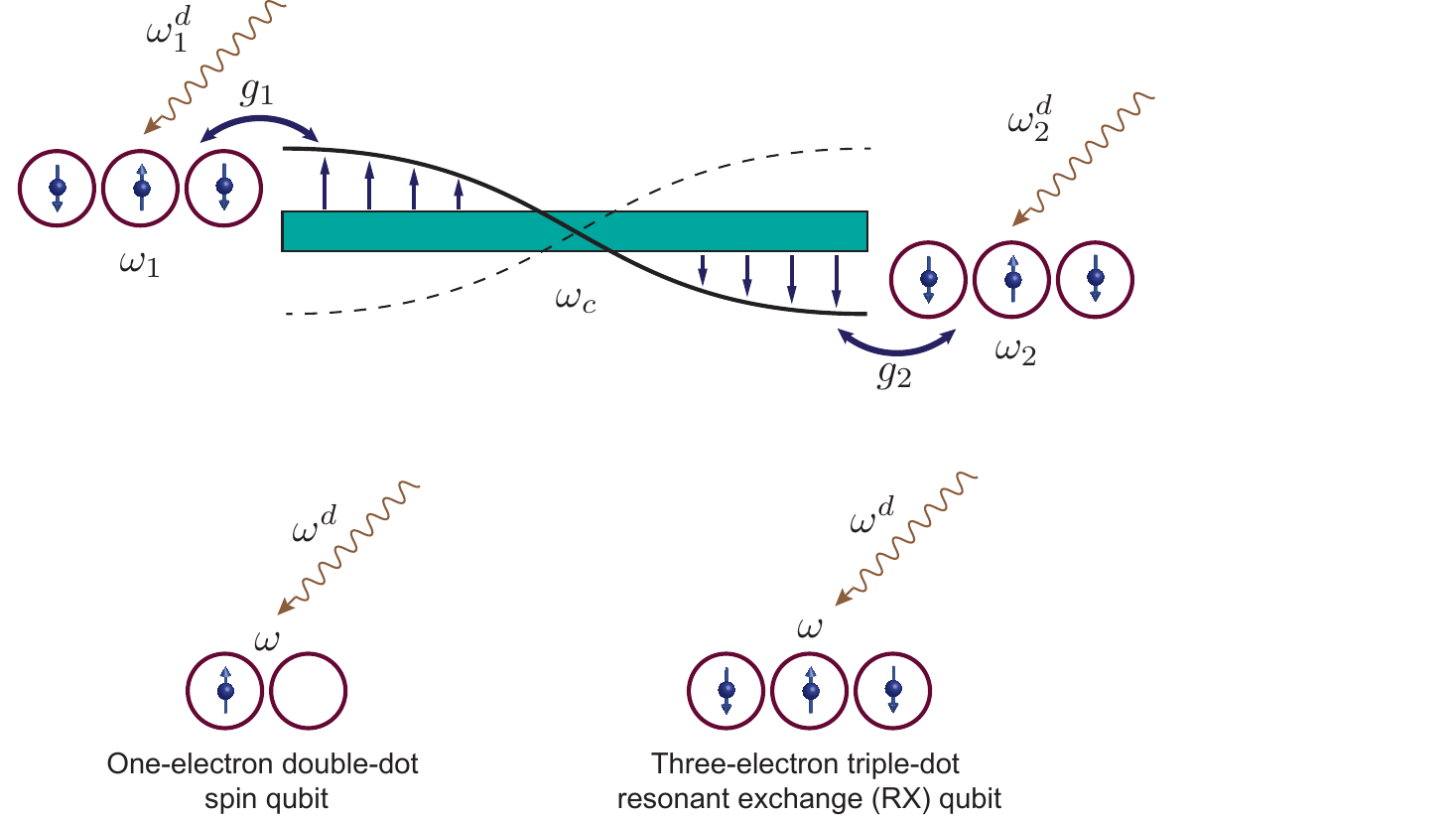}
\par\end{centering}
\caption{\label{fig:dotresdriv}Schematic illustration of system for cavity-mediated
coupling of two parametrically driven quantum dot spin qubits via
sidebands. The qubits have transition frequencies $\omega_{1}$ and
$\omega_{2}$ and are coupled to the fundamental mode of a microwave
transmission line resonator (cavity), which has frequency $\omega_{c},$
with strengths $g_{1}$ and $g_{2},$ respectively. The qubits are
also driven by external microwave fields with frequencies $\omega_{1}^{d}$
and $\omega_{2}^{d}.$ The approach in this work can be applied to
multiple types of qubits that can be sinusoidally driven, including
one-electron spin qubits in double dots and three-electron resonant
exchange (RX) qubits in triple dots.}

\end{figure}

In this work, we consider a pair of qubits based on electron spins
in quantum dots that interact via microwave photons in a superconducting
cavity, and that are also parametrically driven by classical external
electric fields (Fig.~\ref{fig:dotresdriv}). For this system, we
formulate a model for entanglement between the two qubits that incorporates
mutually off-resonant qubit and cavity frequencies and makes use of
the Mollow triplet sidebands of the driven qubits \cite{Mollow1969,Kim2014PRL}
to effectively provide multiple qubit transition frequencies for cavity-mediated
coupling. This approach enables highly tunable qubit-cavity photon
interactions and qubit-qubit entanglement using ac control through
the applied driving fields, without requiring dc tuning of the qubit
frequencies. The spin qubits can therefore be fixed at optimal operation
points that allow for maximal coherence times. The model we develop
can be mapped to a variety of qubit types. Here, we illustrate this
mapping for both single-electron spin qubits in double quantum dots
\cite{Benito2017,Croot2020} and three-electron resonant exchange
(RX) qubits in triple quantum dots \cite{Medford2013,Taylor2013}
in the driven resonant regime \cite{Srinivasa2016}. 

We determine common resonance conditions for the two driven qubits
and the cavity and identify parameters for implementing multiple entangling
gates. In contrast to the sideband-based gates obtained for the driven
resonant regime in prior work \cite{Srinivasa2016,Abadillo-Uriel2021},
entangling gates do not require sequences of multiple sideband pulses
and additionally exhibit suppressed sensitivity to cavity photon occupation,
as we verify through two-qubit gate fidelity calculations. The enhanced
spectral flexibility inherent in the approach we describe provides
a promising route toward scalability and modularity in spin-based
quantum information processing through drive-enabled tunable entanglement
that can also be implemented in micromagnet-free systems for spin-photon
coupling \cite{Landig2018,Landig2019,Yu2023}.

\section{\label{sec:Theoretical-framework}Theoretical framework}

We now develop a general theoretical description of the sideband-based,
cavity-mediated entangling gates between driven qubits discussed in
this work. To illustrate the approach in more concrete terms, we consider
two specific types of quantum dot-based electron spin qubits for which
strong spin-photon coupling has been realized (Fig.~\ref{fig:dotresdriv}
and Appendix \ref{sec:spinqubitHpderiv}): (1) One-electron spin qubits
in double quantum dots with micromagnets for spin-charge coupling
\cite{Benito2017,Mi2018,Samkharadze2018,Harvey-Collard2022}; and
(2) three-electron RX qubits in triple quantum dots, which couple
directly to photons through their intrinsic electric dipole moments
\cite{Medford2013,Taylor2013,Srinivasa2016,Russ2015b,Landig2018,Landig2019}.
Both types of qubits, as well as several other classes of spin qubits
such as two-electron singlet-triplet qubits \cite{Levy2002,Petta2005,Burkard2006,Taylor2006,Taylor2007,Bottcher2022},
quantum dot hybrid qubits \cite{Corrigan2023arxiv}, and hole spin
qubits in silicon and germanium double quantum dots \cite{Yu2023,Scappucci2021},
can be manipulated electrically by parametrically driving the detuning
$\epsilon_{j}$ between the (outer) two dots of qubit $j$ \cite{Benito2019,Croot2020,Taylor2013,Medford2013}.
We write this driving field as
\begin{equation}
\epsilon_{j}\left(t\right)\equiv\epsilon_{0,j}+2\mathcal{F}_{j}\cos\left(\omega_{j}^{d}t+\phi_{j}^{\prime}\right)\label{eq:detdrive}
\end{equation}
in terms of a dc operation point $\epsilon_{0,j},$ as well as the
amplitude $2\mathcal{F}_{j},$ frequency $\omega_{j}^{d},$ and phase
$\phi_{j}^{\prime}$ of an ac drive that sinusoidally modulates the
detuning about $\epsilon_{0,j}$ as a function of time. Here, we fix
the qubits at the ``sweet spot'' detuning operation points $\epsilon_{0,j}=0$
for $j=1,2,$ where the first derivative of the qubit frequency vanishes
for both one-electron spin and symmetric RX qubits, enabling leading-order
protection from charge noise \cite{Koch2007,Houck2009,Medford2013,Taylor2013,Fei2015,Russ2015,Shim2016,Russ2016,Benito2017,Benito2019,Croot2020}.
The operation point $\epsilon_{0,j}=0$ also maximizes the effective
spin-photon coupling strength for one-electron spin qubits \cite{Benito2019,Croot2020},
while optimization of the RX qubit-photon coupling strength involves
multiple parameters (see Appendix \ref{sec:spinqubitHpderiv} for
more details). In what follows, all sums are over the qubit index
$j=1,2$ unless otherwise noted. 

As shown in Appendix \ref{sec:spinqubitHpderiv}, for both one-electron
spin qubits and RX qubits driven according to Eq.~(\ref{eq:detdrive}),
the system Hamiltonian including the cavity and driving fields with
$\epsilon_{0,j}=0$ can be written in the qubit basis as ($\hbar=1$)
\begin{align}
H_{p} & \equiv\omega_{c}a^{\dagger}a+\sum_{j}\frac{\omega_{j}}{2}\sigma_{j}^{z}+\sum_{j}g_{j}\sigma_{j}^{x}\left(a+a^{\dagger}\right)\nonumber \\
 & +\sum_{j}2\Omega_{j}\cos\left(\omega_{j}^{d}t+\phi_{j}\right)\sigma_{j}^{x},\label{eq:Hp}
\end{align}
where $a^{\dagger}$ and $a$ are photon creation and annihilation
operators for the fundamental mode of the cavity with frequency $\omega_{c}$,
$\sigma_{j}^{\alpha}$ with $\alpha=x,y,z$ and $\sigma_{j}^{z}\equiv\ket{1}_{j}\bra{1}-\ket{0}_{j}\bra{0}$
are Pauli operators and $\omega_{j}$ is the transition frequency
for spin qubit $j$, $g_{j}$ is the strength of the coupling between
spin qubit $j$ and photons in the fundamental cavity mode, and $2\Omega_{j}$
is the effective driving amplitude for spin qubit $j$. We see from
Eq.~(\ref{eq:Hp}) that, for both types of qubits, the detuning drive
{[}Eq.~(\ref{eq:detdrive}){]} acting on the electron charge degrees
of freedom is translated into an effective transverse ($\sigma_{x}$)
drive on the qubit. Note that we have redefined the phases $\phi_{j}$
of the drives with respect to Eq.~(\ref{eq:detdrive}) to take into
account sign changes occuring in the derivation of $H_{p}$ (see Appendix
\ref{sec:spinqubitHpderiv} for details). 

To derive sideband-mediated entangling interactions from Eq.~(\ref{eq:Hp}),
we first transform $H_{p}$ to a frame rotating at the frequencies
of both drives and the cavity via 
\begin{equation}
U_{1}=e^{-it\left(\omega_{c}a^{\dagger}a+\sum_{j}\omega_{j}^{d}\sigma_{j}^{z}/2\right)},\label{eq:U1}
\end{equation}
which is equivalent to an interaction picture for resonant driving
of the qubits, $\omega_{j}^{d}=\omega_{j}.$ Defining the cavity-drive
detunings $\Delta_{j}\equiv\omega_{c}-\omega_{j}^{d}$ and the qubit-drive
detunings $\delta_{j}\equiv\omega_{j}-\omega_{j}^{d},$ and making
a rotating wave approximation for $\left|\Delta_{j}\right|\ll\omega_{c}+\omega_{j}^{d},2\omega_{j}^{d},$
we drop rapidly oscillating terms $\sim e^{\pm i\left(\omega_{c}+\omega_{j}^{d}\right)t},\sim e^{\pm2i\omega_{j}^{d}t}$
and find
\begin{align}
H_{p}^{{\rm rf}} & \equiv U_{1}^{\dagger}H_{p}U_{1}-iU_{1}^{\dagger}\dot{U}_{1}\nonumber \\
 & \approx H_{0}+V\left(t\right),\nonumber \\
H_{0} & \equiv\sum_{j}\frac{\delta_{j}}{2}\sigma_{j}^{z}+\sum_{j}\Omega_{j}\left(e^{-i\phi_{j}}\sigma_{j}^{+}+e^{i\phi_{j}}\sigma_{j}^{-}\right),\nonumber \\
V\left(t\right) & \equiv\sum_{j}g_{j}\left(e^{-i\Delta_{j}t}\sigma_{j}^{+}a+e^{i\Delta_{j}t}\sigma_{j}^{-}a^{\dagger}\right),\label{eq:Hprf}
\end{align}
where we assume $g_{j}\ll2\Omega_{j}$ and take $V\left(t\right)$
as a time-dependent perturbation to the other terms in Eq.~(\ref{eq:Hprf}). 

We initially take the limit $g_{j}\rightarrow0$ and diagonalize $H_{0}.$
Choosing the phases of the driving fields to be $\phi_{j}=0$ for
$j=1,2$ to simplify the analysis, we find 
\begin{align}
H_{0} & =\sum_{j}\frac{\delta_{j}}{2}\sigma_{j}^{z}+\sum_{j}\Omega_{j}\sigma_{j}^{x}\nonumber \\
 & \equiv\sum_{j}\frac{W_{j}}{2}\left(\cos\theta_{j}\sigma_{j}^{z}+\sin\theta_{j}\sigma_{j}^{x}\right),\label{eq:H0}
\end{align}
where we have for convenience re-expressed $H_{0}$ in terms of $W_{j}$
and $\theta_{j}$ in the last line of Eq.~(\ref{eq:H0}), which serve
to redefine the Pauli operators $\sigma_{j}^{z}\equiv\ket{e}_{j}\bra{e}-\ket{g}_{j}\bra{g}$
and $\sigma_{j}^{x}$ in terms of the dressed qubit basis $\left\{ \ket{e}_{j},\ket{g}_{j}\right\} .$
We can then diagonalize this Hamiltonian via a rotation around the
$y$ axis for each qubit, described by 
\begin{equation}
U_{q}=e^{-i\sum_{j}\theta_{j}\sigma_{j}^{y}/2}\label{eq:Uq}
\end{equation}
where $\tan\theta_{j}=2\Omega_{j}/\delta_{j},$ which yields 
\begin{align}
H_{0,q} & \equiv U_{q}^{\dagger}H_{0}U_{q}\nonumber \\
 & =\sum_{j}\frac{W_{j}}{2}\sigma_{j}^{z}\label{eq:H0q}
\end{align}
with the dressed qubit frequencies $W_{j}\equiv\sqrt{\delta_{j}^{2}+4\Omega_{j}^{2}}.$
Applying the same rotation $U_{q}$ {[}Eq.~(\ref{eq:Uq}){]} to the
qubit-cavity coupling perturbation $V\left(t\right)$ then yields
the full Hamiltonian $H_{q}\equiv H_{0,q}+V_{q}\left(t\right)$ in
the dressed qubit basis, where $V_{q}\left(t\right)\equiv U_{q}^{\dagger}V\left(t\right)U_{q}.$ 

Finally, we transform to a second frame rotating at the dressed qubit
frequencies $W_{j}$ for the driven qubits via
\begin{equation}
U_{2}=e^{-i\sum_{j}W_{j}\sigma_{j}^{z}/2},\label{eq:U2}
\end{equation}
which is equivalent to the interaction picture with respect to $H_{0,q}$
{[}see Eq.~(\ref{eq:H0q}){]}. In this frame, we find the Hamiltonian
\begin{align}
V_{I}\left(t\right) & \equiv U_{2}^{\dagger}H_{q}U_{2}-iU_{2}^{\dagger}\dot{U}_{2}\nonumber \\
 & =A\left(t\right)a^{\dagger}+A^{\dagger}\left(t\right)a.\label{eq:VI}
\end{align}
In Eq.~(\ref{eq:VI}), we have defined 
\begin{align}
A\left(t\right) & \equiv\sum_{j}A_{j}\nonumber \\
 & \equiv\sum_{j}\frac{g_{j}}{2}e^{i\Delta_{j}t}\left[\sin\theta_{j}\sigma_{j}^{z}-\left(1-\cos\theta_{j}\right)e^{iW_{j}t}\sigma_{j}^{+}\right.\nonumber \\
 & \ \ \ \ \ \ \ \ \ +\left.\left(1+\cos\theta_{j}\right)e^{-iW_{j}t}\sigma_{j}^{-}\right],\label{eq:A}
\end{align}
which represents a sum of time-dependent qubit operators. We note
that $A\left(t\right),$ and therefore $V_{I}\left(t\right),$ have
the three characteristic frequencies $\Delta_{j}$ and $\Delta_{j}\pm W_{j}$
for qubit $j,$ which correspond to the Mollow triplet frequencies
\cite{Mollow1969} consisting of the center and sideband frequencies
of each driven qubit. In the present case, these frequencies are uniformly
shifted by the cavity frequency $\omega_{c}$ due to the rotating-frame
transformation $U_{1}$ {[}Eq.~(\ref{eq:U1}){]}. Driving the qubits
on resonance such that $\omega_{j}^{d}=\omega_{j}$ gives $\delta_{j}=0$
and $W_{j}=2\Omega_{j},$ so that $\sin\theta_{j}=1$ and $\cos\theta_{j}=0.$
In this case, Eq.~(\ref{eq:A}) reduces to 
\begin{align}
A_{r}\left(t\right) & \equiv\sum_{j}\frac{g_{j}}{2}\left[e^{i\Delta_{j}t}\sigma_{j}^{z}-e^{i\left(\Delta_{j}+2\Omega_{j}\right)t}\sigma_{j}^{+}\right.\nonumber \\
 & \ \ \ \ \ \ \ \ \ +\left.e^{-i\left(\Delta_{j}-2\Omega_{j}\right)t}\sigma_{j}^{-}\right],\label{eq:Ares}
\end{align}
and we find that the three characteristic frequencies for driven qubit
$j$ become $\Delta_{j},\Delta_{j}\pm2\Omega_{j}.$ 

In order to determine the gate operations generated by the time-dependent
Hamiltonian $V_{I}\left(t\right)$ in Eq.~(\ref{eq:VI}), we use
the Magnus expansion \cite{Magnus1954} up to second order to approximate
the time evolution operator. For $g_{j}\ll W_{j}$ such that $\lambda\equiv g_{j}/W_{j}$
is a small parameter, we write $U\left(\tau\right)\approx e^{-iH_{{\rm eff}}\tau},$
where ${\rm H}_{{\rm eff}}\left(\tau\right)=\lambda\bar{H}_{1}\left(\tau\right)+\lambda^{2}\bar{H}_{2}\left(\tau\right)$
represents an effective Hamiltonian to $O\left(\lambda^{2}\right)$
with
\begin{align}
\lambda\bar{H}_{1}\left(\tau\right) & \equiv\frac{1}{\tau}\int_{0}^{\tau}dt\ V_{I}\left(t\right),\label{eq:H1avgint}\\
\lambda^{2}\bar{H}_{2}\left(\tau\right) & \equiv\frac{1}{2i\tau}\int_{0}^{\tau}dt\int_{0}^{t}dt^{\prime}\ \left[V_{I}\left(t\right),V_{I}\left(t^{\prime}\right)\right].\label{eq:H2avgint}
\end{align}
We first consider the term $\lambda\bar{H}_{1}\left(\tau\right).$
From Eqs.~(\ref{eq:VI}) and (\ref{eq:A}), we see that the integral
in Eq.~(\ref{eq:H1avgint}) is evaluated via the corresponding integrals
of $A\left(t\right)$ and its Hermitian conjugate. The first-order
term in the effective Hamiltonian is then given by
\begin{align}
\lambda\bar{H}_{1}\left(\tau\right) & =\sum_{j}\frac{g_{j}}{2}\left[f\left(\Delta_{j}\right)\sigma_{j}^{z}-f\left(\Delta_{j}+W_{j}\right)\sigma_{j}^{+}\right.\nonumber \\
 & \ \ \ \ \ \ \ \ \ +\left.f\left(\Delta_{j}-W_{j}\right)\sigma_{j}^{-}a^{\dagger}+\text{{\rm H.c.}}\right],\label{eq:H1avg}
\end{align}
where we have defined the integral
\begin{align}
f\left(\mu\right) & \equiv\frac{1}{\tau}\int_{0}^{\tau}dt\ e^{i\mu t}\label{eq:fjm}
\end{align}
with frequencies $\mu=\Delta_{j},\Delta_{j}\pm W_{j}.$ The first-order
term $\lambda\bar{H}_{1}\left(\tau\right)$ describes the direct ($\sim g_{j}$)
interaction of each qubit with the cavity and includes both red ($\sim\sigma_{j}^{+}a,\sigma_{j}^{-}a^{\dagger}$)
and blue ($\sim\sigma_{j}^{+}a^{\dagger},\sigma_{j}^{-}a$) sideband
terms. These interaction terms can be used to generate entanglement
via sequences of multiple sideband pulses \cite{Childs2000,Blais2007,Wallraff2007,Leek2009,Srinivasa2016,Abadillo-Uriel2021}. 

We now set $\Delta_{j}=p_{j}\eta$ and $W_{j}=q_{j}\eta$ with $p_{j},q_{j}$
integers and $\eta\equiv2\pi/\tau.$ We also assume $\mu\neq0.$ In
this case, we find that $\mu=r_{j}\eta$ with $r_{j}=p_{j},p_{j}\pm q_{j}\neq0$
also an integer, so that all integrals $f\left(\mu\right)$ in Eq.~(\ref{eq:H1avg})
vanish and $\lambda\bar{H}_{1}\left(\tau\right)=0.$ Thus, when $\Delta_{j}$
and $W_{j}$ are both integer multiples of the same frequency $\eta,$
we can completely eliminate the first-order sideband interaction terms
{[}Eq.~(\ref{eq:H1avg}){]} from \textbf{${\rm H}_{{\rm eff}}\left(\tau\right).$} 

To $O\left(\lambda^{2}\right),$ the effective interaction Hamiltonian
generating the gate operation is now given entirely by the second-order
term
\begin{equation}
H_{{\rm eff}}\left(\tau\right)=\lambda^{2}\bar{H}_{2}\left(\tau\right)=\frac{1}{2i\tau}\int_{0}^{\tau}dt\int_{0}^{t}dt^{\prime}\left[V_{I}\left(t\right),V_{I}\left(t^{\prime}\right)\right].\label{eq:HeffH2avg}
\end{equation}
Using Eqs.~(\ref{eq:VI}) and (\ref{eq:A}), we can write the commutator
in Eq.~(\ref{eq:HeffH2avg}) as 
\begin{align}
\left[V_{I}\left(t\right),V_{I}\left(t^{\prime}\right)\right] & =\left[V_{I},V_{I}^{\prime}\right]_{1}+\left[V_{I},V_{I}^{\prime}\right]_{2},\nonumber \\
\left[V_{I},V_{I}^{\prime}\right]_{1} & \equiv\sum_{j}\left\{ \left[A_{j},A_{j}^{\prime}\right]a^{\dagger2}+\left[A_{j}^{\dagger},A_{j}^{\prime\dagger}\right]a^{2}\right.\nonumber \\
 & \ \ \ \ \ \ +\left.\left(\left[A_{j},A_{j}^{\prime\dagger}\right]+\left[A_{j}^{\dagger},A_{j}^{\prime}\right]\right)a^{\dagger}a\right\} ,\nonumber \\
\left[V_{I},V_{I}^{\prime}\right]_{2} & \equiv A^{\dagger}A^{\prime}-A^{\prime\dagger}A,\label{eq:VIcomm}
\end{align}
where $\left[V_{I},V_{I}^{\prime}\right]_{1}$ contains terms involving
only one-qubit operators and $\left[V_{I},V_{I}^{\prime}\right]_{2}$
contains all two-qubit operator terms along with some additional one-qubit
operator terms, as we show below. We note in particular that $\left[V_{I},V_{I}^{\prime}\right]_{2}$
does not involve any photon operators. 

The form of the qubit-qubit interaction terms in the effective Hamiltonian,
and thus the generated two-qubit gate, is determined by $\left[V_{I},V_{I}^{\prime}\right]_{2}=A^{\dagger}A^{\prime}-A^{\prime\dagger}A.$
For $j,k=1,2,$ we can write this commutator as
\begin{align}
\left[V_{I},V_{I}^{\prime}\right]_{2} & =A^{\dagger}A^{\prime}-A^{\prime\dagger}A\nonumber \\
 & =\sum_{j,k}\left(A_{j}^{\dagger}A_{k}^{\prime}-A_{j}^{\prime\dagger}A_{k}\right)\nonumber \\
 & =\sum_{j}\left(A_{j}^{\dagger}A_{j}^{\prime}-A_{j}^{\prime\dagger}A_{j}\right)\nonumber \\
 & +A_{1}^{\dagger}A_{2}^{\prime}-A_{1}^{\prime\dagger}A_{2}+A_{2}^{\dagger}A_{1}^{\prime}-A_{2}^{\prime\dagger}A_{1}.\label{eq:VIcomm2}
\end{align}
The terms $A_{j}^{\dagger}A_{j}^{\prime}-A_{j}^{\prime\dagger}A_{j}$
in $\left[V_{I},V_{I}^{\prime}\right]_{2}$ lead to one-qubit terms
in $H_{{\rm eff}}\left(\tau\right),$ while the qubit-qubit interaction
terms arise entirely from the last line of Eq.~(\ref{eq:VIcomm2}).
In Appendix \ref{sec:Onequbitterms}, we show that for $\Delta_{j},W_{j}\neq0$
and $W_{j}\neq\left|\Delta_{j}\right|,\left|2\Delta_{j}\right|,$
the complete one-qubit contribution to $H_{{\rm eff}}\left(\tau\right)$
appearing at second order and originating from $\left[V_{I},V_{I}^{\prime}\right]_{1}$
and $\left[V_{I},V_{I}^{\prime}\right]_{2}$ reduces to 
\begin{align}
\Lambda & =-\sum_{j}g_{j}^{2}\left[\frac{\delta_{j}^{2}+2\Delta_{j}\delta_{j}+W_{j}^{2}}{2W_{j}\left(\Delta_{j}^{2}-W_{j}^{2}\right)}\right]\sigma_{j}^{z}\left(a^{\dagger}a+\frac{1}{2}\right).\label{eq:Lambda}
\end{align}
These terms represent parametric driving-induced dispersive shifts
that can be tuned via the drive amplitudes and frequencies, and are
well defined in the absence of decay provided $W_{j}\neq\left|\Delta_{j}\right|.$
Such shifts can be harnessed for drive-tunable qubit measurement \cite{Noh2023}.
Two specific cases of interest are $\delta_{j}=0$ and $\delta_{j}\neq0,$
corresponding to resonant and off-resonant driving of the qubits,
respectively. For $\delta_{j}=0$ and $W_{j}\neq0,$ we find from
Eq.~(\ref{eq:Lambda}) that $\Lambda\neq0$ and the dispersive shift
terms persist. In this case, the qubit frequencies are $W_{j}=2\Omega_{j}$
and the dispersive shift terms in Eq.~(\ref{eq:Lambda}) become 
\begin{align}
\emph{\ensuremath{\Lambda_{r}}} & =\sum_{j}\chi_{j}\sigma_{j}^{z}\left(a^{\dagger}a+\frac{1}{2}\right),\nonumber \\
\chi_{j} & \equiv-\frac{g_{j}^{2}\Omega_{j}}{\Delta_{j}^{2}-4\Omega_{j}^{2}}.\label{eq:Lambdar}
\end{align}
As we show in Appendix \ref{sec:Elimdispshiftdyn} for multiple example
cases, the effects of $\emph{\ensuremath{\Lambda_{r}}}$ on the dynamics
for $\delta_{j}=0$ can effectively be eliminated in specific situations
of interest by an appropriate choice of parameters and operation times.
On the other hand, for $\delta_{j}\neq0,$ we can choose $\Delta_{j}$
and $W_{j}$ such that $\Lambda=0$ (see Appendix \ref{sec:Elimdispshiftdyn}
for further details). The effective Hamiltonian $H_{{\rm eff}}\left(\tau\right)$
then consists purely of qubit-qubit interaction terms, which arise
from the terms with $j\neq k$ in Eq.~(\ref{eq:VIcomm2}). 

\begin{table*}
\caption{\label{tab:Vqqresterms}Resonance conditions, corresponding qubit-qubit
interaction terms $V_{qq}\equiv V_{qq}\left(\tau\right)$ {[}Eq.~(\ref{eq:Vqq}){]}
in the second-order effective Hamiltonian {[}Eq.~(\ref{eq:HeffH2avg}){]},
and generated universal entangling gates for resonant driving of the
qubits ($\delta_{j}=0$ for $j=1,2$), such that $W_{j}=2\Omega_{j}.$
The coupling strength for each interaction is given by $\mathcal{J}=g_{1}g_{2}/4\Delta,$
where for notational simplicity we use $\Delta$ to represent the
distinct resonant detuning $\Delta_{\left(R\right)}$ for each resonance
condition $R,$ i.e., $\Delta\equiv\Delta_{\left(R\right)}$ where
$\Delta_{\left(R\right)}$ is defined in the second column {[}$\Delta_{\left(1\right)}\equiv\Delta_{1}=\Delta_{2},$
$\Delta_{\left(2\right)}\equiv\Delta_{1}=\Delta_{2}^{+},\ldots${]}
with $\Delta_{j}^{\pm}\equiv\Delta_{j}\pm W_{j}.$ See Appendix \ref{sec:Qubitqubitintterms}
for details. }

\begin{ruledtabular}
\begin{centering}
\begin{tabular}{ccccc}
$R$ & \textcolor{black}{Resonance condition }{[}$\Delta\equiv\Delta_{\left(R\right)}${]} & \textcolor{black}{Constraints} & \textcolor{black}{Interaction $V_{qq}$ ($\delta_{1}=\delta_{2}=0$)} & Entangling gate\tabularnewline
\hline 
\textcolor{black}{1} & \textcolor{black}{$\Delta_{1}=\Delta_{2}$} & \textcolor{black}{{} $W_{1}\neq\pm W_{2}$} & $-2\mathcal{J}\sigma_{1}^{z}\sigma_{2}^{z}$  & Controlled-phase ($U_{\varphi}$)\tabularnewline
\textcolor{black}{2} & \textcolor{black}{$\Delta_{1}=\Delta_{2}^{+}$} & \textcolor{black}{$W_{1}\neq\pm W_{2},\pm2W_{2}$} & $\mathcal{J}\sigma_{1}^{z}\sigma_{2}^{x}$ & CNOT\tabularnewline
\textcolor{black}{3} & \textcolor{black}{$\Delta_{1}=\Delta_{2}^{-}$} & \textcolor{black}{$W_{1}\neq\pm W_{2},\pm2W_{2}$} & $-\mathcal{J}\sigma_{1}^{z}\sigma_{2}^{x}$ & CNOT\tabularnewline
\textcolor{black}{4} & \textcolor{black}{$\Delta_{1}^{+}=\Delta_{2}$} & \textcolor{black}{$W_{1}\neq\pm W_{2},\pm W_{2}/2$} & $\mathcal{J}\sigma_{1}^{x}\sigma_{2}^{z}$ & CNOT\tabularnewline
\textcolor{black}{5} & \textcolor{black}{$\Delta_{1}^{-}=\Delta_{2}$} & \textcolor{black}{$W_{1}\neq\pm W_{2},\pm W_{2}/2$} & $-\mathcal{J}\sigma_{1}^{x}\sigma_{2}^{z}$ & CNOT\tabularnewline
\textcolor{black}{6} & \textcolor{black}{$\Delta_{1}^{+}=\Delta_{2}^{+}$} & \textcolor{black}{$W_{1}\neq W_{2},2W_{2},W_{2}/2$} & $-\mathcal{J}\left(\sigma_{1}^{+}\sigma_{2}^{-}+\sigma_{1}^{-}\sigma_{2}^{+}\right)$ & \emph{i}SWAP ($U_{i{\rm SW}}$)\tabularnewline
\textcolor{black}{7} & \textcolor{black}{$\Delta_{1}^{-}=\Delta_{2}^{-}$} & \textcolor{black}{$W_{1}\neq W_{2},2W_{2},W_{2}/2$} & \textcolor{black}{$-\mathcal{J}\left(\sigma_{1}^{+}\sigma_{2}^{-}+\sigma_{1}^{-}\sigma_{2}^{+}\right)$} & \emph{i}SWAP ($U_{i{\rm SW}}$)\tabularnewline
\textcolor{black}{8} & \textcolor{black}{$\Delta_{1}^{+}=\Delta_{2}^{-}$} & \textcolor{black}{$W_{1}\neq-W_{2},-2W_{2},-W_{2}/2$} & \textcolor{black}{$\mathcal{J}\left(\sigma_{1}^{+}\sigma_{2}^{+}+\sigma_{1}^{-}\sigma_{2}^{-}\right)$} & Double-excitation ($U_{i{\rm DE}}$)\tabularnewline
\textcolor{black}{9} & \textcolor{black}{$\Delta_{1}^{-}=\Delta_{2}^{+}$} & \textcolor{black}{$W_{1}\neq-W_{2},-2W_{2},-W_{2}/2$} & \textcolor{black}{$\mathcal{J}\left(\sigma_{1}^{+}\sigma_{2}^{+}+\sigma_{1}^{-}\sigma_{2}^{-}\right)$} & Double-excitation ($U_{i{\rm DE}}$)\tabularnewline
\end{tabular}
\par\end{centering}
\end{ruledtabular}
\end{table*}

\section{\label{sec:sidebandgates}Drive-tailored entangling gates via sideband
resonances}

We next focus on the qubit-qubit interaction terms in $H_{{\rm eff}}\left(\tau\right)$,
which are given by 
\begin{align}
V_{qq}\left(\tau\right) & \equiv\frac{1}{2i\tau}\int_{0}^{\tau}dt\int_{0}^{t}dt^{\prime}\left(-A_{1}A_{2}^{\prime\dagger}+A_{1}^{\prime}A_{2}^{\dagger}-{\rm H.c.}\right).\label{eq:Vqq}
\end{align}
Using Eqs.~(\ref{eq:A}) and (\ref{eq:hjnutau}) in Appendix \ref{sec:Onequbitterms}
to write Eq.~(\ref{eq:Vqq}) in terms of functions $h\left(\mu_{1},\mu_{2}\right),$
$h\left(\mu_{2},\mu_{1}\right),$ and their complex conjugates, we
can identify resonance conditions $\mu_{1}=\mu_{2}$ that each give
rise to specific qubit-qubit terms in $V_{qq}\left(\tau\right).$
Since $\mu_{1}\in\left\{ \Delta_{1},\Delta_{1}+W_{1},\Delta_{1}-W_{1}\right\} $
and $\mu_{2}\in\left\{ \Delta_{2},\Delta_{2}+W_{2},\Delta_{2}-W_{2}\right\} ,$
there are nine resonance conditions. Each condition corresponds to
resonance between a center or sideband frequency of qubit 1 and a
center or sideband frequency of qubit 2. These conditions and the
corresponding qubit-qubit terms appearing in the effective Hamiltonian
$H_{{\rm eff}}\left(\tau\right)$ are derived in Appendix \ref{sec:Qubitqubitintterms}
and summarized in Table \ref{tab:Vqqresterms}, where we have defined
$\Delta_{j}^{\pm}\equiv\Delta_{j}\pm W_{j}.$ 

We now specifically consider the case of resonant qubit driving ($\delta_{j}=0$
or $\omega_{j}^{d}=\omega_{j}$), so that $W_{j}=2\Omega_{j}.$ The
Hamiltonian for this case is given by $V_{I}\left(t\right)$ {[}Eq.~(\ref{eq:VI}){]}
with $A\left(t\right)=A_{r}\left(t\right)$ as given in Eq.~(\ref{eq:Ares}).
The characteristic frequencies are therefore $\Delta_{j},\Delta_{j}+2\Omega_{j},$
and $\Delta_{j}-2\Omega_{j},$ corresponding to center, red sideband,
and blue sideband frequencies, respectively, for driven qubit $j$
(shifted with respect to the cavity frequency $\omega_{c}$). Assuming
that we choose parameters such that the effects of the drive-induced
dispersive shift terms $\emph{\ensuremath{\Lambda_{r}}}$ in Eq.~(\ref{eq:Lambdar})
can be neglected (see Appendix \ref{sec:Elimdispshiftdyn} for details),
the evolution generated by the effective Hamiltonian $H_{{\rm eff}}\left(\tau\right)$
reduces to that generated by the qubit-qubit interaction $V_{qq}\left(\tau\right)$
in Eq.~(\ref{eq:Vqq}). Thus, we consider only the dynamics generated
by $V_{qq}\left(\tau\right)$ in what follows. 

For each resonance condition, Table \ref{tab:Vqqresterms} gives the
form of the interaction $V_{qq}\equiv V_{qq}\left(\tau\right)$ for
resonant qubit driving (second-to-last column), assuming that the
dressed qubit frequencies $W_{1}$ and $W_{2}$ satisfy the associated
constraints (third column). These constraints are based on Eq.~(\ref{eq:hjnutau})
and are obtained for each resonance condition by applying the condition
$\mu_{1}\neq\mu_{2}$ to the remaining resonance conditions in Table
\ref{tab:Vqqresterms}, so that all qubit-qubit interaction terms
except for the specified interaction $V_{qq}$ vanish (see Appendix
\ref{sec:Qubitqubitintterms}). We define a corresponding ideal two-qubit
operation generated by $V_{qq}$ as $U_{m}\equiv U\left(\tau_{m}\right)=e^{-iV_{qq}\tau_{m}},$
where $\tau_{m}\equiv m\tau=2\pi m/\eta$ with $m=0,1,2,\dots$ represents
the corresponding gate time and is an integer multiple of $\tau$.
By adjusting the drive amplitudes $2\Omega_{j}$ and frequencies $\omega_{j}^{d}$
to tune to a particular resonance condition and set $W_{1}$ and $W_{2}$
appropriately, it is then possible to select desired qubit-qubit interaction
terms and two-qubit entangling gates. Examples of universal entangling
gates are given in the last column of Table \ref{tab:Vqqresterms}.
The drives can also be used to switch off a given interaction by tuning
the sidebands away from the corresponding resonance condition. We
emphasize that: (1) multiple two-qubit interaction terms exist even
with mutually off-resonant frequencies (i.e., for $j\neq k,$ $\omega_{j}\neq\omega_{k}\neq\omega_{c}$);
and (2) the cavity photon operators $a,a^{\dagger}$ do not appear
explicitly in $V_{qq},$ suggesting suppression of errors due to cavity
photon decay in the sideband-based entangling gate approach we propose
in this work. Our analysis of expected gate performance in Sec.~\ref{sec:gateperfdec}
quantitatively demonstrates the presence of this suppressed cavity
photon sensitivity. 

\begin{figure*}
\begin{centering}
\includegraphics[viewport=320bp -20bp 800bp 330bp,width=0.8\columnwidth]{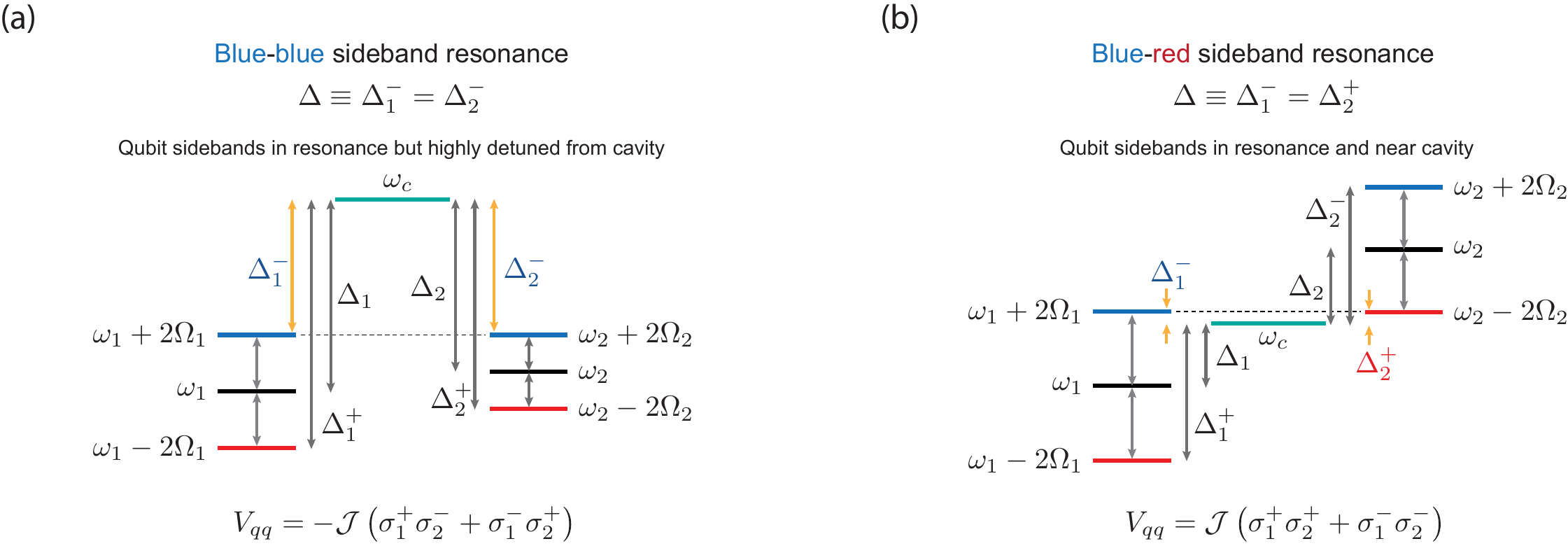}
\par\end{centering}
\caption{\label{fig:sbres} Examples of qubit-qubit interactions enabled by
cavity-mediated coupling via resonant sidebands of parametrically
driven qubits. (a) Blue-blue sideband resonance, described by $\Delta_{1}^{-}=\Delta_{2}^{-}$
(resonance condition 7 in Table \ref{tab:Vqqresterms}). The diagram
shown depicts a case in which the resonant sidebands are highly detuned
from the cavity frequency $\omega_{c}.$ (b) Blue-red sideband resonance,
described by $\Delta=\Delta_{1}^{-}=\Delta_{2}^{+}$ (resonance condition
9 in Table \ref{tab:Vqqresterms}). The diagram depicts a case in
which the resonant sidebands are near the cavity frequency $\omega_{c.}$
In both cases, the qubit frequencies $\omega_{1,2}$ and the cavity
frequency $\omega_{c}$ are mutually off-resonant, i.e., $\omega_{1}\protect\neq\omega_{2}\protect\neq\omega_{c}$. }
\end{figure*}

To more concretely illustrate the approach, we now consider the effective
qubit-qubit interaction and entangling gates generated between resonantly
driven qubits ($\delta_{j}=0,$ $W_{j}=2\Omega_{j}$) for the specific
resonance conditions 7 and 9 in Table \ref{tab:Vqqresterms}. First,
we consider $\Delta_{1}^{-}=\Delta_{2}^{-}$ (resonance condition
7), which is equivalent to $\Delta_{1}-W_{1}=\Delta_{2}-W_{2}.$ As
described in Appendix \ref{sec:Qubitqubitintterms}, the resonance
condition can also be written as $\omega_{1}+2\Omega_{1}=\omega_{2}+2\Omega_{2}$
for resonantly driven qubits and describes resonance between the blue
sidebands of both qubits {[}Fig.~\ref{fig:sbres}(a){]}. Setting
$\Delta_{j}=p_{j}\eta$ and $W_{j}=q_{j}\eta$ with $p_{j},q_{j}$
integers leads to the equivalent condition $p_{1}-q_{1}=p_{2}-q_{2}.$
We also define the integer $w\equiv p_{1}-q_{1}=p_{2}-q_{2}$ such
that $\Delta\equiv\Delta_{1}^{-}=\Delta_{2}^{-}=w\eta.$ From Table
\ref{tab:Vqqresterms}, the constraints associated with resonance
condition 7 are $W_{1}\neq W_{2},2W_{2},W_{2}/2.$ Assuming these
constraints are satisfied, the qubit-qubit interaction takes the form
$V_{qq}=-\mathcal{J}\left(\sigma_{1}^{+}\sigma_{2}^{-}+\sigma_{1}^{-}\sigma_{2}^{+}\right)$
with coupling strength $\mathcal{J}\equiv g_{1}g_{2}/4\Delta.$ 

This interaction can be used to generate the \emph{$i{\rm SWAP}$}
gate, which together with single-qubit rotations constitutes a universal
set of quantum gates \cite{Schuch2003} and also represents a key
element of recently proposed improvements to the implementation of
quantum error correction using the surface code \cite{McEwen2023arxiv}.
Since $V_{qq}$ is independent of the photon operators $a,a^{\dagger},$
$U_{m}\equiv e^{-iV_{qq}\tau_{m}}$ acts nontrivially only on the
qubits and we can work in a subspace of fixed photon number $n.$
Thus, we now project all states and operators into the subspace defined
by $P_{n}\equiv\ket{n}\bra{n}.$ For notational simplicity, we use
$U_{m}$ to denote the evolution operators within the $n$-photon
two-qubit subspace $\left\{ \ket{ee,n},\ket{eg,n},\ket{ge,n},\ket{gg,n}\right\} $
in the remainder of this work unless otherwise specified. Defining
$\Sigma_{x}\equiv\ket{eg}\bra{ge}+\ket{ge}\bra{eg}$ (where we have
suppressed the photon number state $\ket{n}$ for convenience), we
can write $V_{qq}=-\mathcal{J}\Sigma_{x}$ and $U_{m}=e^{i\mathcal{J}\tau_{m}\Sigma_{x}}.$
In the full two-qubit dressed basis $\left\{ \ket{ee},\ket{eg},\ket{ge},\ket{gg}\right\} ,$
$U_{m}$ takes the form
\begin{align}
U_{m} & =\left(\begin{array}{cccc}
1 & 0 & 0 & 0\\
0 & \cos(\mathcal{J}\text{\ensuremath{\tau}}_{m}) & i\sin(\mathcal{J}\text{\ensuremath{\tau}}_{m}) & 0\\
0 & i\sin(\mathcal{J}\text{\ensuremath{\tau}}_{m}) & \cos(\mathcal{J}\text{\ensuremath{\tau}}_{m}) & 0\\
0 & 0 & 0 & 1
\end{array}\right).\label{eq:UmRC7}
\end{align}
In order to obtain an \emph{$i{\rm SWAP}$} gate $U_{i{\rm SW}}$,
we set $\tau_{m}=m\tau=\pi/2\mathcal{J}.$ We choose the initial state
$\ket{\psi_{i}}=\ket{eg}$ for our analysis. To cancel the dynamics
due to the dispersive shift terms in Eq.~(\ref{eq:Lambdar}) for
this case, we also set $\chi_{1}=\chi_{2}.$ As shown in Appendix
\ref{sec:Elimdispshiftdyn}, both the generation of the \emph{$i{\rm SWAP}$}
gate and the drive-induced dispersive shift cancellation can be achieved
for resonance condition 7 by choosing parameters that satisfy the
constraints in Eqs. (\ref{eq:RC7constra}) and (\ref{eq:RC7constrbc}).
These relations are satisfied for multiple sets of parameters. For
the analysis in this work, we choose the set of parameters shown for
resonance condition 7 in Table \ref{tab:parameters}. The evolution
time unit in the Magnus expansion becomes $\tau=2\pi/\eta=20\ {\rm ns},$
yielding an \emph{$i{\rm SWAP}$} gate time $\tau_{m}=m\tau=800\ {\rm ns}.$
The ideal gate evolution generated by $V_{qq}$ at time $\tau_{m}$
for resonance condition 7 and these parameters approximates well the
numerical evolution directly due to Eq.~(\ref{eq:VI}) according
to the time-dependent Schrodinger equation and in the absence of qubit
and cavity decay, with a fidelity $F_{0}\approx0.998$ (we choose
the subspace with $n=0,1,2$ and the initial state $\ket{\psi_{i}}=\ket{eg,0}$
for the numerical analysis; see Appendix \ref{sec:Elimdispshiftdyn}). 

\begin{table*}
\caption{\label{tab:parameters}Parameter values used in the effective interaction
Hamiltonian and entangling gate analysis for resonance conditions
7 and 9 in Table \ref{tab:Vqqresterms}, assuming resonantly driven
qubits ($\delta_{j}\equiv\omega_{j}-\omega_{j}^{d}=0$ for $j=1,2$)
such that $W_{j}=2\Omega_{j}.$ Each set of parameters satisfies the
associated resonance condition with $W_{j}=2\Omega_{j}=q_{j}\eta,$
$\Delta_{j}\equiv\omega_{c}-\omega_{j}^{d}=p_{j}\eta,$ $\Delta_{j}^{\pm}\equiv\Delta_{j}\pm W_{j},$
$\mathcal{J}=g_{1}g_{2}/4\Delta,$ $\chi_{j}$ as given in Eq.~(\ref{eq:Lambdar}),
and the constraints in Eqs. (\ref{eq:RC7constra}) and (\ref{eq:RC7constrbc})
{[}Eqs. (\ref{eq:RC9constra}) and (\ref{eq:RC9constrbc}){]} for
resonance condition 7 (9). }

\begin{ruledtabular}
\begin{centering}
\begin{tabular}{ccc}
\textcolor{black}{Parameter } & \textcolor{black}{Resonance condition 7: $\Delta\equiv\Delta_{1}^{-}=\Delta_{2}^{-}$} & \textcolor{black}{Resonance condition 9: $\Delta\equiv\Delta_{1}^{-}=\Delta_{2}^{+}$}\tabularnewline
\hline 
$q_{1}$ & 7 & 12\tabularnewline
\textcolor{black}{$q_{2}$} & 4 & 11\tabularnewline
\textcolor{black}{$p_{1}$} & 20 & 10\tabularnewline
\textcolor{black}{$p_{2}$} & 17 & -13\tabularnewline
\textcolor{black}{$w$} & 13 & -2\tabularnewline
\textcolor{black}{$m$} & 40 & 10\tabularnewline
\textcolor{black}{$\eta/2\pi=1/\tau$} & $0.05\ {\rm GHz}$ & $0.05\ {\rm GHz}$\tabularnewline
\textcolor{black}{$g_{1}/2\pi$} & $26\ {\rm MHz}$ & $21\ {\rm MHz}$\tabularnewline
\textcolor{black}{$g_{2}/2\pi$} & $31\ {\rm MHz}$ & $23\ {\rm MHz}$\tabularnewline
\textcolor{black}{$\omega_{1}/2\pi$} & $6\ {\rm GHz}$ & $5.7\ {\rm GHz}$\tabularnewline
\textcolor{black}{$\omega_{2}/2\pi$} & $6.15\ {\rm GHz}$ & $6.85\ {\rm GHz}$\tabularnewline
\textcolor{black}{$\omega_{c}/2\pi$} & $7\ {\rm GHz}$ & $6.2\ {\rm GHz}$\tabularnewline
$W_{1}/2\pi=2\Omega_{1}/2\pi$ & $0.35\ {\rm GHz}$ & $0.6\ {\rm GHz}$\tabularnewline
$W_{2}/2\pi=2\Omega_{2}/2\pi$ & $0.2\ {\rm GHz}$ & $0.55\ {\rm GHz}$\tabularnewline
$\Delta_{1}/2\pi$ & $1\ {\rm GHz}$ & $0.5\ {\rm GHz}$\tabularnewline
$\Delta_{2}/2\pi$ & $0.85\ {\rm GHz}$ & -$0.65\ {\rm GHz}$\tabularnewline
$\Delta_{1}^{+}/2\pi$ & $1.35\ {\rm GHz}$ & $1.35\ {\rm GHz}$\tabularnewline
$\Delta_{1}^{-}/2\pi$ & $0.65\ {\rm GHz}$ & $-0.1\ {\rm GHz}$\tabularnewline
$\Delta_{2}^{+}/2\pi$ & $1.05\ {\rm GHz}$ & $-0.1\ {\rm GHz}$\tabularnewline
$\Delta_{2}^{-}/2\pi$ & $0.65\ {\rm GHz}$ & $-1.2\ {\rm GHz}$\tabularnewline
$\mathcal{J}/2\pi$ & $0.31\ {\rm MHz}$ & $-1.25\ {\rm MHz}$\tabularnewline
$\chi_{1}/2\pi$ & $-0.14\ {\rm MHz}$ & $1.25\ {\rm MHz}$\tabularnewline
$\chi_{2}/2\pi$ & $-0.14\ {\rm MHz}$ & $-1.25\ {\rm MHz}$\tabularnewline
\end{tabular}
\par\end{centering}
\end{ruledtabular}
\end{table*}

Resonance condition 9 in Table \ref{tab:Vqqresterms} is given by
$\text{\ensuremath{\Delta\equiv\Delta_{1}^{-}}=\ensuremath{\Delta_{2}^{+}}}.$
For resonantly driven qubits, this condition can also be written as
$\omega_{1}+2\Omega_{1}=\omega_{2}-2\Omega_{2}$ (see Appendix \ref{sec:Qubitqubitintterms})
and describes resonance between the blue sideband of qubit 1 and the
red sideband of qubit 2 {[}Fig.~\ref{fig:sbres}(b){]}. For $\Delta_{j}=p_{j}\eta$
and $W_{j}=q_{j}\eta,$ the resonance condition can also be expressed
as $p_{1}-q_{1}=p_{2}+q_{2}.$ Accordingly, we now define $w\equiv p_{1}-q_{1}=p_{2}+q_{2}$
such that $\Delta\equiv\Delta_{1}^{-}=\Delta_{2}^{+}=w\eta.$ Assuming
the constraints \textcolor{black}{$W_{1}\neq-W_{2},-2W_{2},-W_{2}/2$}
associated with resonance condition 9 are satisfied (note that these
constraints are always satisfied for $W_{1,2}>0$), the qubit-qubit
interaction is $V_{qq}=\mathcal{J}\left(\sigma_{1}^{+}\sigma_{2}^{+}+\sigma_{1}^{-}\sigma_{2}^{-}\right)$
with coupling strength $\mathcal{J}\equiv g_{1}g_{2}/4\Delta.$ 

We again note that $V_{qq}$ is independent of the photon operators
$a,a^{\dagger}$ and generates the gate $U_{m}\equiv e^{-iV_{qq}\tau_{m}}$
within the $n$-photon subspace at time $\tau_{m}=m\tau.$ In terms
of $\Sigma_{x}^{\prime}\equiv\ket{ee}\bra{gg}+\ket{gg}\bra{ee},$
we find $V_{qq}=\mathcal{J}\Sigma_{x}$ and $U_{m}=e^{-i\mathcal{J}\tau_{m}\Sigma_{x}},$
yielding 
\begin{align}
U_{m} & =\left(\begin{array}{cccc}
\cos(\mathcal{J}\text{\ensuremath{\tau}}_{m}) & 0 & 0 & -i\sin(\mathcal{J}\text{\ensuremath{\tau}}_{m})\\
0 & 1 & 0 & 0\\
0 & 0 & 1 & 0\\
-i\sin(\mathcal{J}\text{\ensuremath{\tau}}_{m}) & 0 & 0 & \cos(\mathcal{J}\text{\ensuremath{\tau}}_{m})
\end{array}\right)\label{eq:UmRC9}
\end{align}
in the full two-qubit dressed basis. The gate at $\tau_{m}=m\tau=-\pi/2\mathcal{J}$
is analogous to an \emph{$i{\rm SWAP}$} gate but acts in the subspace
spanned by $\left\{ \ket{ee},\ket{gg}\right\} .$ We denote this gate,
which we refer to as the double-excitation gate, by $U_{i{\rm DE}}.$
As the gate is related to $U_{i{\rm SW}}$ via a rotation of qubit
2, $U_{i{\rm DE}}$ together with single-qubit rotations also constitutes
a universal set of quantum gates. For our analysis, we choose the
initial state $\ket{\psi_{i}}=\ket{ee}$ and also set $\chi_{1}=-\chi_{2}$
to cancel the dynamics due to the dispersive shift terms in Eq.~(\ref{eq:Lambdar})
(Appendix \ref{sec:Elimdispshiftdyn}). Simultaneous generation of
the gate $U_{i{\rm DE}}$ and cancellation of the drive-induced dispersive
shifts for resonance condition 9 is possible by choosing parameters
that satisfy the constraints in Eqs. (\ref{eq:RC9constra}) and (\ref{eq:RC9constrbc}).
As in the case of resonance condition 7, these relations are satisfied
for multiple sets of parameters. Here, we choose the parameters specified
in Table \ref{tab:parameters} for resonance condition 9. The evolution
time unit in the Magnus expansion is again $\tau=2\pi/\eta=20\ {\rm ns},$
yielding a gate time $\tau_{m}=m\tau=200\ {\rm ns}$ for $U_{i{\rm DE}}.$
Comparison of the ideal gate evolution generated by $V_{qq}$ at time
$\tau_{m}$ for resonance condition 9 and these parameters with the
numerical evolution directly due to Eq.~(\ref{eq:VI}) according
to the time-dependent Schrodinger equation and in the absence of qubit
and cavity decay again yields a fidelity $F_{0}\approx0.998$ (as
described in Appendix \ref{sec:Elimdispshiftdyn}, we again choose
the subspace with $n=0,1,2$ along with the initial state $\ket{\psi_{i}}=\ket{ee,0}$
for the numerical analysis). 

We therefore find that, for both resonance conditions 7 and 9 and
appropriately selected parameters, the dynamics due to $V_{I}$ with
cavity photon operators explicitly included are well approximated
by the dynamics generated by just the two-qubit interaction $V_{qq},$
from which cavity photon operators are absent. In the next section,
we show that this absence of explicit cavity photon dependence in
the effective Hamiltonian is manifested in the full dynamics as suppressed
sensitivity of these sideband-based entangling gates to cavity photon
decay. 

\begin{figure*}
\begin{centering}
\includegraphics[viewport=450bp 0bp 860bp 400bp,width=0.6\columnwidth]{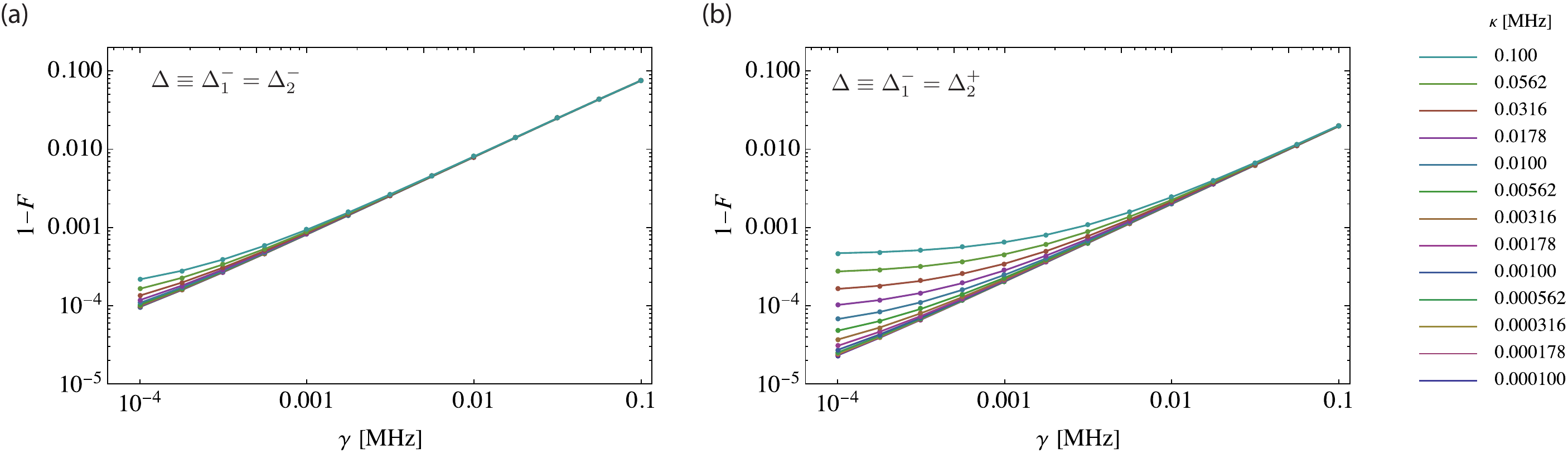}
\par\end{centering}
\caption{\label{fig:Fvsgamkap} Error $1-F$ in entangling gates of duration
$\tau_{m}$ generated by sideband-based cavity-mediated coupling of
parametrically driven qubits, calculated via numerical solution of
the master equation in Eq.~(\ref{eq:meqintpic}) as a function of
the qubit decay rate $\gamma$ and cavity photon decay rate $\kappa.$
(a) Error in \emph{$i{\rm SWAP}$} gate $U_{i{\rm SW}},$ generated
by a blue-blue sideband resonance {[}Fig.~\ref{fig:sbres}(a){]}
with the initial state $\ket{\psi_{i}}=\ket{eg,0}$ and parameters
for resonance condition 7 given in Table \ref{tab:parameters}. (b)
Error in double excitation gate $U_{i{\rm DE}},$ generated by a blue-red
sideband resonance {[}Fig.~\ref{fig:sbres}(b){]} with the initial
state $\ket{\psi_{i}}=\ket{ee,0}$ and parameters for resonance condition
9 given in Table \ref{tab:parameters}. Lines are guides for the eye. }
\end{figure*}

\section{\label{sec:gateperfdec}Sideband gate performance in the presence
of Qubit and cavity decay}

To evaluate the performance of the sideband resonance-based gates
$U_{i{\rm SW}}$ and $U_{i{\rm DE}}$ and quantitatively illustrate
the reduced dependence of the dynamics on cavity photons, we use a
master equation analysis and numerically calculate the gate fidelity
for resonance conditions 7 and 9 in the presence of qubit and cavity
decay. Here, we assume that dephasing in the original qubit basis
with rate $\gamma_{j}$ for qubit $j$ and cavity photon loss with
rate $\kappa$ are the dominant sources of decoherence, as is relevant
for silicon quantum dots \cite{Srinivasa2016}. We write the master
equation as \cite{Blais2007,Srinivasa2016}
\begin{align}
\dot{\rho} & =-i\left[H_{p},\rho\right]+\sum_{j}\frac{\gamma_{j}}{2}\left(\sigma^{z}\rho\sigma^{z}-\rho\right)\nonumber \\
 & +\frac{\kappa}{2}\left(2a\rho a^{\dagger}-a^{\dagger}a\rho-\rho a^{\dagger}a\right),\label{eq:mastereq}
\end{align}
with $H_{p}$ given by Eq.~(\ref{eq:Hp}). Following steps similar
to those used to obtain the interaction-picture Hamiltonian $V_{I}$
{[}Eq.~(\ref{eq:VI}){]}, we transform the master equation in Eq.~(\ref{eq:mastereq})
to the interaction picture. Moving to a rotating frame via $U_{1}$
{[}Eq.~(\ref{eq:U1}){]}, making a rotating wave approximation for
$\left|\Delta_{j}\right|\ll\omega_{c}+\omega_{j}^{d},2\omega_{j}^{d},$
choosing $\phi_{j}=0,$ applying $U_{q}$ {[}Eq.~(\ref{eq:Uq}){]}
to change to the dressed qubit basis, and moving to the interaction
picture via $U_{2}$ yields, after setting $\delta_{j}=0$ for resonant
qubit driving and dropping rapidly oscillating terms $\sim e^{\pm2iW_{j}t},$
\begin{align}
\dot{\rho}_{I} & =-i\left[V_{I},\rho_{I}\right]+\sum_{j}\frac{\gamma_{j}}{2}\left(\sigma_{j}^{+}\rho_{I}\sigma_{j}^{-}+\sigma_{j}^{-}\rho_{I}\sigma_{j}^{+}-\rho_{I}\right)\nonumber \\
 & +\frac{\kappa}{2}\left(2a\rho_{I}a^{\dagger}-a^{\dagger}a\rho_{I}-\rho_{I}a^{\dagger}a\right),\label{eq:meqintpic}
\end{align}
where $\rho_{I}\equiv U_{2}^{\dagger}U_{q}^{\dagger}U_{1}^{\dagger}\rho U_{1}U_{q}U_{2}.$
Equation (\ref{eq:meqintpic}) is the master equation describing the
dynamics in the interaction picture. For the numerical calculations,
we again work in the photon subspace with $n=0,1,2$ and set $\gamma_{1}=\gamma_{2}\equiv\gamma$
for simplicity. To analyze the effects of qubit and cavity decay on
the performance of the entangling gates, we calculate the fidelity
\begin{equation}
F\left(\tau_{m}\right)\equiv{\rm Tr}\left[\rho_{I}^{{\rm \left(0\right)}}\left(\tau_{m}\right)\rho_{I}\left(\tau_{m}\right)\right],\label{eq:F}
\end{equation}
where $\rho_{I}\left(\tau_{m}\right)$ denotes the final state at
time $\tau_{m}$ for the evolution obtained by numerically integrating
Eq.~(\ref{eq:meqintpic}) and $\rho_{I}^{\left(0\right)}\left(\tau_{m}\right)$
denotes the final state for the ideal evolution given by $\gamma=\kappa=0.$
We calculate this fidelity as a function of $\gamma$ and $\kappa$
for the resonance conditions 7 and 9 using the parameter sets in Table
\ref{tab:parameters} and the initial states chosen above for the
ideal gates $U_{i{\rm SW}}$ and $U_{i{\rm DE}}.$ The initial state
is $\rho_{I}\left(0\right)=\ket{\psi_{i}}\bra{\psi_{i}}=\ket{eg,0}\bra{eg,0}$
for resonance condition 7 and $\rho_{I}\left(0\right)=\ket{\psi_{i}}\bra{\psi_{i}}=\ket{ee,0}\bra{ee,0}$
for resonance condition 9. 

In Fig.~\ref{fig:Fvsgamkap}, we plot the error $1-F\left(\tau_{m}\right)$
for the two resonance conditions and corresponding two-qubit entangling
gates. We find theoretical entangling gate fidelities $F>0.995$ for
the full range of $\kappa$ shown (up to $\kappa=100\ {\rm kHz}$)
and $\gamma\lesssim1\ {\rm kHz}$ ($\gamma\lesssim10\ {\rm kHz}$)
for resonance condition 7 (9). While $\gamma$ and $\kappa$ are varied
over three orders of magnitude in both cases, the error for both resonance
conditions depends more strongly on qubit decay $\gamma$ than on
cavity decay $\kappa.$ This reduced dependence on $\kappa$ is expected
for the dispersive regime, in which the cavity virtually mediates
the qubit-qubit interaction in the absence of direct qubit-photon
interaction, and was also found in the results of Ref.~\cite{Srinivasa2016}
for the dispersive regimes relative to the driven resonant regime
of cavity-mediated coupling. We note, however, that the gates derived
here are based on interactions via sideband resonances, as in the
driven resonant regime. 

For resonance condition 7, given by $\Delta_{1}^{-}=\Delta_{2}^{-}$
{[}Fig.~\ref{fig:Fvsgamkap}(a){]}, the resonant blue sidebands of
the driven qubits are highly detuned from the cavity frequency $\omega_{c}$,
with $\Delta/2\pi=\Delta_{1}^{-}/2\pi=\Delta_{2}^{-}/2\pi=0.65\ {\rm GHz}$
{[}see Fig.~\ref{fig:sbres}(a){]}. We see that the error varies
over approximately three orders of magnitude with $\gamma$ but over
less than one order of magnitude with $\kappa.$ The suppressed sensitivity
to $\kappa$ is expected given the large detuning between the qubit
sidebands and the cavity. On the other hand, for resonance condition
9, given by $\Delta_{1}^{-}=\Delta_{2}^{+}$ {[}Fig.~\ref{fig:Fvsgamkap}(b){]},
the resonant sidebands -- the blue sideband of qubit 1 and the red
sideband of qubit 2 -- are close to $\omega_{c}$. Here, $\Delta/2\pi=\Delta_{1}^{-}/2\pi=\Delta_{2}^{+}/2\pi=-0.1\ {\rm GHz}$
{[}see Fig.~\ref{fig:sbres}(b){]}. In this case, we see that the
error again varies over approximately three orders of magnitude with
$\gamma,$ but varies over less than two orders of magnitude with
$\kappa.$ The increased variation with $\kappa$ relative to resonance
condition 7 reflects the smaller detuning $\Delta$ of the resonant
qubit sidebands from the cavity. However, even in this case, we find
that the sensitivity of the error to $\kappa$ is suppressed relative
to the sideband-based two-qubit gates in the driven resonant regime
(compare Fig.~7 in Ref.~\cite{Srinivasa2016}). 

While the reduced sensitivity to cavity decay is consistent with the
dispersive regime, it also reflects the absence of explicit cavity
dependence in the effective interaction Hamiltonian $V_{qq}$ generating
the two-qubit entangling gates {[}Eq.~(\ref{eq:Vqq}) and Table \ref{tab:Vqqresterms}{]}.
We have seen (Sec.~\ref{sec:sidebandgates}) that for appropriately
chosen parameters, the gates generated by $V_{qq}$ closely approximate
the dynamics due to the full interaction-picture Hamiltonian $V_{I}$
in Eq.~(\ref{eq:meqintpic}), where $a$ and $a^{\dagger}$ are explicitly
present in general {[}see Eq. (\ref{eq:VI}){]}. Thus, the effective
Hamiltonian we derive here illustrates that tuning the parametric
drive frequencies and amplitudes with the remaining parameters set
appropriately effectively enables suppression of the sensitivity to
cavity decay. 

\section{Conclusions}

In this work, we have developed an approach for achieving long-range
interactions between a pair of driven spin qubits via cavity-mediated
coupling combined with sideband resonances. Our approach is applicable
to a variety of qubit types that can be controlled via parametric
driving, including one-electron spin qubits in double quantum dots,
three-electron RX qubits in triple quantum dots, and hole spin qubits,
and enables highly tunable qubit-qubit interactions whose nature can
be tailored via the driving fields. The interactions can also be switched
on and off using only ac control, without requiring dc tuning of the
qubits away from optimal operation points and thus allowing for improved
qubit coherence relative to resonant and standard dispersive approaches
for cavity-mediated qubit coupling. 

We note that the approach we describe here is based on the driving
of inherently nonlinear effective two-level systems (i.e., qubits),
whereas commonly used off-resonant coupling approaches for weakly
anharmonic superconducting qubits such as transmons effectively involve
driving multilevel systems \cite{Blais2021}. As transmon anharmonicities
are typically no more than a few hundred MHz, rapid and high-fidelity
gates in these systems are limited by leakage for sufficiently large
drive amplitudes and pulse bandwidths such that transitions to states
outside the qubit space can be excited, in the absence of additional
techniques that compensate for weak anharmonicity \cite{Gambetta2011,Werninghaus2021,Babu2021}.
Limits to the entangling fidelity and scalability of cross-resonance
\cite{Rigetti2010,Chow2011} and FLICFORQ \cite{Rigetti2005,Blais2007}
gates also exist due to small anharmonicities, required qubit frequency
spacing, available bandwidth for control, and spurious interaction
terms $\sim\sigma_{1}^{z}\sigma_{2}^{z}$ that cannot be eliminated
for conventional transmons \cite{Blais2021}. By contrast, spin qubit
systems such as those considered in this work are characterized by
highly nonlinear spectra in which differences of $>1\ {\rm GHz}$
in transition frequencies are routinely realized, including within
hybrid cQED systems \cite{Mi2018,Samkharadze2018,Landig2018}. Therefore,
spin qubits should in principle allow for greater flexibility in the
choice of amplitudes, frequencies, and gate times for achieving desired
high-fidelity gates via parametric driving, without requiring the
added complexity of low-anharmonicity compensation techniques. 

For our parametrically driven, sideband-based approach, we expect
that limits on the driving amplitudes and entangling rates for implementing
high-fidelity gates will instead arise primarily from the requirements
that $\Delta_{j}$ and $W_{j}$ are both integer multiples of the
same frequency $\eta=2\pi/\tau$ to eliminate the first-order interaction
in Eq.~(\ref{eq:H1avg}), which sets a lower bound on $\tau$ and
thus $\tau_{m}$ since $\eta\leq\left|\Delta_{j}\right|,W_{j},$ together
with the conditions $g_{j}\ll W_{j}$ required for the validity of
the effective Hamiltonian, the tuning of the drive amplitudes and
frequencies to a desired resonance condition and interaction (Table
\ref{tab:Vqqresterms}), and the constraints for eliminating other
interaction terms and dynamics due to the parametric drive-induced
dispersive shifts. As we have shown in this work, multiple sets of
experimentally relevant \cite{Zhang2023arxiv} parameters exist for
which these requirements can be simultaneously satisfied (see, e.g.,
Table \ref{tab:parameters}) in order to select desired and eliminate
undesired interaction terms. 

We have analyzed specific examples of sideband-based entangling gates
that include a 200-ns double-excitation gate $U_{i{\rm DE}},$ which
is generated via a blue-red sideband resonance and does not exist
for the standard dispersive regime in the absence of driving fields
\cite{Blais2004,Blais2007,Landig2019,Harvey-Collard2022}. Furthermore,
the rates of the entangling gates described in this work are set by
$\mathcal{J}\propto\Delta^{-1},$ in contrast to the typical $\sim\Delta_{j}^{-1}$
scaling for standard dispersive entangling gate rates (where $\Delta_{j}=\omega_{c}-\omega_{j}$
for resonant qubit driving). As it is possible to have $\Delta<\Delta_{j}$
for multiple sideband resonance conditions (see, e.g., Table \ref{tab:parameters}),
the corresponding entangling gates can potentially be more rapid than
those in the standard dispersive regime. As the gates do not require
sequences of multiple qubit-cavity sideband pulses, the potential
also exists for gate speed improvements relative to the sideband-based
gates in the driven resonant regime considered in prior work \cite{Srinivasa2016,Abadillo-Uriel2021}. 

Unlike the resonant and standard dispersive approaches, realizing
cavity-mediated entangling interactions via the sideband resonance
method we describe here does not rely on simultaneous mutual resonance
among multiple qubit and cavity photon frequencies. Instead, several
sideband resonance conditions are available for generating entanglement
between dressed qubits even with all qubit and cavity frequencies
mutually off-resonant, providing enhanced spectral flexibility. As
a result, the sideband resonance-based approach represents a potential
alternative to the challenging tuning required to bring single-spin
qubit frequencies into simultaneous resonance via precisely oriented
micromagnets that has been essential to spin-spin coupling demonstrations
in silicon to date \cite{Borjans2020,Harvey-Collard2022}. Together
with the suppressed sensitivity to cavity decay expected from our
analysis of example entangling gates, these features render the approach
we present in this work favorable for scaling and promising for the
implementation of modular quantum information processing with spin
qubits. 
\begin{acknowledgments}
Supported by Army Research Office Grants W911NF-15-1-0149 and W911NF-23-1-0104. 
\end{acknowledgments}

\appendix

\section{\label{sec:spinqubitHpderiv}Hamiltonian for cavity-coupled driven
spin qubits }

The Hamiltonian $H_{p}$ in Eq.~(\ref{eq:Hp}) describes parametrically
driven qubits coupled via the fundamental mode of microwave cavity
photons and forms the basis for the sideband-based cavity-mediated
entangling gates derived in this work. Here, we show how we obtain
$H_{p}$ for the two specific examples of driven spin qubits illustrated
in Fig.~\ref{fig:dotresdriv}.

\subsection{Driven one-electron spin qubits in double quantum dots}

We first consider two one-electron spin qubits in double quantum dots
(DQDs) coupled via a microwave cavity \cite{Borjans2020,Harvey-Collard2022}.
In the following analysis, we take into account only the lowest-energy
orbital level in each dot. The charge dipole of the electron in each
DQD couples to the electric field of a microwave cavity photon \cite{Childress2004},
and the spin of the electron couples to the charge via spin-orbit
coupling and/or a magnetic field gradient \cite{Srinivasa2013}. We
focus on electrons occupying the lowest-energy valley states within
silicon quantum dots, for which spin-charge coupling is achieved through
gradient fields produced by a micromagnet \cite{Benito2017,Mi2018,Samkharadze2018,Borjans2020,Harvey-Collard2022}. 

Assuming coupling to only the fundamental cavity photon mode with
frequency $\omega_{c},$ we write the system Hamiltonian including
parametrically driven detuning as 
\begin{align}
H_{s} & =\omega_{c}a^{\dagger}a+H_{d}+\sum_{j=1,2}g_{c,j}\tau_{j}^{z}\left(a+a^{\dagger}\right),\label{eq:Hs}\\
H_{d} & \equiv\frac{1}{2}\sum_{j=1,2}\left[\epsilon_{j}\left(t\right)\tau_{j}^{z}+2t_{j}\tau_{j}^{x}+B_{j}^{z}s_{j}^{z}+B_{j}^{x}\tau_{j}^{z}s_{j}^{x}\right],\label{eq:Hd}
\end{align}
where $\tau_{j}^{z}\equiv\ket{L}_{j}\bra{L}-\ket{R}_{j}\bra{R}$ with
$\ket{L}_{j}$ and $\ket{R}_{j}$ the lowest-energy orbital in the
left and right dots of DQD $j,$ respectively, $s_{j}^{z}\equiv\ket{\uparrow}_{j}\bra{\uparrow}-\ket{\downarrow}_{j}\bra{\downarrow}$
is the Pauli $z$ operator for the electron spin in DQD $j,$ and
the other Pauli orbital (charge) and spin operators are defined accordingly.
The remaining parameters in Eqs. (\ref{eq:Hs}) and (\ref{eq:Hd})
are the detuning $\epsilon_{j}$ between the orbital levels in the
left and right dot, the interdot tunnel coupling $2t_{j},$ the Zeeman
splittings $B_{j}^{z}$ and $B_{j}^{x}$ due to the net magnetic field
components along the $z$ and $x$ axes, respectively (due to both
the external and micromagnet fields \cite{Mi2018}), and the charge-cavity
coupling strength $g_{c,j}$ for DQD $j.$ 

We describe sinusoidal (ac) driving of the detuning for each DQD via
Eq.~(\ref{eq:detdrive}) and choose the ``sweet spot'' detuning
operation points $\epsilon_{0,j}=0$ for $j=1,2$ where the first
derivative of the charge qubit frequency vanishes, enabling leading-order
protection from charge noise \cite{Petersson2010,Benito2017,Benito2019,Croot2020}.
As in the main text, all sums are over the qubit index $j=1,2$ unless
otherwise noted. We first apply the rotation 
\begin{equation}
U_{c}=e^{-i\left(\pi/4\right)\sum_{j}\tau_{j}^{y}}\label{eq:Uc}
\end{equation}
to the charge subspace. The system Hamiltonian $H_{s}$ becomes
\begin{align}
H_{s}^{\prime} & =U_{c}^{\dagger}H_{s}U_{c}\nonumber \\
 & =\omega_{c}a^{\dagger}a+H_{d}^{\prime}-\sum_{j}g_{c,j}\tau_{j}^{x}\left(a+a^{\dagger}\right)\nonumber \\
 & -\sum_{j}\mathcal{F}_{j}\cos\left(\omega_{j}^{d}t+\phi_{j}^{\prime}\right)\tau_{j}^{x},\label{eq:Hspr}\\
H_{d}^{\prime} & =\frac{1}{2}\sum_{j=1,2}\left(2t_{j}\tau_{j}^{z}+B_{j}^{z}s_{j}^{z}-B_{j}^{x}\tau_{j}^{x}s_{j}^{x}\right).\label{eq:Hdpr}
\end{align}
Writing the transformed DQD Hamiltonian $H_{d}^{\prime}$ in the rotated
charge-spin product basis $\left\{ \ket{+,\uparrow}_{j},\ket{-,\downarrow}_{j},\ket{+,\downarrow}_{j},\ket{-,\uparrow}_{j}\right\} ,$
where $\ket{\pm}_{j}=\left(\ket{L}_{j}\pm\ket{R}_{j}\right)/\sqrt{2}$
are the double-dot charge eigenstates for $\epsilon_{0,j}=0,$ reveals
a block-diagonal structure with two decoupled subspaces that we label
as $\mathcal{H}_{a,j}$ and $\mathcal{H}_{b,j}$ and that are spanned
by $\left\{ \ket{+,\uparrow}_{j},\ket{-,\downarrow}_{j}\right\} $
and $\left\{ \ket{+,\downarrow}_{j},\ket{-,\uparrow}_{j}\right\} ,$
respectively \cite{Benito2017,Srinivasa2013}. 

Full diagonalization of the DQD low-energy space including spin for
$\epsilon_{0,j}=0$ is then achieved by applying 
\begin{equation}
U_{d}=e^{i\sum_{j}\left(\Phi_{a,j}\hat{\alpha}_{j}^{y}+\Phi_{b,j}\hat{\beta}_{j}^{y}\right)/2},\label{eq:Ud}
\end{equation}
where $\hat{\alpha}_{j}^{y}\equiv-i\left(\ket{+,\uparrow}_{j}\bra{-,\downarrow}-\ket{-,\downarrow}_{j}\bra{+,\uparrow}\right),$
$\hat{\beta}_{j}^{y}\equiv-i\left(\ket{+,\downarrow}_{j}\bra{-,\uparrow}-\ket{-,\uparrow}_{j}\bra{+,\downarrow}\right),$
and $\tan\Phi_{a\left(b\right),j}=B_{j}^{x}/\left(2t_{j}\pm B_{j}^{z}\right).$
For $\left|2t_{j}-B_{j}^{z}\right|\ll2t_{j}+B_{j}^{z},$ $\tan\Phi_{a,j}\ll\tan\Phi_{b,j}$
and the degree of mixing between $\ket{-,\downarrow}_{j}$ and $\ket{+,\uparrow}_{j}$
is much smaller than that between $\ket{-,\uparrow}_{j}$ and $\ket{+,\downarrow}_{j}.$
The eigenstates in the subspace $\mathcal{H}_{a,j}$ can then be approximated
as \cite{Benito2017} $\ket{0}_{j}\approx\ket{-,\downarrow}_{j}$
and $\ket{3}_{j}\approx\ket{+,\uparrow}_{j},$ and the corresponding
eigenvalues (for $\hbar=1$) are $\omega_{0,j}=-\mathcal{W}_{j}/2$
and $\omega_{3,j}=\mathcal{W}_{j}/2$ with $\mathcal{W}_{j}\equiv\sqrt{\left(2t_{j}+B_{j}^{z}\right)^{2}+\left(B_{j}^{x}\right)^{2}}.$
Setting $\Phi_{j}\equiv\Phi_{b,j}$ for notational convenience, the
eigenstates in the subspace $\mathcal{H}_{b,j}$ are given by
\begin{align}
\ket{1}_{j} & \equiv\sin\left(\frac{\Phi_{j}}{2}\right)\ket{+,\downarrow}_{j}+\cos\left(\frac{\Phi_{j}}{2}\right)\ket{-,\uparrow}_{j},\label{eq:1j}\\
\ket{2}_{j} & \equiv\cos\left(\frac{\Phi_{j}}{2}\right)\ket{+,\downarrow}_{j}-\sin\left(\frac{\Phi_{j}}{2}\right)\ket{-,\uparrow}_{j}\label{eq:2j}
\end{align}
and the corresponding eigenvalues are $\omega_{1,j}=-\mathcal{V}_{j}/2$
and $\omega_{2,j}=\mathcal{V}_{j}/2$ with $\mathcal{V}_{j}\equiv\sqrt{\left(2t_{j}-B_{j}^{z}\right)^{2}+\left(B_{j}^{x}\right)^{2}}.$
Defining the operators $\sigma_{j}^{kl}\equiv\ket{k}_{j}\bra{l},$
we find in the DQD eigenstate basis
\begin{align}
\tilde{H}_{s} & =U_{d}^{\dagger}H_{s}^{\prime}U_{d}\nonumber \\
 & =\omega_{c}a^{\dagger}a+\sum_{j}\sum_{k=0}^{3}\omega_{k,j}\sigma_{j}^{kk}\nonumber \\
 & -\sum_{j}\tilde{d}_{j}\left[\mathcal{F}_{j}\cos\left(\omega_{j}^{d}t+\phi_{j}^{\prime}\right)+g_{c,j}\left(a+a^{\dagger}\right)\right],\label{eq:Hstilde}
\end{align}
where we have defined 
\begin{align}
\tilde{d}_{j} & \equiv U_{d}^{\dagger}\tau_{j}^{x}U_{d}\nonumber \\
 & =d_{j}^{01}\sigma_{j}^{01}+d_{j}^{02}\sigma_{j}^{02}+d_{j}^{13}\sigma_{j}^{13}+d_{j}^{23}\sigma_{j}^{23}+{\rm H.c.},\label{eq:dipjtilde}
\end{align}
with $d_{j}^{01}=-d_{j}^{23}\approx\sin\left(\Phi_{j}/2\right)$ and
$d_{j}^{02}=d_{j}^{13}\approx\cos\left(\Phi_{j}/2\right)$ for $\left|2t_{j}-B_{j}^{z}\right|\ll2t_{j}+B_{j}^{z}.$

The Hamiltonian in Eq.~(\ref{eq:Hstilde}) describes each DQD in
terms of a multilevel picture in which the ground state of the electron
is $\ket{0}_{j}\approx\ket{-,\downarrow}_{j}$ and the highest excited
state is $\ket{3}_{j}\approx\ket{+,\uparrow}_{j},$ while the dominant
charge-spin character of $\ket{1}_{j}$ and $\ket{2}_{j}$ for $\epsilon_{0,j}=0$
depends on the relative magnitudes of $2t_{j}$ and $B_{j}^{z}$ \cite{Benito2017}.
To reduce the DQD description to an approximate two-level picture
corresponding to a spin qubit, we choose the case $2t_{j}>B_{j}^{z}$
(illustrated in Fig.~2(a) of Ref.~\cite{Benito2017}) such that
$\ket{1}_{j}\approx\ket{-,\uparrow}_{j}$ is the first excited state
and $\ket{2}_{j}\approx\ket{+,\downarrow}_{j}$ is the second excited
state for each DQD. We also assume $\omega_{1,j}-\omega_{0,j}\ll\omega_{2,j}-\omega_{1,j},$
which is equivalent to $\left(\mathcal{W}_{j}-\mathcal{V}_{j}\right)/2\mathcal{V}_{j}\ll1,$
as well as $\mathcal{F}_{j}/\mathcal{V}_{j},g_{c,j}/\mathcal{V}_{j}\ll1.$
To first order in these small parameters, Eq.~(\ref{eq:Hstilde})
can be written in the form
\begin{align}
H_{p}^{\left(1\right)} & \equiv\omega_{c}a^{\dagger}a+\sum_{j}\frac{\omega_{j}}{2}\sigma_{j}^{z}+\sum_{j}g_{j}\sigma_{j}^{x}\left(a+a^{\dagger}\right)\nonumber \\
 & +\sum_{j}2\Omega_{j}\cos\left(\omega_{j}^{d}t+\phi_{j}\right)\sigma_{j}^{x},\label{eq:Hp1}
\end{align}
where we have defined $\sigma_{j}^{z}\equiv\ket{1}_{j}\bra{1}-\ket{0}_{j}\bra{0}$
and the effective spin qubit frequencies $\omega_{j}\equiv\omega_{1,j}-\omega_{0,j}=\left(\mathcal{W}_{j}-\mathcal{V}_{j}\right)/2,$
$2\Omega_{j}\equiv\mathcal{F}_{j}\left|\sin\left(\Phi_{j}/2\right)\right|$
is the effective amplitude of the drive acting on the spin qubit,
$g_{j}\equiv g_{c,j}\left|\sin\left(\Phi_{j}/2\right)\right|$ is
the effective spin-photon coupling strength, and the sign of $\sin\left(\Phi_{j}/2\right)$
has been taken into account by making the replacement $\phi_{j}^{\prime}\rightarrow\phi_{j}$
and redefining the phase of $a$ accordingly in Eq.~(\ref{eq:Hstilde}).
The Hamiltonian $H_{p}^{\left(1\right)}$ is identical in form to
Eq.~(\ref{eq:Hp}). Finally, we note that the Hamiltonian for DQD
charge qubits \cite{Petersson2010} at the operation points $\epsilon_{0,j}=0$
{[}described by setting $B_{j}^{z}=B_{j}^{x}=0$ in Eq.~(\ref{eq:Hd}){]}
can also be written in a form analogous to Eq.~(\ref{eq:Hp1}), with
$\sigma_{j}^{z,x}\rightarrow\tau_{j}^{z,x},$ $\omega_{j}\rightarrow2t_{j},$
$2\Omega_{j}\rightarrow\mathcal{F}_{j},$ and $g_{j}\rightarrow g_{c,j}.$ 

\subsection{Three-electron resonant exchange qubits in triple quantum dots }

We now consider two resonant exchange (RX) qubits, each defined by
the spin states of three electrons in a triple quantum dot \cite{Taylor2013,Medford2013},
that interact via a microwave cavity \cite{Srinivasa2016}. In contrast
to one-electron spin qubits, each RX qubit couples directly to the
electric field of microwave cavity photons via an intrinsic electric
dipole moment. This dipole moment arises from the admixture of polarized
charge states in the qubit states, without requiring spin-orbit coupling,
magnetic gradients, or the fabrication of additional device elements
such as micromagnets. The Hamiltonian given in Eq.~(29) of Ref.~\cite{Srinivasa2016}
that is used to describe RX qubits coupled to a microwave cavity in
the driven resonant regime has the same general form as Eq.~(\ref{eq:Hp}).
Here, we briefly summarize the theory used to derive this Hamiltonian
for the case of RX qubits and adapt it to the case of silicon triple
quantum dots. 

An effective Hamiltonian for each RX qubit can be obtained from a
Hubbard model for electrons confined within a linear triple quantum
dot \cite{Taylor2013,Srinivasa2016}. This model can be used to calculate
a charge stability diagram (Fig.~1(b) of Ref.~\cite{Taylor2013})
that describes the triple dot in terms of the lowest-energy charge
configuration $\left(n_{1},n_{2},n_{3}\right)$ (where $n_{i}$ denotes
the occupation number for dot $i$ and the occupation numbers are
ordered from the left dot to the right dot) as a function of the gate
voltages applied to the three dots, which set the on-site energies
$\left(-\varepsilon_{1},-\varepsilon_{2},-\varepsilon_{3}\right).$
For fixed $V_{{\rm tot}}\equiv\sum_{i}\varepsilon_{i},$ the lowest-energy
charge configuration depends on both the detuning $\epsilon\equiv\left(\varepsilon_{3}-\varepsilon_{1}\right)/2$
and the relative middle dot on-site energy $V_{m}\equiv-\varepsilon_{2}+\frac{1}{2}\left(\varepsilon_{1}+\varepsilon_{3}\right).$
Distinct charge configurations are coupled via the left-center and
center-right interdot tunneling amplitudes $t_{l}$ and $t_{r},$
respectively. 

References \cite{Taylor2013} and \cite{Srinivasa2016} consider a
three-electron system in the subspace of the charge configurations
$\left(1,1,1\right),$ $\left(2,0,1\right),$ and $\left(1,0,2\right).$
The relevant operation point is illustrated in Fig.~1(b) of Ref.~\cite{Taylor2013}.
Here, we focus on a silicon triple dot and assume that a sufficiently
large (${\rm \gtrsim100\ mT}$ \cite{Medford2013}) static magnetic
field is applied to the triple dot. Furthermore, we assume that excited
valley states are well-separated in energy from the lowest-energy
valley manifold (by a valley splitting energy $E_{{\rm V}}\gtrsim100\ \mu{\rm eV}$
\cite{Zwanenburg2013,Yang2013,Hollmann2020,McJunkin2022}), such that
a single-orbital picture is valid. We can then define a spin qubit
using three-electron states in the lower-energy $S=1/2$ spin subspace,
which have a spin quantum number for the total $z$ component $m_{s}=-1/2$
due to the positive electron g-factor of silicon. The relevant subspace
is spanned by the $\left(1,1,1\right)$ states 

\begin{eqnarray}
\ket{e_{0}} & \equiv & \ket{s}_{13}\ket{\downarrow}_{2}\nonumber \\
 & = & \frac{1}{\sqrt{2}}\left(c_{1\uparrow}^{\dagger}c_{2\downarrow}^{\dagger}c_{3\downarrow}^{\dagger}-c_{1\downarrow}^{\dagger}c_{2\downarrow}^{\dagger}c_{3\uparrow}^{\dagger}\right)\ket{\mathrm{vac}},\label{eq:state1}\\
\ket{g_{0}} & \equiv & \sqrt{\frac{1}{3}}\ket{t_{0}}_{13}\ket{\downarrow}_{2}-\sqrt{\frac{2}{3}}\ket{t_{-}}_{13}\ket{\uparrow}_{2}\nonumber \\
 & = & \frac{1}{\sqrt{6}}\left(c_{1\uparrow}^{\dagger}c_{2\downarrow}^{\dagger}c_{3\downarrow}^{\dagger}+c_{1\downarrow}^{\dagger}c_{2\downarrow}^{\dagger}c_{3\uparrow}^{\dagger}-2c_{1\downarrow}^{\dagger}c_{2\uparrow}^{\dagger}c_{3\downarrow}^{\dagger}\right)\ket{\mathrm{vac}},\nonumber \\
\label{eq:state0}
\end{eqnarray}
together with the $\left(2,0,1\right)$ and $\left(1,0,2\right)$
states
\begin{eqnarray}
\ket{s_{1,-1/2}} & \equiv & \ket{s}_{11}\ket{\downarrow}_{3}=c_{1\uparrow}^{\dagger}c_{1\downarrow}^{\dagger}c_{3\downarrow}^{\dagger}\ket{\mathrm{vac}},\label{eq:stateS11}\\
\ket{s_{3,-1/2}} & \equiv & \ket{\downarrow}_{1}\ket{s}_{33}=c_{1\downarrow}^{\dagger}c_{3\uparrow}^{\dagger}c_{3\downarrow}^{\dagger}\ket{\mathrm{vac}},\label{eq:stateS33}
\end{eqnarray}
where $\ket{\rm vac}$ denotes the vacuum. In the basis $\left\{ \ket{e_{0}},\ket{g_{0}},\ket{s_{1,-1/2}},\ket{s_{3,-1/2}}\right\} ,$
the Hubbard Hamiltonian matrix has a form identical to that given
in Eq.~(S7) of Ref.~\cite{Taylor2013} for the case $m_{s}=1/2.$

The resonant exchange (RX) qubit is defined within an effective $\left(1,1,1\right)$
subspace $\left\{ \tilde{\ket{e_{0}}},\tilde{\ket{g_{0}}}\right\} $
obtained by perturbatively eliminating the $\left(2,0,1\right)$ and
$\left(1,0,2\right)$ states via a Schrieffer-Wolff transformation
\cite{Taylor2013,Srinivasa2016}. The resulting effective Hamiltonian
can be written in the form
\begin{equation}
H_{{\rm eff}}^{\left(3\right)}=\frac{J}{2}\tilde{\sigma}^{z}-\frac{\sqrt{3}}{2}j\tilde{\sigma}^{x},\label{eq:Heff3}
\end{equation}
where $\tilde{\sigma}^{z}\equiv\tilde{\ket{e_{0}}}\tilde{\bra{e_{0}}}-\tilde{\ket{g_{0}}}\tilde{\bra{g_{0}}},$
$J\equiv\left(J_{l}+J_{r}\right)/2,$ $j\equiv\left(J_{l}-J_{r}\right)/2,$
and the exchange between the center and left (right) dots is $J_{l}=t_{l}^{2}/\left(\Delta+\epsilon\right)$
$\left[J_{r}=t_{r}^{2}/\left(\Delta-\epsilon\right)\right].$ Here,
$\Delta$ is defined in terms of Hubbard model parameters and $V_{m}$
\cite{Taylor2013,Srinivasa2016}. Diagonalizing $H_{{\rm eff}}^{\left(3\right)}$
yields $H_{0}=\omega\sigma^{z}/2$ with $\sigma^{z}\equiv\left|1\right\rangle \left\langle 1\right|-\left|0\right\rangle \left\langle 0\right|,$
where the eigenstates $\ket{0}$ and $\ket{1}$ define the RX qubit
and $\omega\equiv\sqrt{J^{2}+3j^{2}}$ is the qubit energy splitting
(here, we set $\hbar=1$). As the exchange interactions $J_{l}$ and
$J_{r}$ between dots are controlled using only voltages applied to
the triple quantum dot, the RX qubit is a spin qubit that can be fully
manipulated using electric fields alone \cite{DiVincenzo2000Nature,Medford2013NNano,Taylor2013,Medford2013}. 

In addition to being fully controllable via dc electric field pulses,
the RX qubit couples directly to microwave photons by virtue of the
small admixture of the polarized charge states $(2,0,1)$ and $(1,0,2)$
in the logical qubit states \cite{Taylor2013,Srinivasa2016}. This
feature enables full resonant control of the RX qubit via electric
fields, in direct analogy to resonant control of individual electronic
and nuclear spins via magnetic fields in electron spin resonance (ESR)
and nuclear magnetic resonance (NMR). The same charge admixture also
enables coupling of the qubit to the electric field of photons in
a microwave cavity, with a strength characterized by the charge admixture
parameter $\xi\equiv t/\Delta$ (here, $t\equiv t_{l}=t_{r}$). The
parameter $\xi$ is a measure of the electric dipole moment of the
qubit and is inversely proportional to $\Delta,$ which sets the width
of the $\left(1,1,1\right)$ region and is tunable via $V_{m}.$ 

We can write the Hamiltonian for the RX qubit including electric dipole
coupling as $H_{{\rm RX}}=H_{0}+H_{{\rm int}}^{\prime},$ where $H_{{\rm int}}^{\prime}$
is the dipole interaction in the RX qubit basis. The operation point
$\left(\epsilon_{0},\Delta_{0}\right)$ for the RX qubit determines
the qubit frequency $\omega$ and the specific form of the electric
dipole moment. Variations in both $\epsilon$ and $\Delta$ about
this operation point enable coupling to microwave cavity photons \cite{Taylor2013,Srinivasa2016,Russ2015b,Russ2016,Landig2018}
and are implemented via gate voltage control of the on-site energies
$-\varepsilon_{i}.$ Here, we focus on electric dipole coupling for
small variations in the detuning $\epsilon$ \cite{Taylor2013,Srinivasa2016}.
For this case, the electric dipole moment of the RX qubit is along
the triple dot axis and can be described in terms of the operator
$d=\frac{ew_{0}}{2}\left(n_{1}-n_{3}\right),$ where $e$ is the magnitude
of the electron charge and $w_{0}$ is the size of the triple dot
(i.e., the distance between the centers of the outer dots). The operation
point $\epsilon=\epsilon_{0}$ for the RX qubit is chosen such that
the coupling to variations in the detuning $\epsilon$ {[}see, e.g.,
Eq.~(\ref{eq:detdrive}){]} are perpendicular to the quantization
axis of the RX qubit. In the basis of the RX qubit states, the dipole
interaction of the triple dot with the fundamental cavity mode then
takes the form \cite{Srinivasa2016}
\begin{equation}
H_{{\rm int}}^{\prime}=g\sigma^{x}\left(a+a^{\dagger}\right),\label{eq:Hintpr}
\end{equation}
where the effective qubit-photon coupling strength is given by 
\begin{eqnarray}
g & = & \left.\frac{g_{c}}{2}\sqrt{\left(\partial_{\epsilon}J\right)^{2}+3\left(\partial_{\epsilon}j\right)^{2}}\right|_{\epsilon=\epsilon_{0}}\label{eq:geff}
\end{eqnarray}
and $g_{c}$ is the charge-cavity coupling strength (an expression
for $g_{c}$ is derived from a circuit model in Ref.~\cite{Srinivasa2016}).
For the qubit operation point $\epsilon_{0}=0$ chosen in this work,
$t_{l}=t_{r}\equiv t$ and $g=\sqrt{3}\xi^{2}g_{c}/2.$ Thus, maximizing
$\xi$ maximizes the coupling strength $g.$ 

In the driven resonant regime described in Ref.~\cite{Srinivasa2016},
two RX qubits interact with microwave cavity photons, whose frequency
we here denote as $\omega_{c},$ in the presence of an external driving
field of frequency $\omega^{d}\equiv\nu$ applied to the cavity. A
displacement transformation in the regime of large driving field amplitude
and large cavity-drive detuning $\left|\omega_{c}-\omega^{d}\right|\gg g,\left|\omega-\omega^{d}\right|$
then effectively transfers the drive from the cavity to the qubit
and leads to Eq.~(29) in Ref.~\cite{Srinivasa2016} for each qubit.
We can write this Hamiltonian for two RX qubits as
\begin{align}
H_{p}^{\left(3\right)} & =\omega_{c}a^{\dagger}a+\sum_{j}\frac{\omega_{j}}{2}\sigma_{j}^{z}+\sum_{j}g_{j}\sigma_{j}^{x}\left(a+a^{\dagger}\right)\nonumber \\
 & +\sum_{j}2\Omega_{j}\cos\left(\omega_{j}^{d}t+\phi_{j}\right)\sigma_{j}^{x},\label{eq:Hp3}
\end{align}
where we have redefined the phase $\phi_{j}$ to match the form of
$H_{p}$ in Eq.~(\ref{eq:Hp}). We see that, like $H_{p}^{\left(1\right)}$
in Eq.~(\ref{eq:Hp1}), $H_{p}^{\left(3\right)}$ has a form identical
to $H_{p}.$ We also note that this form for the Hamiltonian can be
obtained for cavity-coupled RX qubits without a displacement transformation
by directly driving the detuning of the RX qubits \cite{Taylor2013,Medford2013}
as described by Eq.~(\ref{eq:detdrive}). Thus, the sideband-based
interactions and associated entangling gates that we derive from $H_{p}$
can be implemented using both RX qubits and one-electron spin qubits. 

\section{\label{sec:Onequbitterms}One-qubit second-order terms in effective
Hamiltonian }

In this appendix, we analyze in more detail the one-qubit terms appearing
at second order in the effective Hamiltonian $H_{{\rm eff}}\left(\tau\right)$
{[}Eq.~(\ref{eq:HeffH2avg}){]}, which originate from the commutators
$\left[V_{I},V_{I}^{\prime}\right]_{1}$ and $\left[V_{I},V_{I}^{\prime}\right]_{2}$
defined in Eqs.~(\ref{eq:VIcomm}) and (\ref{eq:VIcomm2}). The one-qubit
terms arising from $\left[V_{I},V_{I}^{\prime}\right]_{1}$ are given
by

\begin{align}
\Lambda_{1}\equiv & \frac{1}{2i\tau}\int_{0}^{\tau}dt\int_{0}^{t}dt^{\prime}\left[V_{I},V_{I}\right]_{1}\nonumber \\
 & =\frac{1}{2i\tau}\int_{0}^{\tau}dt\int_{0}^{t}dt^{\prime}\sum_{j}\left\{ \left[A_{j},A_{j}^{\prime}\right]a^{\dagger2}+\left[A_{j}^{\dagger},A_{j}^{\prime\dagger}\right]a^{2}\right.\nonumber \\
 & \ \ \ \ \ \ \ \ \ \ \ \ \ \ \ \ \ \ \ \ \ \ +\left.\left(\left[A_{j},A_{j}^{\prime\dagger}\right]+\left[A_{j}^{\dagger},A_{j}^{\prime}\right]\right)a^{\dagger}a\right\} \label{eq:Vint1qcomm1}
\end{align}
and those arising from $\left[V_{I},V_{I}^{\prime}\right]_{2}$ are
\begin{align}
\Lambda_{2} & \equiv\frac{1}{2i\tau}\int_{0}^{\tau}dt\int_{0}^{t}dt^{\prime}\sum_{j}\left(A_{j}^{\dagger}A_{j}^{\prime}-A_{j}^{\prime\dagger}A_{j}\right),\label{eq:Vint1qcomm2}
\end{align}
where $A_{j}\equiv A_{j}\left(t\right)$ and $A_{j}^{\prime}\equiv A_{j}\left(t^{\prime}\right).$
We note that the cavity photon operators $a,a^{\dagger}$ appear in
each term of $\Lambda_{1}$ but are absent from $\Lambda_{2}.$ 

Substituting for $A_{j}$ and $A_{j}^{\prime}$ using Eq.~(\ref{eq:A})
shows that evaluating the double integrals in Eqs.~(\ref{eq:Vint1qcomm1})
and (\ref{eq:Vint1qcomm2}) amounts to evaluating double integrals
of products of exponentials of the form
\begin{align}
h\left(\mu,\mu^{\prime}\right) & \equiv\frac{1}{2i\tau}\int_{0}^{\tau}dt\int_{0}^{t}dt^{\prime}e^{-i\mu t}e^{i\mu^{\prime}t^{\prime}}\nonumber \\
 & =\begin{cases}
0, & \mu\neq\mu^{\prime}\\
-\frac{1}{2\mu} & \mu=\mu^{\prime}
\end{cases}\label{eq:hjnutau}
\end{align}
or its complex conjugate $h^{\ast}\left(\mu,\mu^{\prime}\right)=-h\left(-\mu,-\mu^{\prime}\right).$
Here, $\mu,\mu^{\prime}\in\left\{ \pm\Delta_{j},\pm\left(\Delta_{j}+W_{j}\right),\pm\left(\Delta_{j}-W_{j}\right)\right\} .$
Since we set $\Delta_{j}=p_{j}\eta$ and $W_{j}=q_{j}\eta$ with integers
$p_{j}$ and $q_{j},$ we have $\mu_{j}=r_{j}\eta$ and $\mu_{j}^{\prime}=r_{j}^{\prime}\eta$
where $r_{j}$ and $r_{j}^{\prime}$ are also integers. Equation (\ref{eq:Vint1qcomm1})
then becomes
\begin{align*}
\Lambda_{1} & =\sum_{j}g_{j}^{2}\left(\left\{ \sin\theta_{j}\sin^{2}\left(\frac{\theta_{j}}{2}\right)\left[h\left(-\Delta_{j}-W_{j},\Delta_{j}\right)\right.\right.\right.\\
 & \ \ \ \ \ \ \ \ -\left.h\left(-\Delta_{j},\Delta_{j}+W_{j}\right)\right]\sigma_{j}^{+}\\
 & \ \ \ \ \ \ \ \ +\sin\theta_{j}\cos^{2}\left(\frac{\theta_{j}}{2}\right)\left[h\left(-\Delta_{j}+W_{j},\Delta_{j}\right)\right.\\
 & \ \ \ \ \ \ \ \ -\left.h\left(-\Delta_{j},\Delta_{j}-W_{j}\right)\right]\sigma_{j}^{-}\\
 & \ \ \ \ \ \ \ \ -\sin^{2}\left(\frac{\theta_{j}}{2}\right)\cos^{2}\left(\frac{\theta_{j}}{2}\right)\left[h\left(-\Delta_{j}-W_{j},\Delta_{j}-W_{j}\right)\right.\\
 & \ \ \ \ \ \ \ \ -\left.\left.h\left(-\Delta_{j}+W_{j},\Delta_{j}+W_{j}\right)\right]\sigma_{j}^{z}\right\} a^{\dagger2}\\
 & \ \ \ \ \ \ \ \ +\left\{ \left[\sin\theta_{j}\sin^{2}\left(\frac{\theta_{j}}{2}\right)h\left(-\Delta_{j}-W_{j},-\Delta_{j}\right)\right.\right.\\
 & \ \ \ \ \ \ \ \ +\left.\sin\theta_{j}\cos^{2}\left(\frac{\theta_{j}}{2}\right)h\left(-\Delta_{j},-\Delta_{j}+W_{j}\right)\right]\sigma_{j}^{+}\\
 & \ \ \ \ \ \ \ \ +\left[\sin\theta_{j}\cos^{2}\left(\frac{\theta_{j}}{2}\right)h\left(-\Delta_{j}+W_{j},-\Delta_{j}\right)\right.\\
 & \ \ \ \ \ \ \ \ +\left.\sin\theta_{j}\sin^{2}\left(\frac{\theta_{j}}{2}\right)h\left(-\Delta_{j},-\Delta_{j}-W_{j}\right)\right]\sigma_{j}^{-}
\end{align*}
\begin{align}
 & \ \ \ \ \ \ \ \ +\left[\sin^{4}\left(\frac{\theta_{j}}{2}\right)h\left(-\Delta_{j}-W_{j},-\Delta_{j}-W_{j}\right)\right.\nonumber \\
 & -\left.\left.\cos^{4}\left(\frac{\theta_{j}}{2}\right)h\left(-\Delta_{j}+W_{j},-\Delta_{j}+W_{j}\right)\right]\sigma_{j}^{z}\right\} a^{\dagger}a\nonumber \\
 & \ \ \ +\left.{\rm H.c.}\right)\label{eq:Lambda1}
\end{align}
and Eq.~(\ref{eq:Vint1qcomm2}) becomes 
\begin{align}
\Lambda_{2} & =-\sum_{j}\frac{g_{j}^{2}}{2}\left\{ \left[\sin\theta_{j}\sin^{2}\left(\frac{\theta_{j}}{2}\right)h\left(\Delta_{j},\Delta_{j}+W_{j}\right)\right.\right.\nonumber \\
 & \ \ \ \ \ \ \ \ +\left.\sin\theta_{j}\cos^{2}\left(\frac{\theta_{j}}{2}\right)h\left(\Delta_{j}-W_{j},\Delta_{j}\right)\right]\sigma_{+}\nonumber \\
 & \ \ \ \ \ \ \ \ +\left[\sin\theta_{j}\sin^{2}\left(\frac{\theta_{j}}{2}\right)h\left(\Delta_{j}+W_{j},\Delta_{j}\right)\right.\nonumber \\
 & \ \ \ \ \ \ \ \ +\left.\sin\theta_{j}\cos^{2}\left(\frac{\theta_{j}}{2}\right)h\left(\Delta_{j},\Delta_{j}-W_{j}\right)\right]\sigma_{-}\nonumber \\
 & \ \ \ \ \ \ \ \ +\left[\sin^{4}\left(\frac{\theta_{j}}{2}\right)h\left(\Delta_{j}+W_{j},\Delta_{j}+W_{j}\right)\right.\nonumber \\
 & \ \ \ \ \ \ \ \ -\left.\left.\cos^{4}\left(\frac{\theta_{j}}{2}\right)h\left(\Delta_{j}-W_{j},\Delta_{j}-W_{j}\right)\right]\sigma_{z}\right\} \nonumber \\
 & \ \ \ +\left.{\rm H.c.}\right)\label{eq:Lambda2}
\end{align}
Using Eqs.~(\ref{eq:Lambda1}) and (\ref{eq:Lambda2}) together with
Eq.~(\ref{eq:hjnutau}), we can obtain the resonance conditions $\mu=\mu^{\prime}$
under which $h\left(\mu,\mu^{\prime}\right)\neq0$ and particular
one-qubit terms appear (i.e., are nonzero) in the effective Hamiltonian
$H_{{\rm eff}}\left(\tau\right)$. Note that these conditions are
identical for $h\left(\mu,\mu^{\prime}\right)$ and $h^{\ast}\left(\mu,\mu^{\prime}\right).$ 

The one-qubit resonance conditions and corresponding forms of the
effective Hamiltonian terms are summarized in Table \ref{tab:res1qcommHterms}.
We see that the $a^{2}$ and $a^{\dagger2}$ terms occur only for
$W_{j}=\left|2\Delta_{j}\right|$ or $\Delta_{j}=0.$ Assuming that
we choose an operation point such that these conditions are not satisfied,
the $a^{2}$ and $a^{\dagger2}$ terms can be eliminated from the
Hamiltonian. Additionally, the case $W_{j}=0$ corresponds to a vanishing
gap for dressed qubit $j$ and thus does not have well-defined physical
meaning. We can therefore also ignore the $\sigma_{j}^{+}a^{\dagger}a,$
$\sigma_{j}^{-}a^{\dagger}a,$ $\sigma_{j}^{+},$ and $\sigma_{j}^{-}$
terms. As a result, the only remaining one-qubit terms appearing at
second order in the effective Hamiltonian are those of the form $\sigma_{j}^{z}a^{\dagger}a$
in Eq.~(\ref{eq:Lambda1}) and $\sigma_{j}^{z}$ in Eq.~(\ref{eq:Lambda2}).
These terms exist for any values of $W_{j}$ and $\Delta_{j}.$ Finally,
we note that Eq.~(\ref{eq:hjnutau}) is well defined only for $\mu,\mu^{\prime}\neq0,$
which leads to the constraints $\Delta_{j}\neq0$ and $W_{j}\neq\left|\Delta_{j}\right|.$
The full one-qubit contribution to $H_{{\rm eff}}\left(\tau\right)$
for $\Delta_{j},W_{j}\neq0$ and $W_{j}\neq\left|\Delta_{j}\right|,\left|2\Delta_{j}\right|$
is thus
\begin{align}
\Lambda & \equiv\Lambda_{1}+\Lambda_{2}\nonumber \\
 & =\sum_{j}g_{j}^{2}\left[\frac{\sin^{4}\left(\frac{\theta_{j}}{2}\right)}{\Delta_{j}+W_{j}}-\frac{\cos^{4}\left(\frac{\theta_{j}}{2}\right)}{\Delta_{j}-W_{j}}\right]\sigma_{j}^{z}\left(a^{\dagger}a+\frac{1}{2}\right)\nonumber \\
 & =-\sum_{j}g_{j}^{2}\left[\frac{\delta_{j}^{2}+2\Delta_{j}\delta_{j}+W_{j}^{2}}{2W_{j}\left(\Delta_{j}^{2}-W_{j}^{2}\right)}\right]\sigma_{j}^{z}\left(a^{\dagger}a+\frac{1}{2}\right),\label{eq:Lambdafull}
\end{align}
where we have used $\cos\theta_{j}=\delta_{j}/W_{j}$ {[}see Eq.~(\ref{eq:H0}){]}.
Equation (\ref{eq:Lambdafull}) represents the dispersive shift terms
induced by the parametric driving \cite{Noh2023}. 

\begin{table}
\caption{\label{tab:res1qcommHterms}Resonance conditions for the appearance
of specific one-qubit terms in the second-order effective Hamiltonian
of Eq.~(\ref{eq:HeffH2avg}).}

\begin{ruledtabular}
\begin{centering}
\begin{tabular}{cc}
\textbf{\textcolor{black}{$\ \ $}}\textcolor{black}{Resonance condition}\textbf{\textcolor{black}{$\ \ $}} & \textbf{\textcolor{black}{$\ \ $}}\textcolor{black}{Hamiltonian terms}\textbf{\textcolor{black}{$\ \ $}}\tabularnewline
\hline 
$W_{j}=2\Delta_{j}$ & $\sigma_{j}^{+}a^{2},\ \sigma_{j}^{-}a^{\dagger2}$\tabularnewline
$W_{j}=-2\Delta_{j}$ & $\sigma_{j}^{+}a^{\dagger2},\ \sigma_{j}^{-}a^{2}$\tabularnewline
$\Delta_{j}=0$ & $\sigma_{j}^{z}a^{2},\ \sigma_{j}^{z}a^{\dagger2}$\tabularnewline
$W_{j}=0$ & $\sigma_{j}^{+}a^{\dagger}a,\ \sigma_{j}^{-}a^{\dagger}a,\ \sigma_{j}^{+},\ \sigma_{j}^{-}$\tabularnewline
Any $W_{j},\Delta_{j}$ & $\sigma_{j}^{z}a^{\dagger}a,\ \sigma_{j}^{z}$\tabularnewline
\end{tabular}
\par\end{centering}
\end{ruledtabular}
\end{table}

\section{\label{sec:Qubitqubitintterms}Qubit-qubit interaction terms in effective
Hamiltonian}

The qubit-qubit interaction terms in $H_{{\rm eff}}\left(\tau\right)$
{[}Eq.~(\ref{eq:HeffH2avg}){]}, and therefore the generated two-qubit
gates $U\left(\tau\right)\approx e^{-iH_{{\rm eff}}\tau},$ are given
by Eq.~(\ref{eq:Vqq}). Here, we derive the form of $V_{qq}\left(\tau\right)$
for each of the nine resonance conditions $\mu_{1}=\mu_{2},$ where
$\mu_{1}\in\left\{ \Delta_{1},\Delta_{1}+W_{1},\Delta_{1}-W_{1}\right\} $
and $\mu_{2}\in\left\{ \Delta_{2},\Delta_{2}+W_{2},\Delta_{2}-W_{2}\right\} .$
As noted in the main text, these resonance conditions are obtained
by substituting Eq.~(\ref{eq:A}) into $V_{qq}\left(\tau\right),$
which yields terms with frequencies $\mu_{1}-\mu_{2},$ and subsequently
applying Eq.~(\ref{eq:hjnutau}) to identify nonzero terms. In what
follows, we use the abbreviated notation $\Delta_{j}^{\pm}\equiv\Delta_{j}\pm W_{j}$
and refer to the resonance conditions by the numbers given in Table
\ref{tab:Vqqresterms}. 

First, we consider the resonance condition $\Delta_{1}=\Delta_{2}$
(resonance condition 1), which is equivalent to $\omega_{1}^{d}=\omega_{2}^{d}$
and therefore describes the driving of both qubits at the same frequency.
In order to select a specific form for the two-qubit interaction,
we also assume that the constraints $W_{1}\neq\pm W_{2}$ are satisfied
so that terms in $V_{qq}\left(\tau\right)$ with frequencies $\mu_{1}-\mu_{2}\neq\Delta_{1}-\Delta_{2}$
remain nonzero and vanish according to Eq.~(\ref{eq:hjnutau}). In
terms of the resonant detuning $\Delta\equiv\Delta_{1}=\Delta_{2},$
we find that Eq.~(\ref{eq:Vqq}) becomes
\begin{align}
V_{qq}\left(\tau\right) & =\frac{g_{1}g_{2}}{4}\left[2h\left(\Delta,\Delta\right)-2h\left(-\Delta,-\Delta\right)\right]\nonumber \\
 & \ \ \ \ \times\sin\theta_{1}\sin\theta_{2}\sigma_{1}^{z}\sigma_{2}^{z}\nonumber \\
 & =-\frac{g_{1}g_{2}}{2\Delta}\sin\theta_{1}\sin\theta_{2}\sigma_{1}^{z}\sigma_{2}^{z}\nonumber \\
 & =-2\mathcal{J}\sin\theta_{1}\sin\theta_{2}\sigma_{1}^{z}\sigma_{2}^{z},\label{eq:Vqq1}
\end{align}
where we have defined $\mathcal{J}\equiv g_{1}g_{2}/4\Delta.$ For
resonant qubit driving such that $\delta_{1}=\delta_{2}=0,$ $W_{1}=2\Omega_{1},$
$W_{2}=2\Omega_{2},$ and $\sin\theta_{1}=\sin\theta_{2}=1.$ In this
case, $\Delta_{1}=\Delta_{2}$ corresponds to $\omega_{1}=\omega_{2}.$
In other words, since the detunings of both qubit frequencies from
the cavity frequency $\omega_{c}$ are aligned, the center frequencies
of the driven qubits are also resonant. Equation~(\ref{eq:Vqq1})
then reduces to
\begin{equation}
V_{qq}\left(\tau\right)=-2\mathcal{J}\sigma_{1}^{z}\sigma_{2}^{z},\label{eq:Vqq1res}
\end{equation}
which can be used to generate a controlled-phase gate (see Appendix
\ref{sec:Elimdispshiftdyn}). 

In a similar way, we can derive the form of the two-qubit interaction
and corresponding constraints for each of the remaining resonance
conditions. Resonance condition 2 (3), given by $\Delta_{1}=\Delta_{2}^{+}$
($\Delta_{1}=\Delta_{2}^{-}$), is equivalent to $\omega_{1}^{d}=\omega_{2}^{d}-W_{2}$
($\omega_{1}^{d}=\omega_{2}^{d}+W_{2}$). For $\Delta\equiv\Delta_{1}=\Delta_{2}^{\pm},$
we find
\begin{align}
V_{qq}\left(\tau\right) & =\mp\frac{g_{1}g_{2}}{4}\left[h\left(\Delta,\Delta\right)-h\left(-\Delta,-\Delta\right)\right]\nonumber \\
 & \ \ \ \ \times\sin\theta_{1}\left(1\mp\cos\theta_{2}\right)\sigma_{1}^{z}\sigma_{2}^{x}\nonumber \\
 & =\pm\mathcal{J}\sin\theta_{1}\left(1\mp\cos\theta_{2}\right)\sigma_{1}^{z}\sigma_{2}^{x},\label{eq:Vqq23}
\end{align}
with the same set of constraints $W_{1}\neq\pm W_{2},\pm2W_{2}$ for
each resonance condition. In the case of resonant qubit driving, the
resonance condition $\Delta_{1}=\Delta_{2}^{+}$ ($\Delta_{1}=\Delta_{2}^{-}$)
corresponds to $\omega_{1}=\omega_{2}-2\Omega_{2}$ ($\omega_{1}=\omega_{2}+2\Omega_{2}$)
such that the center frequency of qubit 1 is aligned with the red
(blue) sideband of qubit 2. In this case, $\sin\theta_{1}=1$ and
$\cos\theta_{2}=0$, which reduces Eq.~(\ref{eq:Vqq23}) to
\begin{equation}
V_{qq}\left(\tau\right)=\pm\mathcal{J}\sigma_{1}^{z}\sigma_{2}^{x}\label{eq:Vqq23res}
\end{equation}
for $\Delta_{1}=\Delta_{2}^{\pm}.$

The resonance conditions $\Delta_{1}^{+}=\Delta_{2}$ and $\Delta_{1}^{-}=\Delta_{2}$
(resonance conditions 4 and 5) describe cases in which the roles of
qubits 1 and 2 are reversed relative to $\Delta_{1}=\Delta_{2}^{+}$
and $\Delta_{1}=\Delta_{2}^{-},$ respectively. Accordingly, $\Delta_{1}^{+}=\Delta_{2}$
($\Delta_{1}^{-}=\Delta_{2}$) is equivalent to $\omega_{1}^{d}-W_{1}=\omega_{2}^{d}$
($\omega_{1}^{d}+W_{1}=\omega_{2}^{d}$) and the qubit-qubit interaction
takes the form 
\begin{align}
V_{qq}\left(\tau\right) & =\pm\mathcal{J}\left(1\mp\cos\theta_{1}\right)\sin\theta_{2}\sigma_{1}^{x}\sigma_{2}^{z}\label{eq:Vqq45}
\end{align}
for $\Delta\equiv\Delta_{1}^{\pm}=\Delta_{2}$ and constraints such
that $W_{1}\leftrightarrow W_{2}$ relative to the constraints for
resonance conditions 2 and 3. In the case of resonantly driven qubits,
$\Delta_{1}^{+}=\Delta_{2}$ ($\Delta_{1}^{-}=\Delta_{2}$) corresponds
to $\omega_{1}-2\Omega_{1}=\omega_{2}$ ($\omega_{1}+2\Omega_{1}=\omega_{2}$)
so that the center frequency of qubit 2 is aligned with the red (blue)
sideband of qubit 1, and Eq.~(\ref{eq:Vqq45}) becomes 
\begin{equation}
V_{qq}\left(\tau\right)=\pm\mathcal{J}\sigma_{1}^{x}\sigma_{2}^{z}\label{eq:Vqq45res}
\end{equation}
for $\Delta_{1}^{\pm}=\Delta_{2}.$ 

We next consider the resonance conditions $\Delta_{1}^{+}=\Delta_{2}^{+}$
and $\Delta_{1}^{-}=\Delta_{2}^{-}$ (resonance conditions 6 and 7),
which are equivalent to $\omega_{1}^{d}-W_{1}=\omega_{2}^{d}-W_{2}$
and $\omega_{1}^{d}+W_{1}=\omega_{2}^{d}+W_{2},$ respectively. The
form of the effective qubit-qubit interaction for $\Delta\equiv\Delta_{1}^{\pm}=\Delta_{2}^{\pm}$
is
\begin{align}
V_{qq}\left(\tau\right) & =-\mathcal{J}\left(1\mp\cos\theta_{1}\right)\left(1\mp\cos\theta_{2}\right)\nonumber \\
 & \ \ \ \ \times\left(\sigma_{1}^{+}\sigma_{2}^{-}+\sigma_{1}^{-}\sigma_{2}^{+}\right),\label{eq:Vqq67}
\end{align}
with an identical set of constraints $W_{1}\neq W_{2},2W_{2},W_{2}/2$
for each resonance condition. For resonant driving of both qubits,
$\Delta_{1}^{+}=\Delta_{2}^{+}$ ($\Delta_{1}^{-}=\Delta_{2}^{-}$)
corresponds to $\omega_{1}-2\Omega_{1}=\omega_{2}-2\Omega_{2}$ ($\omega_{1}+2\Omega_{1}=\omega_{2}+2\Omega_{2}$),
such that the red (blue) sidebands of both qubits are in resonance,
and Eq.~(\ref{eq:Vqq67}) becomes 
\begin{equation}
V_{qq}\left(\tau\right)=-\mathcal{J}\left(\sigma_{1}^{+}\sigma_{2}^{-}+\sigma_{1}^{-}\sigma_{2}^{+}\right)\label{eq:Vqq67res}
\end{equation}
for both $\Delta=\Delta_{1}^{+}=\Delta_{2}^{+}$ and $\Delta=\Delta_{1}^{-}=\Delta_{2}^{-}.$

Finally, the resonance conditions $\Delta_{1}^{+}=\Delta_{2}^{-}$
and $\Delta_{1}^{-}=\Delta_{2}^{+}$ (resonance conditions 8 and 9)
are equivalent to $\omega_{1}^{d}-W_{1}=\omega_{2}^{d}+W_{2}$ and
$\omega_{1}^{d}+W_{1}=\omega_{2}^{d}-W_{2},$ respectively. The effective
qubit-qubit interaction for $\Delta\equiv\Delta_{1}^{\pm}=\Delta_{2}^{\mp}$
takes the form
\begin{align}
V_{qq}\left(\tau\right) & =\mathcal{J}\left(1\mp\cos\theta_{1}\right)\left(1\pm\cos\theta_{2}\right)\nonumber \\
 & \ \ \ \ \times\left(\sigma_{1}^{+}\sigma_{2}^{+}+\sigma_{1}^{-}\sigma_{2}^{-}\right),\label{eq:Vqq89}
\end{align}
again with an identical set of constraints $W_{1}\neq-W_{2},-2W_{2},-W_{2}/2$
for each resonance condition. Note that, for the physically relevant
case $W_{1,2}>0,$ these constraints are always satisfied. For resonant
driving of both qubits, $\Delta_{1}^{+}=\Delta_{2}^{-}$ ($\Delta_{1}^{-}=\Delta_{2}^{+}$)
corresponds to $\omega_{1}-2\Omega_{1}=\omega_{2}+2\Omega_{2}$ ($\omega_{1}+2\Omega_{1}=\omega_{2}-2\Omega_{2}$),
such that the red (blue) sideband of qubit 1 is in resonance with
the blue (red) sideband of qubit 2. Equation~(\ref{eq:Vqq89}) becomes
\begin{equation}
V_{qq}\left(\tau\right)=\mathcal{J}\left(\sigma_{1}^{+}\sigma_{2}^{+}+\sigma_{1}^{-}\sigma_{2}^{-}\right)\label{eq:Vqq89res}
\end{equation}
for both $\Delta=\Delta_{1}^{+}=\Delta_{2}^{-}$ and $\Delta=\Delta_{1}^{-}=\Delta_{2}^{+}.$

The above nine resonance conditions, their associated constraints,
and the corresponding forms of the effective qubit-qubit interaction
$V_{qq}\left(\tau\right)$ for resonant qubit driving are summarized
in Table \ref{tab:Vqqresterms} of the main text. 

\section{\label{sec:Elimdispshiftdyn}Elimination of drive-induced dispersive
shift dynamics from effective Hamiltonian}

As shown in Appendix \ref{sec:Onequbitterms}, the parametric drive-induced
dispersive shift terms $\Lambda$ {[}Eq.~(\ref{eq:Lambdafull}){]}
in general exist for any $W_{j}$ and $\Delta_{j}$ (provided $W_{j}\neq\left|\Delta_{j}\right|).$
Nevertheless, it is possible to effectively eliminate the dynamics
generated by $\Lambda$ in specific cases by appropriately choosing
parameters and operation times such that the gate in the presence
of $\Lambda$ is equivalent to that for $\Lambda=0.$ Here, we first
illustrate this method for the resonant qubit driving case $\delta_{j}=0$
and the resonance conditions 7 and 9 in Table \ref{tab:Vqqresterms}
and Fig.~\ref{fig:sbres}. We then consider the effects of $\Lambda$
for additional resonance conditions, as well as the off-resonant qubit
driving case $\delta_{j}\neq0$ for which we can choose parameters
such that $\Lambda=0.$

We first consider resonant qubit driving, described by $\delta_{j}=0$
such that $W_{j}=2\Omega_{j},$ $\Lambda=\emph{\ensuremath{\Lambda_{r}}},$
and $\chi_{j}\neq0.$ From Eqs.~(\ref{eq:HeffH2avg}), (\ref{eq:Lambdar}),
and (\ref{eq:Vqq}), the effective Hamiltonian for $\Delta_{j}=p_{j}\eta$
and $W_{j}=q_{j}\eta$ with $p_{j},q_{j}$ integers, $\Delta_{j},W_{j}\neq0,$
and $W_{j}\neq\left|\Delta_{j}\right|,\left|2\Delta_{j}\right|$ is
given by
\begin{align}
H_{{\rm eff}}\left(\tau\right) & =\lambda^{2}\bar{H}_{2}\left(\tau\right)\nonumber \\
 & =V_{qq}\left(\tau\right)+\emph{\ensuremath{\Lambda_{r}}}.\label{eq:HeffVqqLambda0}
\end{align}
As discussed in the main text, the desired ideal two-qubit evolution
is generated by the qubit-qubit interaction term $V_{qq}$ according
to $U_{m}\equiv e^{-iV_{qq}\tau_{m}},$ where $\tau_{m}\equiv m\tau=2\pi m/\eta$
is the corresponding gate time. We now also define an evolution operator
with the dispersive shift terms included as $U_{m}^{\prime}\equiv e^{-iH_{{\rm eff}}\tau_{m}},$
where $H_{{\rm eff}}$ is given by Eq.~(\ref{eq:HeffVqqLambda0}). 

From Eq.~(\ref{eq:Vqq}), we note that $V_{qq}$ is independent of
photon operators $a,a^{\dagger}.$ We also see from Eq.~(\ref{eq:Lambdar})
that $\emph{\ensuremath{\Lambda_{r}}}$ is diagonal in the photon
number $n.$ We can therefore write 
\begin{equation}
\Lambda_{r}=\sum_{n}\Lambda_{n},\label{eq:LambdarPn}
\end{equation}
where $\Lambda_{n}\equiv P_{n}\emph{\ensuremath{\Lambda_{r}}}P_{n}=\bra{n}\emph{\ensuremath{\Lambda_{r}}}\ket{n}P_{n}$
is the operator describing the dispersive shift terms within the $n$-photon
subspace and $P_{n}\equiv\ket{n}\bra{n}$ is the corresponding subspace
projector, and work in a subspace of fixed $n.$ Noting that $V_{qq}\left(\tau\right)$
consists only of qubit operators {[}see Eq.~(\ref{eq:Vqq}){]}, we
find $H_{{\rm eff}}^{\left(n\right)}\equiv P_{n}H_{{\rm eff}}P_{n}=P_{n}\left(V_{qq}+\emph{\ensuremath{\Lambda_{r}}}\right)P_{n}=V_{qq}+\Lambda_{n},$
where we now use $V_{qq}$ to denote the $n$-photon subspace operator.
In the basis $\left\{ \ket{ee,n},\ket{eg,n},\ket{ge,n},\ket{gg,n}\right\} ,$
$\Lambda_{n}$ takes the form 
\begin{align}
\Lambda_{n} & =\left(n+\frac{1}{2}\right)\nonumber \\
\times & \left(\begin{array}{cccc}
\chi_{1}+\chi_{2}\\
 & \chi_{1}-\chi_{2}\\
 &  & -\chi_{1}+\chi_{2}\\
 &  &  & -\chi_{1}-\chi_{2}
\end{array}\right),\label{eq:Lambdan}
\end{align}
while the form of $V_{qq}$ depends on the particular resonance condition
according to Table \ref{tab:Vqqresterms}. The evolution with the
dispersive shift terms included is described by $P_{n}U_{m}^{\prime}P_{n}=e^{-iH_{{\rm eff}}^{\left(n\right)}\tau_{m}}=e^{-i\left(V_{qq}+\emph{\ensuremath{\Lambda_{n}}}\right)\tau_{m}}.$
As in the main text, we use $U_{m}$ and $U_{m}^{\prime}$ to denote
the evolution operators within the $n$-photon two-qubit subspace
and suppress the photon number state $\ket{n}$ in what follows. 

We first consider resonance condition 7, which is defined by $\Delta\equiv\Delta_{1}^{-}=\Delta_{2}^{-}.$
In the main text, we noted that the ideal evolution within the $n$-photon
subspace can be written as $U_{m}=e^{i\mathcal{J}\tau_{m}\Sigma_{x}},$
where $\Sigma_{x}\equiv\ket{eg}\bra{ge}+\ket{ge}\bra{eg}$. The ideal
operation $U_{m}$ therefore acts nontrivially only within the two-dimensional
subspace $\left\{ \ket{eg},\ket{ge}\right\} ,$ and $\Sigma_{x}$
has an action analogous to the Pauli operator $\sigma_{x}$ in this
subspace. 

From Eq.~(\ref{eq:Lambdan}) and the form of $V_{qq}$ for resonance
condition 7 {[}Eq.~(\ref{eq:Vqq67res}){]}, we find that $H_{{\rm eff}}^{\left(n\right)}$
is block-diagonal. In this case, we can write $H_{{\rm eff}}^{\left(n\right)}=PH_{{\rm eff}}^{\left(n\right)}P+QH_{{\rm eff}}^{\left(n\right)}Q,$
where we have defined the two-qubit subspace projectors 
\begin{align}
P & \equiv\ket{ee}\bra{ee}+\ket{gg}\bra{gg},\nonumber \\
Q & \equiv\ket{eg}\bra{eg}+\ket{ge}\bra{ge}.\label{eq:PQ}
\end{align}
We can then characterize the effect of $\emph{\ensuremath{\Lambda_{n}}}$
on the dynamics for a given initial state $\ket{\psi_{i}}$ via a
fidelity 
\begin{align}
F_{s} & \equiv\left|\bra{\psi_{i}}U_{m}^{\dagger}U_{m}^{\prime}\ket{\psi_{i}}\right|^{2}\nonumber \\
 & =\left|\bra{\psi_{i}}P\left(PU_{m}^{\dagger}P\right)\left(PU_{m}^{\prime}P\right)P\ket{\psi_{i}}\right.\nonumber \\
 & \ \ \ \ +\left.\bra{\psi_{i}}Q\left(QU_{m}^{\dagger}Q\right)\left(QU_{m}^{\prime}Q\right)Q\ket{\psi_{i}}\right|^{2},\label{eq:Fs}
\end{align}
where we have used $U_{m}=PU_{m}P+QU_{m}Q$ and $U_{m}^{\prime}=PU_{m}^{\prime}P+QU_{m}^{\prime}Q.$

For resonance condition 7, we find 
\begin{align}
PH_{{\rm eff}}^{\left(n\right)}P & =\left(n+\frac{1}{2}\right)\left(\chi_{1}+\chi_{2}\right)\Sigma_{z},\nonumber \\
QH_{{\rm eff}}^{\left(n\right)}Q & =\left(n+\frac{1}{2}\right)\left(\chi_{1}-\chi_{2}\right)\Sigma_{z}-\mathcal{J}\Sigma_{x},\label{eq:PQHeffsubRC7}
\end{align}
where we have defined the operator $\Sigma_{z}\equiv\ket{eg}\bra{eg}-\ket{ge}\bra{ge}$
in analogy to the Pauli operator $\sigma_{z}.$ For an initial state
$\ket{\psi_{i}}$ in the subspace associated with $Q,$ evolution
for $\Lambda_{n}\neq0$ remains within this space such that $U_{m}^{\prime}\ket{\psi_{i}}=QU_{m}^{\prime}Q\ket{\psi_{i}}.$
In the main text, we consider the initial state $\ket{\psi_{i}}=\ket{eg}.$
For the \emph{$i{\rm SWAP}$} gate $U_{i{\rm SW}},$ which is equivalent
to $U_{m}$ at time $\tau_{m}=\pi/2\mathcal{J}$ {[}see Eq.~(\ref{eq:UmRC7}){]},
the ideal final state is $\ket{\psi_{f}}=U_{i{\rm SW}}\ket{\psi_{i}}=U_{i{\rm SW}}\ket{eg}=i\ket{ge}.$ 

We now set $n=0$ for simplicity {[}in what follows, generalization
of the expressions to any $n$ using Eq.~(\ref{eq:Lambdan}) is straightforward
and amounts to the replacements $\left(\chi_{1}\pm\chi_{2}\right)/2\rightarrow\left(n+\frac{1}{2}\right)\left(\chi_{1}\pm\chi_{2}\right)${]}.
In the presence of $\Lambda_{0}$ and defining the unit vector $\hat{\mathbf{u}}\equiv\left[\left(\chi_{1}-\chi_{2}\right)\hat{z}-2\mathcal{J}\hat{x}\right]/\tilde{\Omega}$
with $\tilde{\Omega}\equiv\sqrt{\left(\chi_{1}-\chi_{2}\right)^{2}+4\mathcal{J}^{2}},$
the action of the gate is modified to 
\begin{align}
QU_{m}^{\prime}Q\ket{\psi_{i}} & =e^{-i\frac{\tilde{\Omega}\tau_{m}}{2}\hat{\mathbf{u}}\cdot\boldsymbol{\Sigma}}\ket{eg}\nonumber \\
 & =\left[\cos\left(\frac{\tilde{\Omega}\tau_{m}}{2}\right)\right.\nonumber \\
 & \left.-i\sin\left(\frac{\tilde{\Omega}\tau_{m}}{2}\right)\left(\frac{\chi_{1}-\chi_{2}}{\tilde{\Omega}}\right)\right]\ket{eg}\nonumber \\
 & +i\sin\left(\frac{\tilde{\Omega}\tau_{m}}{2}\right)\left(\frac{2\mathcal{J}}{\tilde{\Omega}}\right)\ket{ge}.\label{eq:QUmprQpsii}
\end{align}
Using Eqs.~(\ref{eq:Fs}) and (\ref{eq:QUmprQpsii}) along with $PU_{m}P\ket{\psi_{i}}=PU_{m}^{\prime}P\ket{\psi_{i}}=0,$
we find 
\begin{align}
F_{s} & =\left|\bra{\psi_{i}}Q\left(QU_{m}^{\dagger}Q\right)\left(QU_{m}^{\prime}Q\right)Q\ket{\psi_{i}}\right|^{2}\nonumber \\
 & =\frac{4\mathcal{J}^{2}}{\tilde{\Omega}^{2}}\sin^{2}\left(\frac{\tilde{\Omega}\tau_{m}}{2}\right),\label{eq:FsRC7}
\end{align}
which oscillates with the modified frequency $\tilde{\Omega}$ instead
of $\mathcal{J}$ in the ideal case. Nevertheless, we can recover
the ideal dynamics and obtain $F_{s}=1$ for the \emph{$i{\rm SWAP}$}
gate time $\tau_{m}=\pi/2\mathcal{J}$ by choosing parameters for
which $\chi_{1}=\chi_{2},$ such that $QH_{{\rm eff}}^{\left(n\right)}Q=V_{qq}$
and $QU_{m}^{\prime}Q=QU_{m}Q.$ The constraints that must be satisfied
by the parameters for resonance condition 7 (in addition to those
listed in Table \ref{tab:Vqqresterms}) are therefore (a) $\tau_{m}=\pi/2\mathcal{J},$
(b) $\chi_{1}=\chi_{2},$ and (c) $w\equiv p_{1}-q_{1}=p_{2}-q_{2}$
from the resonance condition itself From constraint (a) and using
$\Delta=w\eta,$ we find $\tau_{m}\equiv2\pi m/\eta=\pi/2\mathcal{J}=2\pi w\eta/g_{1}g_{2},$
or 
\begin{equation}
g_{1}g_{2}=\frac{w}{m}\eta^{2},\label{eq:RC7constra}
\end{equation}
while constraint (b) becomes, using Eq.~(\ref{eq:Lambdar}) with
$\Delta_{j}=p_{j}\eta,$ $W_{j}=2\Omega_{j}=q_{j}\eta,$ and $\Delta=\Delta_{1}^{-}=\Delta_{2}^{-}=w\eta$
and incorporating constraint (c),
\begin{equation}
\frac{g_{2}^{2}}{g_{1}^{2}}=\frac{2+\frac{w}{q_{2}}}{2+\frac{w}{q_{1}}}.\label{eq:RC7constrbc}
\end{equation}
We also note that $q_{1},q_{2},p_{1},p_{2},w,$ and $m$ must all
be integers. 

There are multiple sets of parameter values that satisfy these constraints,
and we choose one set (given in Table \ref{tab:parameters}) for the
gate fidelity analysis described in this work. To show that this approach
effectively restores the ideal \emph{$i{\rm SWAP}$} evolution in
the absence of the dispersive shift terms, we choose $n=0$ and compare
the ideal state at time $\tau_{m}$ for $\emph{\ensuremath{\Lambda_{r}}}=0$
with the numerical solution to the time-dependent Schrödinger equation
for the density matrix in the interaction picture, $\dot{\rho}_{I}=-i\left[V_{I},\rho_{I}\right],$
for $V_{I}$ given by Eq.~(\ref{eq:VI}), the initial state $\rho_{I}\left(0\right)=\ket{\psi_{i}}\bra{\psi_{i}}=\ket{eg,0}\bra{eg,0},$
and the chosen set of parameters. For the numerical calculations,
we work in the photon subspace with $n=0,1,2.$ The ideal final state
is given by 
\begin{align}
\rho_{I}^{f}\left(\tau_{m}\right) & =\ket{\psi_{f}}\bra{\psi_{f}}\nonumber \\
 & =U_{i{\rm SW}}\rho_{I}\left(0\right)U_{i{\rm SW}}^{\dagger}\nonumber \\
 & =\ket{ge,0}\bra{ge,0}.\label{eq:rhofidRC7}
\end{align}
 For the comparison, we calculate the fidelity
\begin{equation}
F_{0}\left(\tau_{m}\right)\equiv{\rm Tr}\left[\rho_{I}^{f}\left(\tau_{m}\right)\rho_{I}^{{\rm \left(0\right)}}\left(\tau_{m}\right)\right],\label{eq:F0}
\end{equation}
where $\rho_{I}^{{\rm \left(0\right)}}\left(\tau_{m}\right)$ denotes
the numerical solution in the absence of decay, i.e., for $\gamma_{j}=\kappa=0$
in the master equation of Eq.~(\ref{eq:meqintpic}). For the chosen
parameters, we find $F_{0}\approx0.998,$ confirming that setting
$\chi_{1}=\chi_{2}$ effectively eliminates the modification to the
dynamics induced by the dispersive shift terms $\emph{\ensuremath{\Lambda_{r}}}$
and restores the evolution expected for the \emph{$i{\rm SWAP}$}
gate. 

We now apply this approach to resonance condition 9, which from Table
\ref{tab:Vqqresterms} is given by $\Delta\equiv\Delta_{1}^{-}=\Delta_{2}^{+}.$
From the form of $V_{qq}$ {[}Eq.~(\ref{eq:Vqq89res}){]}, the effective
Hamiltonian for this case is again block-diagonal and can be written
as $H_{{\rm eff}}^{\left(n\right)}=PH_{{\rm eff}}^{\left(n\right)}P+QH_{{\rm eff}}^{\left(n\right)}Q.$
As described in the main text, ideal evolution within the $n$-photon
subspace can be expressed as $U_{m}=e^{-i\mathcal{J}\tau_{m}\Sigma_{x}^{\prime}},$
where $\Sigma_{x}^{\prime}\equiv\ket{ee}\bra{gg}+\ket{gg}\bra{ee}$.
The ideal evolution in this case occurs nontrivially only within the
two-dimensional subspace associated with $P$ in Eq.~(\ref{eq:PQ}),
and we find
\begin{align}
PH_{{\rm eff}}^{\left(n\right)}P & =\left(n+\frac{1}{2}\right)\left(\chi_{1}+\chi_{2}\right)\Sigma_{z}^{\prime}+\mathcal{J}\Sigma_{x}^{\prime},\nonumber \\
QH_{{\rm eff}}^{\left(n\right)}Q & =\left(n+\frac{1}{2}\right)\left(\chi_{1}-\chi_{2}\right)\Sigma_{z}^{\prime},\label{eq:PQHeffsubRC9}
\end{align}
where $\Sigma_{z}^{\prime}\equiv\ket{ee}\bra{ee}-\ket{gg}\bra{gg}.$
If we now assume that the initial state $\ket{\psi_{i}}$ lies within
the subspace associated with $P,$ evolution for $\Lambda_{n}\neq0$
remains within this space such that $U_{m}^{\prime}\ket{\psi_{i}}=PU_{m}^{\prime}P\ket{\psi_{i}}.$
Here, we consider the initial state $\ket{\psi_{i}}=\ket{ee}.$ For
the double-excitation gate $U_{i{\rm DE}}$ described in the main
text, which is equivalent to $U_{m}$ at time $\tau_{m}=-\pi/2\mathcal{J}$
{[}see Eq.~(\ref{eq:UmRC9}){]} the ideal final state is $\ket{\psi_{f}}=U_{i{\rm DE}}\ket{\psi_{i}}=U_{i{\rm DE}}\ket{ee}=i\ket{gg}.$

Again, we set $n=0$ for simplicity {[}with straightforward generalization
of expressions to any $n$ via $\left(\chi_{1}\pm\chi_{2}\right)/2\rightarrow\left(n+\frac{1}{2}\right)\left(\chi_{1}\pm\chi_{2}\right)${]}.
In terms of $\hat{\mathbf{u}}^{\prime}\equiv\left[\left(\chi_{1}+\chi_{2}\right)\hat{z}+2\mathcal{J}\hat{x}\right]/\Omega^{\prime}$
with $\Omega^{\prime}\equiv\sqrt{\left(\chi_{1}+\chi_{2}\right)^{2}+4\mathcal{J}^{2},}$
the modified action of the gate in the presence of $\emph{\ensuremath{\Lambda}}_{r}$
is given by
\begin{align}
PU_{m}^{\prime}P\ket{\psi_{i}} & =e^{-i\frac{\Omega^{\prime}\tau_{m}}{2}\hat{\mathbf{u}}^{\prime}\cdot\boldsymbol{\Sigma}^{\prime}}\ket{ee}\nonumber \\
 & =\left[\cos\left(\frac{\Omega^{\prime}\tau_{m}}{2}\right)\right.\nonumber \\
 & \left.-i\sin\left(\frac{\Omega^{\prime}\tau_{m}}{2}\right)\left(\frac{\chi_{1}+\chi_{2}}{\Omega^{\prime}}\right)\right]\ket{ee}\nonumber \\
 & -i\sin\left(\frac{\Omega^{\prime}\tau_{m}}{2}\right)\left(\frac{2\mathcal{J}}{\Omega^{\prime}}\right)\ket{gg}.\label{eq:PUmprPpsii}
\end{align}
Since $QU_{m}Q\ket{\psi_{i}}=QU_{m}^{\prime}Q\ket{\psi_{i}}=0,$ Eq.~(\ref{eq:Fs})
then reduces to
\begin{align}
F_{s} & =\left|\bra{\psi_{i}}P\left(PU_{m}^{\dagger}P\right)\left(PU_{m}^{\prime}P\right)P\ket{\psi_{i}}\right|^{2}\nonumber \\
 & =\frac{4\mathcal{J}^{2}}{\Omega^{\prime}{}^{2}}\sin^{2}\left(\frac{\Omega^{\prime}\tau_{m}}{2}\right).\label{eq:FsRC9}
\end{align}
$F_{s}$ now oscillates with the modified frequency $\Omega^{\prime}$
instead of $\mathcal{J}$ in the ideal case. We can again recover
the ideal dynamics and obtain $F_{s}=1$ for the gate time $\tau_{m}=-\pi/2\mathcal{J}$
by choosing parameters such that $\chi_{1}=-\chi_{2},$ which yields
$PH_{{\rm eff}}^{\left(n\right)}P=V_{qq}$ and $PU_{m}^{\prime}P=PU_{m}P.$
Thus, the constraints that must be satisfied by the parameters for
resonance condition 9 (in addition to those listed in Table \ref{tab:Vqqresterms})
are (a) $\tau_{m}=-\pi/2\mathcal{J},$ (b) $\chi_{1}=-\chi_{2},$
and (c) $w\equiv p_{1}-q_{1}=p_{2}+q_{2}$ from the resonance condition.
From constraint (a) and using $\Delta=w\eta,$ we find $\tau_{m}\equiv2\pi m/\eta=-\pi/2\mathcal{J}=-2\pi w\eta/g_{1}g_{2},$
which gives 
\begin{equation}
g_{1}g_{2}=-\frac{w}{m}\eta^{2}.\label{eq:RC9constra}
\end{equation}
On the other hand, constraint (b) with constraint (c) incorporated
becomes {[}again using Eq.~(\ref{eq:Lambdar}) with $\Delta_{j}=p_{j}\eta,$
$W_{j}=2\Omega_{j}=q_{j}\eta,$ and $\Delta=\Delta_{1}^{-}=\Delta_{2}^{-}=w\eta${]},
\begin{equation}
\frac{g_{2}^{2}}{g_{1}^{2}}=\frac{2-\frac{w}{q_{2}}}{2+\frac{w}{q_{1}}}.\label{eq:RC9constrbc}
\end{equation}
As before, $q_{1},q_{2},p_{1},p_{2},w,$ and $m$ must all be integers. 

There are once again multiple sets of parameter values that satisfy
these constraints, and we choose one set (given in Table \ref{tab:parameters})
for the analysis described in this work. We now show that the ideal
evolution described by $U_{i{\rm DE}}$ in the absence of the dispersive
shift terms is restored for $\chi_{1}=-\chi_{2}$ and the chosen parameters.
Working in the photon subspace with $n=0,1,2$ and numerically solving
$\dot{\rho}_{I}=-i\left[V_{I},\rho_{I}\right]$ for resonance condition
9 with the initial state $\rho_{I}\left(0\right)=\ket{\psi_{i}}\bra{\psi_{i}}=\ket{ee,0}\bra{ee,0},$
we compare the $\gamma_{j}=\kappa=0$ solution $\rho_{I}^{{\rm \left(0\right)}}\left(\tau_{m}\right)$
at time $\tau_{m}$ with the ideal ($\emph{\ensuremath{\Lambda_{r}}}=0$)
final state
\begin{align}
\rho_{I}^{f}\left(\tau_{m}\right) & =\ket{\psi_{f}}\bra{\psi_{f}}\nonumber \\
 & =U_{i{\rm DE}}\rho_{I}\left(0\right)U_{i{\rm DE}}^{\dagger}\nonumber \\
 & =\ket{gg,0}\bra{gg,0}\label{eq:rhofidRC9}
\end{align}
via the fidelity in Eq.~(\ref{eq:F0}). We find $F_{0}\approx0.998,$
showing that setting $\chi_{1}=-\chi_{2}$ effectively eliminates
the modification to the dynamics induced by the dispersive shift terms
and restores the expected gate evolution generated by $V_{qq}$ for
resonance condition 9. The small error $1-F_{0}$ found for both resonance
conditions 7 and 9 also serves as a measure of the validity of the
model that we develop in this work. 

By choosing parameters that satisfy additional constraints, a similar
approach can be used to eliminate dispersive shift dynamics for general
initial states $\ket{\psi_{i}}$ with arbitrary photon number $n,$
as well as for other resonance conditions and the corresponding interactions
and two-qubit gates listed in Table \ref{tab:Vqqresterms}. For a
general initial state
\begin{align}
\ket{\psi_{i}} & =\left(c_{ee}\ket{ee,n}+c_{eg}\ket{eg,n}\right.\nonumber \\
 & \left.+c_{ge}\ket{ge,n}+c_{gg}\ket{gg,n}\right),\label{eq:Psiigen}
\end{align}
both the subspaces associated with $P$ and $Q$ must be taken into
account. For resonance condition 7, we find
\begin{align}
PU_{m}^{\prime}P\ket{\psi_{i}} & =Pe^{-i\emph{\ensuremath{\Lambda_{n}}}\tau_{m}}P\nonumber \\
 & =e^{-i\left(n+\frac{1}{2}\right)\left(\chi_{1}+\chi_{2}\right)\tau_{m}}\ket{ee,n}\bra{ee,n}\nonumber \\
 & +e^{i\left(n+\frac{1}{2}\right)\left(\chi_{1}+\chi_{2}\right)\tau_{m}}\ket{gg,n}\bra{gg,n}\label{eq:PUmprPRC7}
\end{align}
and $PU_{m}P\ket{\psi_{i}}=P\ket{\psi_{i}},$ while for resonance
condition 9, 
\begin{align}
QU_{m}^{\prime}Q\ket{\psi_{i}} & =Qe^{-i\emph{\ensuremath{\Lambda_{n}}}\tau_{m}}Q\nonumber \\
 & =e^{-i\left(n+\frac{1}{2}\right)\left(\chi_{1}-\chi_{2}\right)\tau_{m}}\ket{eg,n}\bra{eg,n}\nonumber \\
 & +e^{i\left(n+\frac{1}{2}\right)\left(\chi_{1}-\chi_{2}\right)\tau_{m}}\ket{ge,n}\bra{ge,n}\label{eq:QUmprQRC9}
\end{align}
and $QU_{m}Q\ket{\psi_{i}}=Q\ket{\psi_{i}}.$ From the forms of Eqs.~(\ref{eq:PUmprPRC7})
and (\ref{eq:QUmprQRC9}), we see that setting $\chi_{j}=2\pi k_{j}$
with $k_{j}$ an integer for $j=1,2,$ along with $k_{1}=k_{2}$ ($k_{1}=-k_{2}$)
for resonance condition 7 (9) to satisfy $\chi_{1}=\chi_{2}$ ($\chi_{1}=-\chi_{2}$)
eliminates the dynamical phases due to $\emph{\ensuremath{\Lambda_{r}}}$
and restores the ideal evolution due to $V_{qq}$ within a subspace
of fixed $n,$ or for an integer average photon number $\left\langle a^{\dagger}a\right\rangle .$ 

We now consider the other resonance conditions in Table \ref{tab:Vqqresterms}.
Resonance condition 1 is given by $\Delta\equiv\Delta_{1}=\Delta_{2}$
and (assuming the associated constraints listed in the table are satisfied
by $W_{1}$ and $W_{2}$) yields $V_{qq}=-2\mathcal{J}\sigma_{1}^{z}\sigma_{2}^{z}.$
This interaction is equivalent to that which generates a Mølmer-Sørensen
gate in the original qubit basis \cite{Sorensen1999} and can be used
to construct a controlled-phase gate in the dressed qubit basis considered
here. Using $U_{m}=e^{-iV_{qq}\tau_{m}}=e^{i2\mathcal{J}\tau_{m}\sigma_{1}^{z}\sigma_{2}^{z}},$
a sequence for a controlled-phase gate is given by 
\begin{align}
U_{\varphi} & =e^{i2\mathcal{J}\tau_{m}}e^{-i2\mathcal{J}\tau_{m}\sigma_{1}^{z}}e^{-i2\mathcal{J}\tau_{m}\sigma_{2}^{z}}U_{m}\nonumber \\
 & =\left(\begin{array}{cccc}
1\\
 & 1\\
 &  & 1\\
 &  &  & e^{i\varphi}
\end{array}\right),\label{eq:Uphi}
\end{align}
where $\varphi\equiv8\mathcal{J}\tau_{m},$ which shows that $U_{m}$
for $\Delta\equiv\Delta_{1}=\Delta_{2}$ is equivalent to a controlled-phase
gate up to single-qubit rotations. For $\varphi=\left(2l+1\right)\pi$
or $\tau_{m}=\left(2l+1\right)\pi/8\mathcal{J}$ with $l$ an integer,
$U_{\varphi}$ represents a controlled-Z gate $U_{CZ}.$ 

To determine conditions for eliminating the dispersive shift dynamics
in the case of resonance condition 1, we can compare the action of
$U_{m}$ and $U_{m}^{\prime}=e^{-i\left(V_{qq}+\emph{\ensuremath{\Lambda_{n}}}\right)\tau_{m}}=e^{-i\emph{\ensuremath{\Lambda_{n}}}\tau_{m}}U_{m},$
both of which now act in the full two-qubit space, on a general state
$\ket{\psi_{i}}$ in this space {[}Eq.~(\ref{eq:Psiigen}){]} via
\begin{align}
F_{s} & =\left|\bra{\psi_{i}}U_{m}^{\dagger}U_{m}^{\prime}\ket{\psi_{i}}\right|^{2}\nonumber \\
 & =\left|\bra{\psi_{i}}e^{-i\emph{\ensuremath{\Lambda_{n}}}\tau_{m}}\ket{\psi_{i}}\right|^{2}\nonumber \\
 & =\left|\left|c_{ee}\right|^{2}e^{-i\left(n+\frac{1}{2}\right)\left(\chi_{1}+\chi_{2}\right)\tau_{m}}\right.\nonumber \\
 & \ \ \ \ +\left|c_{eg}\right|^{2}e^{-i\left(n+\frac{1}{2}\right)\left(\chi_{1}-\chi_{2}\right)\tau_{m}}\nonumber \\
 & \ \ \ \ +\left|c_{ge}\right|^{2}e^{i\left(n+\frac{1}{2}\right)\left(\chi_{1}-\chi_{2}\right)\tau_{m}}\nonumber \\
 & \ \ \ \ \left.+\left|c_{gg}\right|^{2}e^{i\left(n+\frac{1}{2}\right)\left(\chi_{1}+\chi_{2}\right)\tau_{m}}\right|^{2}.\label{eq:eq:FsRC1}
\end{align}
In this case, we can in principle recover the ideal dynamics and obtain
$F_{s}=1$ for any state $\ket{\psi_{i}}$ by choosing parameters
that satisfy $\chi_{j}=2\pi k_{j}$ with $k_{j}$ an integer for $j=1,2,$
and that also satisfy $\tau_{m}=-\varphi/8\mathcal{J},$ $p_{1}=p_{2}$
from the resonance condition, and $q_{1}\neq\pm q_{2}$ from Table
\ref{tab:Vqqresterms} to obtain a controlled-phase gate $U_{\varphi}$
according to the construction in Eq.~(\ref{eq:Uphi}). For $U_{CZ},$
$\tau_{m}=\left(2l+1\right)\pi/8\mathcal{J},$ and we can proceed
to identify potential parameter sets by choosing, e.g., $l=0$ and
setting $\tau_{m}\equiv2\pi m/\eta=-\pi/8\mathcal{J}$ in analogy
to the analysis described above for resonance conditions 7 and 9. 

The remaining physically distinct case in Table \ref{tab:Vqqresterms}
is that of resonance condition 4, given by $\Delta\equiv\Delta_{1}^{+}=\Delta_{2}$
and corresponding to $V_{qq}=\mathcal{J}\sigma_{1}^{x}\sigma_{2}^{z}$
when $W_{1}$ and $W_{2}$ satisfy the listed constraints. This interaction
can be used to construct a controlled-NOT (CNOT) gate \cite{Chow2011}.
We note that $V_{qq},$ and therefore $U_{m}$ and $U_{m}^{\prime},$
are block-diagonal with the associated subspace projectors $P^{\prime}=\ket{ee}\bra{ee}+\ket{ge}\bra{ge}$
and $Q^{\prime}=\ket{eg}\bra{eg}+\ket{gg}\bra{gg}.$ If we choose
$\ket{\psi_{i}}=\ket{eg}$ such that $P^{\prime}U_{m}P^{\prime}\ket{\psi_{i}}=P^{\prime}U_{m}^{\prime}P^{\prime}\ket{\psi_{i}}=0,$
we find from a calculation similar to that in Eqs.~(\ref{eq:Fs})
and (\ref{eq:FsRC7}),
\begin{align}
F_{s} & =\left|\bra{\psi_{i}}U_{m}^{\dagger}U_{m}^{\prime}\ket{\psi_{i}}\right|^{2}\nonumber \\
 & =\frac{4\mathcal{J}^{2}}{\bar{\Omega}^{2}}\sin^{2}\left(\frac{\bar{\Omega}\tau_{m}}{2}\right)\label{eq:FsRC4}
\end{align}
with $\bar{\Omega}\equiv\sqrt{\chi_{1}+4\mathcal{J}^{2}}.$ In this
case, recovering the dynamics for $\emph{\ensuremath{\Lambda_{r}}}=0$
requires $\chi_{1}=0,$ which is not possible for $\delta_{j}=0$
and $g_{1},2\Omega_{1}\neq0$ {[}see Eq.~(\ref{eq:Lambdar}){]}.
Similar results hold for other initial states $\ket{\psi_{i}}.$ While
$F_{s}=1$ is not achievable exactly for resonance condition 4 when
$\delta_{j}=0,$ choosing appropriate parameters to minimize $\chi_{j}$
in principle allows $F_{s}$ to be made arbitrarily close to unity,
enabling the dynamics for $\emph{\ensuremath{\Lambda_{r}}}=0$ and
thus the gate generated by $V_{qq}=\mathcal{J}\sigma_{1}^{x}\sigma_{2}^{z}$
to be well approximated. The dispersive shift dynamics can also be
eliminated in this case for $\delta_{j}\neq0,$ as we describe below.
The other resonance conditions in Table \ref{tab:Vqqresterms} lead
to effective interactions $V_{qq}$ that are either identical in form
to those considered above or have the roles of qubits 1 and 2 reversed. 

Finally, we consider the conditions under which the one-qubit contribution
to $H_{{\rm eff}}\left(\tau\right),$ including the dispersive shift
terms $\Lambda,$ can be made to vanish for each qubit individually.
In this case, $H_{{\rm eff}}\left(\tau\right)=V_{qq}\left(\tau\right)$
consists solely of qubit-qubit interaction terms. Setting $\Lambda=0$
in Eq.~(\ref{eq:Lambda}) yields the condition $\delta_{j}^{2}+2\Delta_{j}\delta_{j}+W_{j}^{2}=0,$
or equivalently $\delta_{j}^{2}+2p_{j}\eta\delta_{j}+q_{j}^{2}\eta^{2}=0.$
We note that for resonant driving, described by $\delta_{j}=0,$ the
one-qubit contribution $\Lambda\neq0$ unless $W_{j}=0,$ which is
not physically meaningful (see Appendix \ref{sec:Onequbitterms}).
Thus, provided the drive frequency is not on resonance with the qubit
frequency $\left(\omega_{j}^{d}\neq\omega_{j}\right),$ the dispersive
shift terms can be eliminated individually for each qubit by an appropriate
choice of parameters. 

\bibliography{sqparcp}

\begin{thebibliography}{95}%
\makeatletter
\providecommand \@ifxundefined [1]{%
 \@ifx{#1\undefined}
}%
\providecommand \@ifnum [1]{%
 \ifnum #1\expandafter \@firstoftwo
 \else \expandafter \@secondoftwo
 \fi
}%
\providecommand \@ifx [1]{%
 \ifx #1\expandafter \@firstoftwo
 \else \expandafter \@secondoftwo
 \fi
}%
\providecommand \natexlab [1]{#1}%
\providecommand \enquote  [1]{``#1''}%
\providecommand \bibnamefont  [1]{#1}%
\providecommand \bibfnamefont [1]{#1}%
\providecommand \citenamefont [1]{#1}%
\providecommand \href@noop [0]{\@secondoftwo}%
\providecommand \href [0]{\begingroup \@sanitize@url \@href}%
\providecommand \@href[1]{\@@startlink{#1}\@@href}%
\providecommand \@@href[1]{\endgroup#1\@@endlink}%
\providecommand \@sanitize@url [0]{\catcode `\\12\catcode `\$12\catcode
  `\&12\catcode `\#12\catcode `\^12\catcode `\_12\catcode `\%12\relax}%
\providecommand \@@startlink[1]{}%
\providecommand \@@endlink[0]{}%
\providecommand \url  [0]{\begingroup\@sanitize@url \@url }%
\providecommand \@url [1]{\endgroup\@href {#1}{\urlprefix }}%
\providecommand \urlprefix  [0]{URL }%
\providecommand \Eprint [0]{\href }%
\providecommand \doibase [0]{https://doi.org/}%
\providecommand \selectlanguage [0]{\@gobble}%
\providecommand \bibinfo  [0]{\@secondoftwo}%
\providecommand \bibfield  [0]{\@secondoftwo}%
\providecommand \translation [1]{[#1]}%
\providecommand \BibitemOpen [0]{}%
\providecommand \bibitemStop [0]{}%
\providecommand \bibitemNoStop [0]{.\EOS\space}%
\providecommand \EOS [0]{\spacefactor3000\relax}%
\providecommand \BibitemShut  [1]{\csname bibitem#1\endcsname}%
\let\auto@bib@innerbib\@empty
\bibitem [{\citenamefont {DiVincenzo}(2000)}]{DiVincenzo2000FortschrPhys}%
  \BibitemOpen
  \bibfield  {author} {\bibinfo {author} {\bibfnamefont {D.~P.}\ \bibnamefont
  {DiVincenzo}},\ }\bibfield  {title} {\bibinfo {title} {{{The Physical
  Implementation of Quantum Computation}}},\ }\href@noop {} {\bibfield
  {journal} {\bibinfo  {journal} {Fortschr. Phys.}\ }\textbf {\bibinfo {volume}
  {48}},\ \bibinfo {pages} {771} (\bibinfo {year} {2000})}\BibitemShut
  {NoStop}%
\bibitem [{\citenamefont {Ladd}\ \emph {et~al.}(2010)\citenamefont {Ladd},
  \citenamefont {Jelezko}, \citenamefont {Laflamme}, \citenamefont {Nakamura},
  \citenamefont {Monroe},\ and\ \citenamefont {O'Brien}}]{Ladd2010}%
  \BibitemOpen
  \bibfield  {author} {\bibinfo {author} {\bibfnamefont {T.~D.}\ \bibnamefont
  {Ladd}}, \bibinfo {author} {\bibfnamefont {F.}~\bibnamefont {Jelezko}},
  \bibinfo {author} {\bibfnamefont {R.}~\bibnamefont {Laflamme}}, \bibinfo
  {author} {\bibfnamefont {Y.}~\bibnamefont {Nakamura}}, \bibinfo {author}
  {\bibfnamefont {C.}~\bibnamefont {Monroe}},\ and\ \bibinfo {author}
  {\bibfnamefont {J.~L.}\ \bibnamefont {O'Brien}},\ }\bibfield  {title}
  {\bibinfo {title} {{Quantum computers}},\ }\href@noop {} {\bibfield
  {journal} {\bibinfo  {journal} {Nature}\ }\textbf {\bibinfo {volume} {464}},\
  \bibinfo {pages} {45} (\bibinfo {year} {2010})}\BibitemShut {NoStop}%
\bibitem [{\citenamefont {Taylor}\ \emph {et~al.}(2005)\citenamefont {Taylor},
  \citenamefont {Engel}, \citenamefont {Dur}, \citenamefont {Yacoby},
  \citenamefont {Marcus}, \citenamefont {Zoller},\ and\ \citenamefont
  {Lukin}}]{Taylor2005}%
  \BibitemOpen
  \bibfield  {author} {\bibinfo {author} {\bibfnamefont {J.~M.}\ \bibnamefont
  {Taylor}}, \bibinfo {author} {\bibfnamefont {H.~A.}\ \bibnamefont {Engel}},
  \bibinfo {author} {\bibfnamefont {W.}~\bibnamefont {Dur}}, \bibinfo {author}
  {\bibfnamefont {A.}~\bibnamefont {Yacoby}}, \bibinfo {author} {\bibfnamefont
  {C.~M.}\ \bibnamefont {Marcus}}, \bibinfo {author} {\bibfnamefont
  {P.}~\bibnamefont {Zoller}},\ and\ \bibinfo {author} {\bibfnamefont {M.~D.}\
  \bibnamefont {Lukin}},\ }\bibfield  {title} {\bibinfo {title}
  {{Fault-tolerant architecture for quantum computation using electrically
  controlled semiconductor spins}},\ }\href@noop {} {\bibfield  {journal}
  {\bibinfo  {journal} {Nat. Phys.}\ }\textbf {\bibinfo {volume} {1}},\
  \bibinfo {pages} {177} (\bibinfo {year} {2005})}\BibitemShut {NoStop}%
\bibitem [{\citenamefont {Jiang}\ \emph {et~al.}(2007)\citenamefont {Jiang},
  \citenamefont {Taylor}, \citenamefont {S\o{}rensen},\ and\ \citenamefont
  {Lukin}}]{Jiang2007}%
  \BibitemOpen
  \bibfield  {author} {\bibinfo {author} {\bibfnamefont {L.}~\bibnamefont
  {Jiang}}, \bibinfo {author} {\bibfnamefont {J.~M.}\ \bibnamefont {Taylor}},
  \bibinfo {author} {\bibfnamefont {A.~S.}\ \bibnamefont {S\o{}rensen}},\ and\
  \bibinfo {author} {\bibfnamefont {M.~D.}\ \bibnamefont {Lukin}},\ }\bibfield
  {title} {\bibinfo {title} {{Distributed quantum computation based on small
  quantum registers}},\ }\href@noop {} {\bibfield  {journal} {\bibinfo
  {journal} {Phys. Rev. A}\ }\textbf {\bibinfo {volume} {76}},\ \bibinfo
  {pages} {062323} (\bibinfo {year} {2007})}\BibitemShut {NoStop}%
\bibitem [{\citenamefont {Monroe}\ \emph {et~al.}(2014)\citenamefont {Monroe},
  \citenamefont {Raussendorf}, \citenamefont {Ruthven}, \citenamefont {Brown},
  \citenamefont {Maunz}, \citenamefont {Duan},\ and\ \citenamefont
  {Kim}}]{Monroe2014}%
  \BibitemOpen
  \bibfield  {author} {\bibinfo {author} {\bibfnamefont {C.}~\bibnamefont
  {Monroe}}, \bibinfo {author} {\bibfnamefont {R.}~\bibnamefont {Raussendorf}},
  \bibinfo {author} {\bibfnamefont {A.}~\bibnamefont {Ruthven}}, \bibinfo
  {author} {\bibfnamefont {K.~R.}\ \bibnamefont {Brown}}, \bibinfo {author}
  {\bibfnamefont {P.}~\bibnamefont {Maunz}}, \bibinfo {author} {\bibfnamefont
  {L.-M.}\ \bibnamefont {Duan}},\ and\ \bibinfo {author} {\bibfnamefont
  {J.}~\bibnamefont {Kim}},\ }\bibfield  {title} {\bibinfo {title}
  {{Large-scale modular quantum-computer architecture with atomic memory and
  photonic interconnects}},\ }\href@noop {} {\bibfield  {journal} {\bibinfo
  {journal} {Phys. Rev. A}\ }\textbf {\bibinfo {volume} {89}},\ \bibinfo
  {pages} {022317} (\bibinfo {year} {2014})}\BibitemShut {NoStop}%
\bibitem [{\citenamefont {Vandersypen}\ \emph {et~al.}(2017)\citenamefont
  {Vandersypen}, \citenamefont {Bluhm}, \citenamefont {Clarke}, \citenamefont
  {Dzurak}, \citenamefont {Ishihara}, \citenamefont {Morello}, \citenamefont
  {Reilly}, \citenamefont {Schreiber},\ and\ \citenamefont
  {Veldhorst}}]{Vandersypen2017}%
  \BibitemOpen
  \bibfield  {author} {\bibinfo {author} {\bibfnamefont {L.~M.~K.}\
  \bibnamefont {Vandersypen}}, \bibinfo {author} {\bibfnamefont
  {H.}~\bibnamefont {Bluhm}}, \bibinfo {author} {\bibfnamefont {J.~S.}\
  \bibnamefont {Clarke}}, \bibinfo {author} {\bibfnamefont {A.~S.}\
  \bibnamefont {Dzurak}}, \bibinfo {author} {\bibfnamefont {R.}~\bibnamefont
  {Ishihara}}, \bibinfo {author} {\bibfnamefont {A.}~\bibnamefont {Morello}},
  \bibinfo {author} {\bibfnamefont {D.~J.}\ \bibnamefont {Reilly}}, \bibinfo
  {author} {\bibfnamefont {L.~R.}\ \bibnamefont {Schreiber}},\ and\ \bibinfo
  {author} {\bibfnamefont {M.}~\bibnamefont {Veldhorst}},\ }\bibfield  {title}
  {\bibinfo {title} {Interfacing spin qubits in quantum dots and donors---hot,
  dense, and coherent},\ }\href@noop {} {\bibfield  {journal} {\bibinfo
  {journal} {npj Quantum Inf.}\ }\textbf {\bibinfo {volume} {3}},\ \bibinfo
  {pages} {34} (\bibinfo {year} {2017})}\BibitemShut {NoStop}%
\bibitem [{\citenamefont {Tosi}\ \emph {et~al.}(2017)\citenamefont {Tosi},
  \citenamefont {Mohiyaddin}, \citenamefont {Schmitt}, \citenamefont {Tenberg},
  \citenamefont {Rahman}, \citenamefont {Klimeck},\ and\ \citenamefont
  {Morello}}]{Tosi2017}%
  \BibitemOpen
  \bibfield  {author} {\bibinfo {author} {\bibfnamefont {G.}~\bibnamefont
  {Tosi}}, \bibinfo {author} {\bibfnamefont {F.~A.}\ \bibnamefont
  {Mohiyaddin}}, \bibinfo {author} {\bibfnamefont {V.}~\bibnamefont {Schmitt}},
  \bibinfo {author} {\bibfnamefont {S.}~\bibnamefont {Tenberg}}, \bibinfo
  {author} {\bibfnamefont {R.}~\bibnamefont {Rahman}}, \bibinfo {author}
  {\bibfnamefont {G.}~\bibnamefont {Klimeck}},\ and\ \bibinfo {author}
  {\bibfnamefont {A.}~\bibnamefont {Morello}},\ }\bibfield  {title} {\bibinfo
  {title} {Silicon quantum processor with robust long-distance qubit
  couplings},\ }\href@noop {} {\bibfield  {journal} {\bibinfo  {journal} {Nat.
  Commun.}\ }\textbf {\bibinfo {volume} {8}},\ \bibinfo {pages} {450} (\bibinfo
  {year} {2017})}\BibitemShut {NoStop}%
\bibitem [{\citenamefont {Jnane}\ \emph {et~al.}(2022)\citenamefont {Jnane},
  \citenamefont {Undseth}, \citenamefont {Cai}, \citenamefont {Benjamin},\ and\
  \citenamefont {Koczor}}]{Jnane2022}%
  \BibitemOpen
  \bibfield  {author} {\bibinfo {author} {\bibfnamefont {H.}~\bibnamefont
  {Jnane}}, \bibinfo {author} {\bibfnamefont {B.}~\bibnamefont {Undseth}},
  \bibinfo {author} {\bibfnamefont {Z.}~\bibnamefont {Cai}}, \bibinfo {author}
  {\bibfnamefont {S.~C.}\ \bibnamefont {Benjamin}},\ and\ \bibinfo {author}
  {\bibfnamefont {B.}~\bibnamefont {Koczor}},\ }\bibfield  {title} {\bibinfo
  {title} {Multicore quantum computing},\ }\href@noop {} {\bibfield  {journal}
  {\bibinfo  {journal} {Phys. Rev. Applied}\ }\textbf {\bibinfo {volume}
  {18}},\ \bibinfo {pages} {044064} (\bibinfo {year} {2022})}\BibitemShut
  {NoStop}%
\bibitem [{\citenamefont {Loss}\ and\ \citenamefont
  {DiVincenzo}(1998)}]{Loss1998}%
  \BibitemOpen
  \bibfield  {author} {\bibinfo {author} {\bibfnamefont {D.}~\bibnamefont
  {Loss}}\ and\ \bibinfo {author} {\bibfnamefont {D.~P.}\ \bibnamefont
  {DiVincenzo}},\ }\bibfield  {title} {\bibinfo {title} {{Quantum computation
  with quantum dots}},\ }\href@noop {} {\bibfield  {journal} {\bibinfo
  {journal} {Phys. Rev. A}\ }\textbf {\bibinfo {volume} {57}},\ \bibinfo
  {pages} {120} (\bibinfo {year} {1998})}\BibitemShut {NoStop}%
\bibitem [{\citenamefont {Kane}(1998)}]{Kane1998}%
  \BibitemOpen
  \bibfield  {author} {\bibinfo {author} {\bibfnamefont {B.~E.}\ \bibnamefont
  {Kane}},\ }\bibfield  {title} {\bibinfo {title} {{A silicon-based nuclear
  spin quantum computer}},\ }\href@noop {} {\bibfield  {journal} {\bibinfo
  {journal} {Nature}\ }\textbf {\bibinfo {volume} {393}},\ \bibinfo {pages}
  {133} (\bibinfo {year} {1998})}\BibitemShut {NoStop}%
\bibitem [{\citenamefont {Hanson}\ \emph {et~al.}(2007)\citenamefont {Hanson},
  \citenamefont {Kouwenhoven}, \citenamefont {Petta}, \citenamefont {Tarucha},\
  and\ \citenamefont {Vandersypen}}]{Hanson2007RMP}%
  \BibitemOpen
  \bibfield  {author} {\bibinfo {author} {\bibfnamefont {R.}~\bibnamefont
  {Hanson}}, \bibinfo {author} {\bibfnamefont {L.~P.}\ \bibnamefont
  {Kouwenhoven}}, \bibinfo {author} {\bibfnamefont {J.~R.}\ \bibnamefont
  {Petta}}, \bibinfo {author} {\bibfnamefont {S.}~\bibnamefont {Tarucha}},\
  and\ \bibinfo {author} {\bibfnamefont {L.~M.~K.}\ \bibnamefont
  {Vandersypen}},\ }\bibfield  {title} {\bibinfo {title} {{Spins in
  few-electron quantum dots}},\ }\href@noop {} {\bibfield  {journal} {\bibinfo
  {journal} {Rev. Mod. Phys.}\ }\textbf {\bibinfo {volume} {79}},\ \bibinfo
  {pages} {1217} (\bibinfo {year} {2007})}\BibitemShut {NoStop}%
\bibitem [{\citenamefont {Zwanenburg}\ \emph {et~al.}(2013)\citenamefont
  {Zwanenburg}, \citenamefont {Dzurak}, \citenamefont {Morello}, \citenamefont
  {Simmons}, \citenamefont {Hollenberg}, \citenamefont {Klimeck}, \citenamefont
  {Rogge}, \citenamefont {Coppersmith},\ and\ \citenamefont
  {Eriksson}}]{Zwanenburg2013}%
  \BibitemOpen
  \bibfield  {author} {\bibinfo {author} {\bibfnamefont {F.~A.}\ \bibnamefont
  {Zwanenburg}}, \bibinfo {author} {\bibfnamefont {A.~S.}\ \bibnamefont
  {Dzurak}}, \bibinfo {author} {\bibfnamefont {A.}~\bibnamefont {Morello}},
  \bibinfo {author} {\bibfnamefont {M.~Y.}\ \bibnamefont {Simmons}}, \bibinfo
  {author} {\bibfnamefont {L.~C.~L.}\ \bibnamefont {Hollenberg}}, \bibinfo
  {author} {\bibfnamefont {G.}~\bibnamefont {Klimeck}}, \bibinfo {author}
  {\bibfnamefont {S.}~\bibnamefont {Rogge}}, \bibinfo {author} {\bibfnamefont
  {S.~N.}\ \bibnamefont {Coppersmith}},\ and\ \bibinfo {author} {\bibfnamefont
  {M.~A.}\ \bibnamefont {Eriksson}},\ }\bibfield  {title} {\bibinfo {title}
  {{Silicon quantum electronics}},\ }\href@noop {} {\bibfield  {journal}
  {\bibinfo  {journal} {Rev. Mod. Phys.}\ }\textbf {\bibinfo {volume} {85}},\
  \bibinfo {pages} {961} (\bibinfo {year} {2013})}\BibitemShut {NoStop}%
\bibitem [{\citenamefont {Petit}\ \emph {et~al.}(2020)\citenamefont {Petit},
  \citenamefont {Eenink}, \citenamefont {Russ}, \citenamefont {Lawrie},
  \citenamefont {Hendrickx}, \citenamefont {Philips}, \citenamefont {Clarke},
  \citenamefont {Vandersypen},\ and\ \citenamefont {Veldhorst}}]{Petit2020}%
  \BibitemOpen
  \bibfield  {author} {\bibinfo {author} {\bibfnamefont {L.}~\bibnamefont
  {Petit}}, \bibinfo {author} {\bibfnamefont {H.~G.~J.}\ \bibnamefont
  {Eenink}}, \bibinfo {author} {\bibfnamefont {M.}~\bibnamefont {Russ}},
  \bibinfo {author} {\bibfnamefont {W.~I.~L.}\ \bibnamefont {Lawrie}}, \bibinfo
  {author} {\bibfnamefont {N.~W.}\ \bibnamefont {Hendrickx}}, \bibinfo {author}
  {\bibfnamefont {S.~G.~J.}\ \bibnamefont {Philips}}, \bibinfo {author}
  {\bibfnamefont {J.~S.}\ \bibnamefont {Clarke}}, \bibinfo {author}
  {\bibfnamefont {L.~M.~K.}\ \bibnamefont {Vandersypen}},\ and\ \bibinfo
  {author} {\bibfnamefont {M.}~\bibnamefont {Veldhorst}},\ }\bibfield  {title}
  {\bibinfo {title} {Universal quantum logic in hot silicon qubits},\
  }\href@noop {} {\bibfield  {journal} {\bibinfo  {journal} {Nature}\ }\textbf
  {\bibinfo {volume} {580}},\ \bibinfo {pages} {355} (\bibinfo {year}
  {2020})}\BibitemShut {NoStop}%
\bibitem [{\citenamefont {Chatterjee}\ \emph {et~al.}(2021)\citenamefont
  {Chatterjee}, \citenamefont {Stevenson}, \citenamefont {De~Franceschi},
  \citenamefont {Morello}, \citenamefont {de~Leon},\ and\ \citenamefont
  {Kuemmeth}}]{Chatterjee2021}%
  \BibitemOpen
  \bibfield  {author} {\bibinfo {author} {\bibfnamefont {A.}~\bibnamefont
  {Chatterjee}}, \bibinfo {author} {\bibfnamefont {P.}~\bibnamefont
  {Stevenson}}, \bibinfo {author} {\bibfnamefont {S.}~\bibnamefont
  {De~Franceschi}}, \bibinfo {author} {\bibfnamefont {A.}~\bibnamefont
  {Morello}}, \bibinfo {author} {\bibfnamefont {N.~P.}\ \bibnamefont
  {de~Leon}},\ and\ \bibinfo {author} {\bibfnamefont {F.}~\bibnamefont
  {Kuemmeth}},\ }\bibfield  {title} {\bibinfo {title} {Semiconductor qubits in
  practice},\ }\href@noop {} {\bibfield  {journal} {\bibinfo  {journal} {Nat.
  Rev. Phys.}\ }\textbf {\bibinfo {volume} {3}},\ \bibinfo {pages} {157}
  (\bibinfo {year} {2021})}\BibitemShut {NoStop}%
\bibitem [{\citenamefont {Noiri}\ \emph {et~al.}(2022)\citenamefont {Noiri},
  \citenamefont {Takeda}, \citenamefont {Nakajima}, \citenamefont {Kobayashi},
  \citenamefont {Sammak}, \citenamefont {Scappucci},\ and\ \citenamefont
  {Tarucha}}]{Noiri2022}%
  \BibitemOpen
  \bibfield  {author} {\bibinfo {author} {\bibfnamefont {A.}~\bibnamefont
  {Noiri}}, \bibinfo {author} {\bibfnamefont {K.}~\bibnamefont {Takeda}},
  \bibinfo {author} {\bibfnamefont {T.}~\bibnamefont {Nakajima}}, \bibinfo
  {author} {\bibfnamefont {T.}~\bibnamefont {Kobayashi}}, \bibinfo {author}
  {\bibfnamefont {A.}~\bibnamefont {Sammak}}, \bibinfo {author} {\bibfnamefont
  {G.}~\bibnamefont {Scappucci}},\ and\ \bibinfo {author} {\bibfnamefont
  {S.}~\bibnamefont {Tarucha}},\ }\bibfield  {title} {\bibinfo {title} {Fast
  universal quantum gate above the fault-tolerance threshold in silicon},\
  }\href@noop {} {\bibfield  {journal} {\bibinfo  {journal} {Nature}\ }\textbf
  {\bibinfo {volume} {601}},\ \bibinfo {pages} {338} (\bibinfo {year}
  {2022})}\BibitemShut {NoStop}%
\bibitem [{\citenamefont {Xue}\ \emph {et~al.}(2022)\citenamefont {Xue},
  \citenamefont {Russ}, \citenamefont {Samkharadze}, \citenamefont {Undseth},
  \citenamefont {Sammak}, \citenamefont {Scappucci},\ and\ \citenamefont
  {Vandersypen}}]{Xue2022}%
  \BibitemOpen
  \bibfield  {author} {\bibinfo {author} {\bibfnamefont {X.}~\bibnamefont
  {Xue}}, \bibinfo {author} {\bibfnamefont {M.}~\bibnamefont {Russ}}, \bibinfo
  {author} {\bibfnamefont {N.}~\bibnamefont {Samkharadze}}, \bibinfo {author}
  {\bibfnamefont {B.}~\bibnamefont {Undseth}}, \bibinfo {author} {\bibfnamefont
  {A.}~\bibnamefont {Sammak}}, \bibinfo {author} {\bibfnamefont
  {G.}~\bibnamefont {Scappucci}},\ and\ \bibinfo {author} {\bibfnamefont
  {L.~M.~K.}\ \bibnamefont {Vandersypen}},\ }\bibfield  {title} {\bibinfo
  {title} {Quantum logic with spin qubits crossing the surface code
  threshold},\ }\href@noop {} {\bibfield  {journal} {\bibinfo  {journal}
  {Nature}\ }\textbf {\bibinfo {volume} {601}},\ \bibinfo {pages} {343}
  (\bibinfo {year} {2022})}\BibitemShut {NoStop}%
\bibitem [{\citenamefont {Mills}\ \emph {et~al.}(2022)\citenamefont {Mills},
  \citenamefont {Guinn}, \citenamefont {Gullans}, \citenamefont {Sigillito},
  \citenamefont {Feldman}, \citenamefont {Nielsen},\ and\ \citenamefont
  {Petta}}]{Mills2022}%
  \BibitemOpen
  \bibfield  {author} {\bibinfo {author} {\bibfnamefont {A.~R.}\ \bibnamefont
  {Mills}}, \bibinfo {author} {\bibfnamefont {C.~R.}\ \bibnamefont {Guinn}},
  \bibinfo {author} {\bibfnamefont {M.~J.}\ \bibnamefont {Gullans}}, \bibinfo
  {author} {\bibfnamefont {A.~J.}\ \bibnamefont {Sigillito}}, \bibinfo {author}
  {\bibfnamefont {M.~M.}\ \bibnamefont {Feldman}}, \bibinfo {author}
  {\bibfnamefont {E.}~\bibnamefont {Nielsen}},\ and\ \bibinfo {author}
  {\bibfnamefont {J.~R.}\ \bibnamefont {Petta}},\ }\bibfield  {title} {\bibinfo
  {title} {Two-qubit silicon quantum processor with operation fidelity
  exceeding 99\%},\ }\href@noop {} {\bibfield  {journal} {\bibinfo  {journal}
  {Sci. Adv.}\ }\textbf {\bibinfo {volume} {8}},\ \bibinfo {pages} {eabn5130}
  (\bibinfo {year} {2022})}\BibitemShut {NoStop}%
\bibitem [{\citenamefont {Weinstein}\ \emph {et~al.}(2023)\citenamefont
  {Weinstein}, \citenamefont {Reed}, \citenamefont {Jones}, \citenamefont
  {Andrews}, \citenamefont {Barnes}, \citenamefont {Blumoff}, \citenamefont
  {Euliss}, \citenamefont {Eng}, \citenamefont {Fong}, \citenamefont {Ha},
  \citenamefont {Hulbert}, \citenamefont {Jackson}, \citenamefont {Jura},
  \citenamefont {Keating}, \citenamefont {Kerckhoff}, \citenamefont {Kiselev},
  \citenamefont {Matten}, \citenamefont {Sabbir}, \citenamefont {Smith},
  \citenamefont {Wright}, \citenamefont {Rakher}, \citenamefont {Ladd},\ and\
  \citenamefont {Borselli}}]{Weinstein2023}%
  \BibitemOpen
  \bibfield  {author} {\bibinfo {author} {\bibfnamefont {A.~J.}\ \bibnamefont
  {Weinstein}}, \bibinfo {author} {\bibfnamefont {M.~D.}\ \bibnamefont {Reed}},
  \bibinfo {author} {\bibfnamefont {A.~M.}\ \bibnamefont {Jones}}, \bibinfo
  {author} {\bibfnamefont {R.~W.}\ \bibnamefont {Andrews}}, \bibinfo {author}
  {\bibfnamefont {D.}~\bibnamefont {Barnes}}, \bibinfo {author} {\bibfnamefont
  {J.~Z.}\ \bibnamefont {Blumoff}}, \bibinfo {author} {\bibfnamefont {L.~E.}\
  \bibnamefont {Euliss}}, \bibinfo {author} {\bibfnamefont {K.}~\bibnamefont
  {Eng}}, \bibinfo {author} {\bibfnamefont {B.~H.}\ \bibnamefont {Fong}},
  \bibinfo {author} {\bibfnamefont {S.~D.}\ \bibnamefont {Ha}}, \bibinfo
  {author} {\bibfnamefont {D.~R.}\ \bibnamefont {Hulbert}}, \bibinfo {author}
  {\bibfnamefont {C.~A.~C.}\ \bibnamefont {Jackson}}, \bibinfo {author}
  {\bibfnamefont {M.}~\bibnamefont {Jura}}, \bibinfo {author} {\bibfnamefont
  {T.~E.}\ \bibnamefont {Keating}}, \bibinfo {author} {\bibfnamefont
  {J.}~\bibnamefont {Kerckhoff}}, \bibinfo {author} {\bibfnamefont {A.~A.}\
  \bibnamefont {Kiselev}}, \bibinfo {author} {\bibfnamefont {J.}~\bibnamefont
  {Matten}}, \bibinfo {author} {\bibfnamefont {G.}~\bibnamefont {Sabbir}},
  \bibinfo {author} {\bibfnamefont {A.}~\bibnamefont {Smith}}, \bibinfo
  {author} {\bibfnamefont {J.}~\bibnamefont {Wright}}, \bibinfo {author}
  {\bibfnamefont {M.~T.}\ \bibnamefont {Rakher}}, \bibinfo {author}
  {\bibfnamefont {T.~D.}\ \bibnamefont {Ladd}},\ and\ \bibinfo {author}
  {\bibfnamefont {M.~G.}\ \bibnamefont {Borselli}},\ }\bibfield  {title}
  {\bibinfo {title} {Universal logic with encoded spin qubits in silicon},\
  }\href@noop {} {\bibfield  {journal} {\bibinfo  {journal} {Nature}\ }\textbf
  {\bibinfo {volume} {615}},\ \bibinfo {pages} {817} (\bibinfo {year}
  {2023})}\BibitemShut {NoStop}%
\bibitem [{\citenamefont {Burkard}\ \emph {et~al.}(2023)\citenamefont
  {Burkard}, \citenamefont {Ladd}, \citenamefont {Pan}, \citenamefont
  {Nichol},\ and\ \citenamefont {Petta}}]{Burkard2023}%
  \BibitemOpen
  \bibfield  {author} {\bibinfo {author} {\bibfnamefont {G.}~\bibnamefont
  {Burkard}}, \bibinfo {author} {\bibfnamefont {T.~D.}\ \bibnamefont {Ladd}},
  \bibinfo {author} {\bibfnamefont {A.}~\bibnamefont {Pan}}, \bibinfo {author}
  {\bibfnamefont {J.~M.}\ \bibnamefont {Nichol}},\ and\ \bibinfo {author}
  {\bibfnamefont {J.~R.}\ \bibnamefont {Petta}},\ }\bibfield  {title} {\bibinfo
  {title} {Semiconductor spin qubits},\ }\href@noop {} {\bibfield  {journal}
  {\bibinfo  {journal} {Rev. Mod. Phys.}\ }\textbf {\bibinfo {volume} {95}},\
  \bibinfo {pages} {025003} (\bibinfo {year} {2023})}\BibitemShut {NoStop}%
\bibitem [{\citenamefont {Childress}\ \emph {et~al.}(2004)\citenamefont
  {Childress}, \citenamefont {S\o{}rensen},\ and\ \citenamefont
  {Lukin}}]{Childress2004}%
  \BibitemOpen
  \bibfield  {author} {\bibinfo {author} {\bibfnamefont {L.}~\bibnamefont
  {Childress}}, \bibinfo {author} {\bibfnamefont {A.~S.}\ \bibnamefont
  {S\o{}rensen}},\ and\ \bibinfo {author} {\bibfnamefont {M.~D.}\ \bibnamefont
  {Lukin}},\ }\bibfield  {title} {\bibinfo {title} {{Mesoscopic cavity quantum
  electrodynamics with quantum dots}},\ }\href@noop {} {\bibfield  {journal}
  {\bibinfo  {journal} {Phys. Rev. A}\ }\textbf {\bibinfo {volume} {69}},\
  \bibinfo {pages} {042302} (\bibinfo {year} {2004})}\BibitemShut {NoStop}%
\bibitem [{\citenamefont {Blais}\ \emph {et~al.}(2004)\citenamefont {Blais},
  \citenamefont {Huang}, \citenamefont {Wallraff}, \citenamefont {Girvin},\
  and\ \citenamefont {Schoelkopf}}]{Blais2004}%
  \BibitemOpen
  \bibfield  {author} {\bibinfo {author} {\bibfnamefont {A.}~\bibnamefont
  {Blais}}, \bibinfo {author} {\bibfnamefont {R.-S.}\ \bibnamefont {Huang}},
  \bibinfo {author} {\bibfnamefont {A.}~\bibnamefont {Wallraff}}, \bibinfo
  {author} {\bibfnamefont {S.~M.}\ \bibnamefont {Girvin}},\ and\ \bibinfo
  {author} {\bibfnamefont {R.~J.}\ \bibnamefont {Schoelkopf}},\ }\bibfield
  {title} {\bibinfo {title} {{Cavity quantum electrodynamics for
  superconducting electrical circuits: An architecture for quantum
  computation}},\ }\href@noop {} {\bibfield  {journal} {\bibinfo  {journal}
  {Phys. Rev. A}\ }\textbf {\bibinfo {volume} {69}},\ \bibinfo {pages} {062320}
  (\bibinfo {year} {2004})}\BibitemShut {NoStop}%
\bibitem [{\citenamefont {Wallraff}\ \emph {et~al.}(2004)\citenamefont
  {Wallraff}, \citenamefont {Schuster}, \citenamefont {Blais}, \citenamefont
  {Frunzio}, \citenamefont {Huang}, \citenamefont {Majer}, \citenamefont
  {Kumar}, \citenamefont {Girvin},\ and\ \citenamefont
  {Schoelkopf}}]{Wallraff2004}%
  \BibitemOpen
  \bibfield  {author} {\bibinfo {author} {\bibfnamefont {A.}~\bibnamefont
  {Wallraff}}, \bibinfo {author} {\bibfnamefont {D.~I.}\ \bibnamefont
  {Schuster}}, \bibinfo {author} {\bibfnamefont {A.}~\bibnamefont {Blais}},
  \bibinfo {author} {\bibfnamefont {L.}~\bibnamefont {Frunzio}}, \bibinfo
  {author} {\bibfnamefont {R.~S.}\ \bibnamefont {Huang}}, \bibinfo {author}
  {\bibfnamefont {J.}~\bibnamefont {Majer}}, \bibinfo {author} {\bibfnamefont
  {S.}~\bibnamefont {Kumar}}, \bibinfo {author} {\bibfnamefont {S.~M.}\
  \bibnamefont {Girvin}},\ and\ \bibinfo {author} {\bibfnamefont {R.~J.}\
  \bibnamefont {Schoelkopf}},\ }\bibfield  {title} {\bibinfo {title} {{Strong
  coupling of a single photon to a superconducting qubit using circuit quantum
  electrodynamics}},\ }\href@noop {} {\bibfield  {journal} {\bibinfo  {journal}
  {Nature}\ }\textbf {\bibinfo {volume} {431}},\ \bibinfo {pages} {162}
  (\bibinfo {year} {2004})}\BibitemShut {NoStop}%
\bibitem [{\citenamefont {Blais}\ \emph {et~al.}(2007)\citenamefont {Blais},
  \citenamefont {Gambetta}, \citenamefont {Wallraff}, \citenamefont {Schuster},
  \citenamefont {Girvin}, \citenamefont {Devoret},\ and\ \citenamefont
  {Schoelkopf}}]{Blais2007}%
  \BibitemOpen
  \bibfield  {author} {\bibinfo {author} {\bibfnamefont {A.}~\bibnamefont
  {Blais}}, \bibinfo {author} {\bibfnamefont {J.}~\bibnamefont {Gambetta}},
  \bibinfo {author} {\bibfnamefont {A.}~\bibnamefont {Wallraff}}, \bibinfo
  {author} {\bibfnamefont {D.~I.}\ \bibnamefont {Schuster}}, \bibinfo {author}
  {\bibfnamefont {S.~M.}\ \bibnamefont {Girvin}}, \bibinfo {author}
  {\bibfnamefont {M.~H.}\ \bibnamefont {Devoret}},\ and\ \bibinfo {author}
  {\bibfnamefont {R.~J.}\ \bibnamefont {Schoelkopf}},\ }\bibfield  {title}
  {\bibinfo {title} {{Quantum-information processing with circuit quantum
  electrodynamics}},\ }\href@noop {} {\bibfield  {journal} {\bibinfo  {journal}
  {Phys. Rev. A}\ }\textbf {\bibinfo {volume} {75}},\ \bibinfo {pages} {032329}
  (\bibinfo {year} {2007})}\BibitemShut {NoStop}%
\bibitem [{\citenamefont {Majer}\ \emph {et~al.}(2007)\citenamefont {Majer},
  \citenamefont {Chow}, \citenamefont {Gambetta}, \citenamefont {Koch},
  \citenamefont {Johnson}, \citenamefont {Schreier}, \citenamefont {Frunzio},
  \citenamefont {Schuster}, \citenamefont {Houck}, \citenamefont {Wallraff},
  \citenamefont {Blais}, \citenamefont {Devoret}, \citenamefont {Girvin},\ and\
  \citenamefont {Schoelkopf}}]{Majer2007}%
  \BibitemOpen
  \bibfield  {author} {\bibinfo {author} {\bibfnamefont {J.}~\bibnamefont
  {Majer}}, \bibinfo {author} {\bibfnamefont {J.~M.}\ \bibnamefont {Chow}},
  \bibinfo {author} {\bibfnamefont {J.~M.}\ \bibnamefont {Gambetta}}, \bibinfo
  {author} {\bibfnamefont {J.}~\bibnamefont {Koch}}, \bibinfo {author}
  {\bibfnamefont {B.~R.}\ \bibnamefont {Johnson}}, \bibinfo {author}
  {\bibfnamefont {J.~A.}\ \bibnamefont {Schreier}}, \bibinfo {author}
  {\bibfnamefont {L.}~\bibnamefont {Frunzio}}, \bibinfo {author} {\bibfnamefont
  {D.~I.}\ \bibnamefont {Schuster}}, \bibinfo {author} {\bibfnamefont {A.~A.}\
  \bibnamefont {Houck}}, \bibinfo {author} {\bibfnamefont {A.}~\bibnamefont
  {Wallraff}}, \bibinfo {author} {\bibfnamefont {A.}~\bibnamefont {Blais}},
  \bibinfo {author} {\bibfnamefont {M.~H.}\ \bibnamefont {Devoret}}, \bibinfo
  {author} {\bibfnamefont {S.~M.}\ \bibnamefont {Girvin}},\ and\ \bibinfo
  {author} {\bibfnamefont {R.~J.}\ \bibnamefont {Schoelkopf}},\ }\bibfield
  {title} {\bibinfo {title} {{Coupling superconducting qubits via a cavity
  bus}},\ }\href@noop {} {\bibfield  {journal} {\bibinfo  {journal} {Nature}\
  }\textbf {\bibinfo {volume} {449}},\ \bibinfo {pages} {443} (\bibinfo {year}
  {2007})}\BibitemShut {NoStop}%
\bibitem [{\citenamefont {Sillanp{\"a}{\"a}}\ \emph {et~al.}(2007)\citenamefont
  {Sillanp{\"a}{\"a}}, \citenamefont {Park},\ and\ \citenamefont
  {Simmonds}}]{Sillanpaa2007}%
  \BibitemOpen
  \bibfield  {author} {\bibinfo {author} {\bibfnamefont {M.~A.}\ \bibnamefont
  {Sillanp{\"a}{\"a}}}, \bibinfo {author} {\bibfnamefont {J.~I.}\ \bibnamefont
  {Park}},\ and\ \bibinfo {author} {\bibfnamefont {R.~W.}\ \bibnamefont
  {Simmonds}},\ }\bibfield  {title} {\bibinfo {title} {{Coherent quantum state
  storage and transfer between two phase qubits via a resonant cavity}},\
  }\href@noop {} {\bibfield  {journal} {\bibinfo  {journal} {Nature}\ }\textbf
  {\bibinfo {volume} {449}},\ \bibinfo {pages} {438} (\bibinfo {year}
  {2007})}\BibitemShut {NoStop}%
\bibitem [{\citenamefont {Blais}\ \emph {et~al.}(2021)\citenamefont {Blais},
  \citenamefont {Grimsmo}, \citenamefont {Girvin},\ and\ \citenamefont
  {Wallraff}}]{Blais2021}%
  \BibitemOpen
  \bibfield  {author} {\bibinfo {author} {\bibfnamefont {A.}~\bibnamefont
  {Blais}}, \bibinfo {author} {\bibfnamefont {A.~L.}\ \bibnamefont {Grimsmo}},
  \bibinfo {author} {\bibfnamefont {S.~M.}\ \bibnamefont {Girvin}},\ and\
  \bibinfo {author} {\bibfnamefont {A.}~\bibnamefont {Wallraff}},\ }\bibfield
  {title} {\bibinfo {title} {Circuit quantum electrodynamics},\ }\href@noop {}
  {\bibfield  {journal} {\bibinfo  {journal} {Rev. Mod. Phys.}\ }\textbf
  {\bibinfo {volume} {93}},\ \bibinfo {pages} {025005} (\bibinfo {year}
  {2021})}\BibitemShut {NoStop}%
\bibitem [{\citenamefont {Clerk}\ \emph {et~al.}(2020)\citenamefont {Clerk},
  \citenamefont {Lehnert}, \citenamefont {Bertet}, \citenamefont {Petta},\ and\
  \citenamefont {Nakamura}}]{Clerk2020}%
  \BibitemOpen
  \bibfield  {author} {\bibinfo {author} {\bibfnamefont {A.~A.}\ \bibnamefont
  {Clerk}}, \bibinfo {author} {\bibfnamefont {K.~W.}\ \bibnamefont {Lehnert}},
  \bibinfo {author} {\bibfnamefont {P.}~\bibnamefont {Bertet}}, \bibinfo
  {author} {\bibfnamefont {J.~R.}\ \bibnamefont {Petta}},\ and\ \bibinfo
  {author} {\bibfnamefont {Y.}~\bibnamefont {Nakamura}},\ }\bibfield  {title}
  {\bibinfo {title} {Hybrid quantum systems with circuit quantum
  electrodynamics},\ }\href@noop {} {\bibfield  {journal} {\bibinfo  {journal}
  {Nat. Phys.}\ }\textbf {\bibinfo {volume} {16}},\ \bibinfo {pages} {257}
  (\bibinfo {year} {2020})}\BibitemShut {NoStop}%
\bibitem [{\citenamefont {Burkard}\ \emph {et~al.}(2020)\citenamefont
  {Burkard}, \citenamefont {Gullans}, \citenamefont {Mi},\ and\ \citenamefont
  {Petta}}]{Burkard2020}%
  \BibitemOpen
  \bibfield  {author} {\bibinfo {author} {\bibfnamefont {G.}~\bibnamefont
  {Burkard}}, \bibinfo {author} {\bibfnamefont {M.~J.}\ \bibnamefont
  {Gullans}}, \bibinfo {author} {\bibfnamefont {X.}~\bibnamefont {Mi}},\ and\
  \bibinfo {author} {\bibfnamefont {J.~R.}\ \bibnamefont {Petta}},\ }\bibfield
  {title} {\bibinfo {title} {Superconductor--semiconductor hybrid-circuit
  quantum electrodynamics},\ }\href@noop {} {\bibfield  {journal} {\bibinfo
  {journal} {Nat. Rev. Phys.}\ }\textbf {\bibinfo {volume} {2}},\ \bibinfo
  {pages} {129} (\bibinfo {year} {2020})}\BibitemShut {NoStop}%
\bibitem [{\citenamefont {Veldhorst}\ \emph {et~al.}(2014)\citenamefont
  {Veldhorst}, \citenamefont {Hwang}, \citenamefont {Yang}, \citenamefont
  {Leenstra}, \citenamefont {de~Ronde}, \citenamefont {Dehollain},
  \citenamefont {Muhonen}, \citenamefont {Hudson}, \citenamefont {Itoh},
  \citenamefont {Morello},\ and\ \citenamefont {Dzurak}}]{Veldhorst2014}%
  \BibitemOpen
  \bibfield  {author} {\bibinfo {author} {\bibfnamefont {M.}~\bibnamefont
  {Veldhorst}}, \bibinfo {author} {\bibfnamefont {J.~C.~C.}\ \bibnamefont
  {Hwang}}, \bibinfo {author} {\bibfnamefont {C.~H.}\ \bibnamefont {Yang}},
  \bibinfo {author} {\bibfnamefont {A.~W.}\ \bibnamefont {Leenstra}}, \bibinfo
  {author} {\bibfnamefont {B.}~\bibnamefont {de~Ronde}}, \bibinfo {author}
  {\bibfnamefont {J.~P.}\ \bibnamefont {Dehollain}}, \bibinfo {author}
  {\bibfnamefont {J.~T.}\ \bibnamefont {Muhonen}}, \bibinfo {author}
  {\bibfnamefont {F.~E.}\ \bibnamefont {Hudson}}, \bibinfo {author}
  {\bibfnamefont {K.~M.}\ \bibnamefont {Itoh}}, \bibinfo {author}
  {\bibfnamefont {A.}~\bibnamefont {Morello}},\ and\ \bibinfo {author}
  {\bibfnamefont {A.~S.}\ \bibnamefont {Dzurak}},\ }\bibfield  {title}
  {\bibinfo {title} {{An addressable quantum dot qubit with fault-tolerant
  control-fidelity}},\ }\href@noop {} {\bibfield  {journal} {\bibinfo
  {journal} {Nat. Nanotechnol.}\ }\textbf {\bibinfo {volume} {9}},\ \bibinfo
  {pages} {981} (\bibinfo {year} {2014})}\BibitemShut {NoStop}%
\bibitem [{\citenamefont {Veldhorst}\ \emph {et~al.}(2015)\citenamefont
  {Veldhorst}, \citenamefont {Yang}, \citenamefont {Hwang}, \citenamefont
  {Huang}, \citenamefont {Dehollain}, \citenamefont {Muhonen}, \citenamefont
  {Simmons}, \citenamefont {Laucht}, \citenamefont {Hudson}, \citenamefont
  {Itoh}, \citenamefont {Morello},\ and\ \citenamefont
  {Dzurak}}]{Veldhorst2015}%
  \BibitemOpen
  \bibfield  {author} {\bibinfo {author} {\bibfnamefont {M.}~\bibnamefont
  {Veldhorst}}, \bibinfo {author} {\bibfnamefont {C.~H.}\ \bibnamefont {Yang}},
  \bibinfo {author} {\bibfnamefont {J.~C.~C.}\ \bibnamefont {Hwang}}, \bibinfo
  {author} {\bibfnamefont {W.}~\bibnamefont {Huang}}, \bibinfo {author}
  {\bibfnamefont {J.~P.}\ \bibnamefont {Dehollain}}, \bibinfo {author}
  {\bibfnamefont {J.~T.}\ \bibnamefont {Muhonen}}, \bibinfo {author}
  {\bibfnamefont {S.}~\bibnamefont {Simmons}}, \bibinfo {author} {\bibfnamefont
  {A.}~\bibnamefont {Laucht}}, \bibinfo {author} {\bibfnamefont {F.~E.}\
  \bibnamefont {Hudson}}, \bibinfo {author} {\bibfnamefont {K.~M.}\
  \bibnamefont {Itoh}}, \bibinfo {author} {\bibfnamefont {A.}~\bibnamefont
  {Morello}},\ and\ \bibinfo {author} {\bibfnamefont {A.~S.}\ \bibnamefont
  {Dzurak}},\ }\bibfield  {title} {\bibinfo {title} {{A two-qubit logic gate in
  silicon}},\ }\href@noop {} {\bibfield  {journal} {\bibinfo  {journal}
  {Nature}\ }\textbf {\bibinfo {volume} {526}},\ \bibinfo {pages} {410}
  (\bibinfo {year} {2015})}\BibitemShut {NoStop}%
\bibitem [{\citenamefont {Yoneda}\ \emph {et~al.}(2018)\citenamefont {Yoneda},
  \citenamefont {Takeda}, \citenamefont {Otsuka}, \citenamefont {Nakajima},
  \citenamefont {Delbecq}, \citenamefont {Allison}, \citenamefont {Honda},
  \citenamefont {Kodera}, \citenamefont {Oda}, \citenamefont {Hoshi},
  \citenamefont {Usami}, \citenamefont {Itoh},\ and\ \citenamefont
  {Tarucha}}]{Yoneda2018}%
  \BibitemOpen
  \bibfield  {author} {\bibinfo {author} {\bibfnamefont {J.}~\bibnamefont
  {Yoneda}}, \bibinfo {author} {\bibfnamefont {K.}~\bibnamefont {Takeda}},
  \bibinfo {author} {\bibfnamefont {T.}~\bibnamefont {Otsuka}}, \bibinfo
  {author} {\bibfnamefont {T.}~\bibnamefont {Nakajima}}, \bibinfo {author}
  {\bibfnamefont {M.~R.}\ \bibnamefont {Delbecq}}, \bibinfo {author}
  {\bibfnamefont {G.}~\bibnamefont {Allison}}, \bibinfo {author} {\bibfnamefont
  {T.}~\bibnamefont {Honda}}, \bibinfo {author} {\bibfnamefont
  {T.}~\bibnamefont {Kodera}}, \bibinfo {author} {\bibfnamefont
  {S.}~\bibnamefont {Oda}}, \bibinfo {author} {\bibfnamefont {Y.}~\bibnamefont
  {Hoshi}}, \bibinfo {author} {\bibfnamefont {N.}~\bibnamefont {Usami}},
  \bibinfo {author} {\bibfnamefont {K.~M.}\ \bibnamefont {Itoh}},\ and\
  \bibinfo {author} {\bibfnamefont {S.}~\bibnamefont {Tarucha}},\ }\bibfield
  {title} {\bibinfo {title} {A quantum-dot spin qubit with coherence limited by
  charge noise and fidelity higher than 99.9{\%}},\ }\href@noop {} {\bibfield
  {journal} {\bibinfo  {journal} {Nat. Nanotechnol.}\ }\textbf {\bibinfo
  {volume} {13}},\ \bibinfo {pages} {102} (\bibinfo {year} {2018})}\BibitemShut
  {NoStop}%
\bibitem [{\citenamefont {Mi}\ \emph {et~al.}(2018)\citenamefont {Mi},
  \citenamefont {Benito}, \citenamefont {Putz}, \citenamefont {Zajac},
  \citenamefont {Taylor}, \citenamefont {Burkard},\ and\ \citenamefont
  {Petta}}]{Mi2018}%
  \BibitemOpen
  \bibfield  {author} {\bibinfo {author} {\bibfnamefont {X.}~\bibnamefont
  {Mi}}, \bibinfo {author} {\bibfnamefont {M.}~\bibnamefont {Benito}}, \bibinfo
  {author} {\bibfnamefont {S.}~\bibnamefont {Putz}}, \bibinfo {author}
  {\bibfnamefont {D.~M.}\ \bibnamefont {Zajac}}, \bibinfo {author}
  {\bibfnamefont {J.~M.}\ \bibnamefont {Taylor}}, \bibinfo {author}
  {\bibfnamefont {G.}~\bibnamefont {Burkard}},\ and\ \bibinfo {author}
  {\bibfnamefont {J.~R.}\ \bibnamefont {Petta}},\ }\bibfield  {title} {\bibinfo
  {title} {A coherent spin--photon interface in silicon},\ }\href@noop {}
  {\bibfield  {journal} {\bibinfo  {journal} {Nature}\ }\textbf {\bibinfo
  {volume} {555}},\ \bibinfo {pages} {599} (\bibinfo {year}
  {2018})}\BibitemShut {NoStop}%
\bibitem [{\citenamefont {Samkharadze}\ \emph {et~al.}(2018)\citenamefont
  {Samkharadze}, \citenamefont {Zheng}, \citenamefont {Kalhor}, \citenamefont
  {Brousse}, \citenamefont {Sammak}, \citenamefont {Mendes}, \citenamefont
  {Blais}, \citenamefont {Scappucci},\ and\ \citenamefont
  {Vandersypen}}]{Samkharadze2018}%
  \BibitemOpen
  \bibfield  {author} {\bibinfo {author} {\bibfnamefont {N.}~\bibnamefont
  {Samkharadze}}, \bibinfo {author} {\bibfnamefont {G.}~\bibnamefont {Zheng}},
  \bibinfo {author} {\bibfnamefont {N.}~\bibnamefont {Kalhor}}, \bibinfo
  {author} {\bibfnamefont {D.}~\bibnamefont {Brousse}}, \bibinfo {author}
  {\bibfnamefont {A.}~\bibnamefont {Sammak}}, \bibinfo {author} {\bibfnamefont
  {U.~C.}\ \bibnamefont {Mendes}}, \bibinfo {author} {\bibfnamefont
  {A.}~\bibnamefont {Blais}}, \bibinfo {author} {\bibfnamefont
  {G.}~\bibnamefont {Scappucci}},\ and\ \bibinfo {author} {\bibfnamefont
  {L.~M.~K.}\ \bibnamefont {Vandersypen}},\ }\bibfield  {title} {\bibinfo
  {title} {Strong spin-photon coupling in silicon},\ }\href@noop {} {\bibfield
  {journal} {\bibinfo  {journal} {Science}\ }\textbf {\bibinfo {volume}
  {359}},\ \bibinfo {pages} {1123} (\bibinfo {year} {2018})}\BibitemShut
  {NoStop}%
\bibitem [{\citenamefont {Landig}\ \emph {et~al.}(2018)\citenamefont {Landig},
  \citenamefont {Koski}, \citenamefont {Scarlino}, \citenamefont {Mendes},
  \citenamefont {Blais}, \citenamefont {Reichl}, \citenamefont {Wegscheider},
  \citenamefont {Wallraff}, \citenamefont {Ensslin},\ and\ \citenamefont
  {Ihn}}]{Landig2018}%
  \BibitemOpen
  \bibfield  {author} {\bibinfo {author} {\bibfnamefont {A.~J.}\ \bibnamefont
  {Landig}}, \bibinfo {author} {\bibfnamefont {J.~V.}\ \bibnamefont {Koski}},
  \bibinfo {author} {\bibfnamefont {P.}~\bibnamefont {Scarlino}}, \bibinfo
  {author} {\bibfnamefont {U.~C.}\ \bibnamefont {Mendes}}, \bibinfo {author}
  {\bibfnamefont {A.}~\bibnamefont {Blais}}, \bibinfo {author} {\bibfnamefont
  {C.}~\bibnamefont {Reichl}}, \bibinfo {author} {\bibfnamefont
  {W.}~\bibnamefont {Wegscheider}}, \bibinfo {author} {\bibfnamefont
  {A.}~\bibnamefont {Wallraff}}, \bibinfo {author} {\bibfnamefont
  {K.}~\bibnamefont {Ensslin}},\ and\ \bibinfo {author} {\bibfnamefont
  {T.}~\bibnamefont {Ihn}},\ }\bibfield  {title} {\bibinfo {title} {Coherent
  spin-photon coupling using a resonant exchange qubit},\ }\href@noop {}
  {\bibfield  {journal} {\bibinfo  {journal} {Nature}\ }\textbf {\bibinfo
  {volume} {560}},\ \bibinfo {pages} {179} (\bibinfo {year}
  {2018})}\BibitemShut {NoStop}%
\bibitem [{\citenamefont {Landig}\ \emph {et~al.}(2019)\citenamefont {Landig},
  \citenamefont {Koski}, \citenamefont {Scarlino}, \citenamefont {M{\"u}ller},
  \citenamefont {Abadillo-Uriel}, \citenamefont {Kratochwil}, \citenamefont
  {Reichl}, \citenamefont {Wegscheider}, \citenamefont {Coppersmith},
  \citenamefont {Friesen}, \citenamefont {Wallraff}, \citenamefont {Ihn},\ and\
  \citenamefont {Ensslin}}]{Landig2019}%
  \BibitemOpen
  \bibfield  {author} {\bibinfo {author} {\bibfnamefont {A.~J.}\ \bibnamefont
  {Landig}}, \bibinfo {author} {\bibfnamefont {J.~V.}\ \bibnamefont {Koski}},
  \bibinfo {author} {\bibfnamefont {P.}~\bibnamefont {Scarlino}}, \bibinfo
  {author} {\bibfnamefont {C.}~\bibnamefont {M{\"u}ller}}, \bibinfo {author}
  {\bibfnamefont {J.~C.}\ \bibnamefont {Abadillo-Uriel}}, \bibinfo {author}
  {\bibfnamefont {B.}~\bibnamefont {Kratochwil}}, \bibinfo {author}
  {\bibfnamefont {C.}~\bibnamefont {Reichl}}, \bibinfo {author} {\bibfnamefont
  {W.}~\bibnamefont {Wegscheider}}, \bibinfo {author} {\bibfnamefont {S.~N.}\
  \bibnamefont {Coppersmith}}, \bibinfo {author} {\bibfnamefont
  {M.}~\bibnamefont {Friesen}}, \bibinfo {author} {\bibfnamefont
  {A.}~\bibnamefont {Wallraff}}, \bibinfo {author} {\bibfnamefont
  {T.}~\bibnamefont {Ihn}},\ and\ \bibinfo {author} {\bibfnamefont
  {K.}~\bibnamefont {Ensslin}},\ }\bibfield  {title} {\bibinfo {title}
  {Virtual-photon-mediated spin-qubit--transmon coupling},\ }\href@noop {}
  {\bibfield  {journal} {\bibinfo  {journal} {Nat. Commun.}\ }\textbf {\bibinfo
  {volume} {10}},\ \bibinfo {pages} {5037} (\bibinfo {year}
  {2019})}\BibitemShut {NoStop}%
\bibitem [{\citenamefont {Yu}\ \emph {et~al.}(2023)\citenamefont {Yu},
  \citenamefont {Zihlmann}, \citenamefont {Abadillo-Uriel}, \citenamefont
  {Michal}, \citenamefont {Rambal}, \citenamefont {Niebojewski}, \citenamefont
  {Bedecarrats}, \citenamefont {Vinet}, \citenamefont {Dumur}, \citenamefont
  {Filippone}, \citenamefont {Bertrand}, \citenamefont {De~Franceschi},
  \citenamefont {Niquet},\ and\ \citenamefont {Maurand}}]{Yu2023}%
  \BibitemOpen
  \bibfield  {author} {\bibinfo {author} {\bibfnamefont {C.~X.}\ \bibnamefont
  {Yu}}, \bibinfo {author} {\bibfnamefont {S.}~\bibnamefont {Zihlmann}},
  \bibinfo {author} {\bibfnamefont {J.}~\bibnamefont {Abadillo-Uriel}},
  \bibinfo {author} {\bibfnamefont {V.~P.}\ \bibnamefont {Michal}}, \bibinfo
  {author} {\bibfnamefont {N.}~\bibnamefont {Rambal}}, \bibinfo {author}
  {\bibfnamefont {H.}~\bibnamefont {Niebojewski}}, \bibinfo {author}
  {\bibfnamefont {T.}~\bibnamefont {Bedecarrats}}, \bibinfo {author}
  {\bibfnamefont {M.}~\bibnamefont {Vinet}}, \bibinfo {author} {\bibfnamefont
  {{\'E}.}~\bibnamefont {Dumur}}, \bibinfo {author} {\bibfnamefont
  {M.}~\bibnamefont {Filippone}}, \bibinfo {author} {\bibfnamefont
  {B.}~\bibnamefont {Bertrand}}, \bibinfo {author} {\bibfnamefont
  {S.}~\bibnamefont {De~Franceschi}}, \bibinfo {author} {\bibfnamefont {Y.-M.}\
  \bibnamefont {Niquet}},\ and\ \bibinfo {author} {\bibfnamefont
  {R.}~\bibnamefont {Maurand}},\ }\bibfield  {title} {\bibinfo {title} {Strong
  coupling between a photon and a hole spin in silicon},\ }\href@noop {}
  {\bibfield  {journal} {\bibinfo  {journal} {Nat. Nanotechnol.}\ } (\bibinfo
  {year} {2023})}\BibitemShut {NoStop}%
\bibitem [{\citenamefont {Borjans}\ \emph {et~al.}(2020)\citenamefont
  {Borjans}, \citenamefont {Croot}, \citenamefont {Mi}, \citenamefont
  {Gullans},\ and\ \citenamefont {Petta}}]{Borjans2020}%
  \BibitemOpen
  \bibfield  {author} {\bibinfo {author} {\bibfnamefont {F.}~\bibnamefont
  {Borjans}}, \bibinfo {author} {\bibfnamefont {X.~G.}\ \bibnamefont {Croot}},
  \bibinfo {author} {\bibfnamefont {X.}~\bibnamefont {Mi}}, \bibinfo {author}
  {\bibfnamefont {M.~J.}\ \bibnamefont {Gullans}},\ and\ \bibinfo {author}
  {\bibfnamefont {J.~R.}\ \bibnamefont {Petta}},\ }\bibfield  {title} {\bibinfo
  {title} {Resonant microwave-mediated interactions between distant electron
  spins},\ }\href@noop {} {\bibfield  {journal} {\bibinfo  {journal} {Nature}\
  }\textbf {\bibinfo {volume} {577}},\ \bibinfo {pages} {195} (\bibinfo {year}
  {2020})}\BibitemShut {NoStop}%
\bibitem [{\citenamefont {Harvey-Collard}\ \emph {et~al.}(2022)\citenamefont
  {Harvey-Collard}, \citenamefont {Dijkema}, \citenamefont {Zheng},
  \citenamefont {Sammak}, \citenamefont {Scappucci},\ and\ \citenamefont
  {Vandersypen}}]{Harvey-Collard2022}%
  \BibitemOpen
  \bibfield  {author} {\bibinfo {author} {\bibfnamefont {P.}~\bibnamefont
  {Harvey-Collard}}, \bibinfo {author} {\bibfnamefont {J.}~\bibnamefont
  {Dijkema}}, \bibinfo {author} {\bibfnamefont {G.}~\bibnamefont {Zheng}},
  \bibinfo {author} {\bibfnamefont {A.}~\bibnamefont {Sammak}}, \bibinfo
  {author} {\bibfnamefont {G.}~\bibnamefont {Scappucci}},\ and\ \bibinfo
  {author} {\bibfnamefont {L.~M.~K.}\ \bibnamefont {Vandersypen}},\ }\bibfield
  {title} {\bibinfo {title} {Coherent spin-spin coupling mediated by virtual
  microwave photons},\ }\href@noop {} {\bibfield  {journal} {\bibinfo
  {journal} {Phys. Rev. X}\ }\textbf {\bibinfo {volume} {12}},\ \bibinfo
  {pages} {021026} (\bibinfo {year} {2022})}\BibitemShut {NoStop}%
\bibitem [{\citenamefont {Benito}\ \emph
  {et~al.}(2019{\natexlab{a}})\citenamefont {Benito}, \citenamefont {Petta},\
  and\ \citenamefont {Burkard}}]{Benito2019b}%
  \BibitemOpen
  \bibfield  {author} {\bibinfo {author} {\bibfnamefont {M.}~\bibnamefont
  {Benito}}, \bibinfo {author} {\bibfnamefont {J.~R.}\ \bibnamefont {Petta}},\
  and\ \bibinfo {author} {\bibfnamefont {G.}~\bibnamefont {Burkard}},\
  }\bibfield  {title} {\bibinfo {title} {Optimized cavity-mediated dispersive
  two-qubit gates between spin qubits},\ }\href@noop {} {\bibfield  {journal}
  {\bibinfo  {journal} {Phys. Rev. B}\ }\textbf {\bibinfo {volume} {100}},\
  \bibinfo {pages} {081412} (\bibinfo {year} {2019}{\natexlab{a}})}\BibitemShut
  {NoStop}%
\bibitem [{\citenamefont {Warren}\ \emph {et~al.}(2019)\citenamefont {Warren},
  \citenamefont {Barnes},\ and\ \citenamefont {Economou}}]{Warren2019}%
  \BibitemOpen
  \bibfield  {author} {\bibinfo {author} {\bibfnamefont {A.}~\bibnamefont
  {Warren}}, \bibinfo {author} {\bibfnamefont {E.}~\bibnamefont {Barnes}},\
  and\ \bibinfo {author} {\bibfnamefont {S.~E.}\ \bibnamefont {Economou}},\
  }\bibfield  {title} {\bibinfo {title} {Long-distance entangling gates between
  quantum dot spins mediated by a superconducting resonator},\ }\href@noop {}
  {\bibfield  {journal} {\bibinfo  {journal} {Phys. Rev. B}\ }\textbf {\bibinfo
  {volume} {100}},\ \bibinfo {pages} {161303} (\bibinfo {year}
  {2019})}\BibitemShut {NoStop}%
\bibitem [{\citenamefont {Rigetti}\ \emph {et~al.}(2005)\citenamefont
  {Rigetti}, \citenamefont {Blais},\ and\ \citenamefont
  {Devoret}}]{Rigetti2005}%
  \BibitemOpen
  \bibfield  {author} {\bibinfo {author} {\bibfnamefont {C.}~\bibnamefont
  {Rigetti}}, \bibinfo {author} {\bibfnamefont {A.}~\bibnamefont {Blais}},\
  and\ \bibinfo {author} {\bibfnamefont {M.}~\bibnamefont {Devoret}},\
  }\bibfield  {title} {\bibinfo {title} {{Protocol for Universal Gates in
  Optimally Biased Superconducting Qubits}},\ }\href@noop {} {\bibfield
  {journal} {\bibinfo  {journal} {Phys. Rev. Lett.}\ }\textbf {\bibinfo
  {volume} {94}},\ \bibinfo {pages} {240502} (\bibinfo {year}
  {2005})}\BibitemShut {NoStop}%
\bibitem [{\citenamefont {Wallraff}\ \emph {et~al.}(2007)\citenamefont
  {Wallraff}, \citenamefont {Schuster}, \citenamefont {Blais}, \citenamefont
  {Gambetta}, \citenamefont {Schreier}, \citenamefont {Frunzio}, \citenamefont
  {Devoret}, \citenamefont {Girvin},\ and\ \citenamefont
  {Schoelkopf}}]{Wallraff2007}%
  \BibitemOpen
  \bibfield  {author} {\bibinfo {author} {\bibfnamefont {A.}~\bibnamefont
  {Wallraff}}, \bibinfo {author} {\bibfnamefont {D.~I.}\ \bibnamefont
  {Schuster}}, \bibinfo {author} {\bibfnamefont {A.}~\bibnamefont {Blais}},
  \bibinfo {author} {\bibfnamefont {J.~M.}\ \bibnamefont {Gambetta}}, \bibinfo
  {author} {\bibfnamefont {J.}~\bibnamefont {Schreier}}, \bibinfo {author}
  {\bibfnamefont {L.}~\bibnamefont {Frunzio}}, \bibinfo {author} {\bibfnamefont
  {M.~H.}\ \bibnamefont {Devoret}}, \bibinfo {author} {\bibfnamefont {S.~M.}\
  \bibnamefont {Girvin}},\ and\ \bibinfo {author} {\bibfnamefont {R.~J.}\
  \bibnamefont {Schoelkopf}},\ }\bibfield  {title} {\bibinfo {title} {{Sideband
  Transitions and Two-Tone Spectroscopy of a Superconducting Qubit Strongly
  Coupled to an On-Chip Cavity}},\ }\href@noop {} {\bibfield  {journal}
  {\bibinfo  {journal} {Phys. Rev. Lett.}\ }\textbf {\bibinfo {volume} {99}},\
  \bibinfo {pages} {050501} (\bibinfo {year} {2007})}\BibitemShut {NoStop}%
\bibitem [{\citenamefont {Leek}\ \emph {et~al.}(2009)\citenamefont {Leek},
  \citenamefont {Filipp}, \citenamefont {Maurer}, \citenamefont {Baur},
  \citenamefont {Bianchetti}, \citenamefont {Fink}, \citenamefont {G\"oppl},
  \citenamefont {Steffen},\ and\ \citenamefont {Wallraff}}]{Leek2009}%
  \BibitemOpen
  \bibfield  {author} {\bibinfo {author} {\bibfnamefont {P.~J.}\ \bibnamefont
  {Leek}}, \bibinfo {author} {\bibfnamefont {S.}~\bibnamefont {Filipp}},
  \bibinfo {author} {\bibfnamefont {P.}~\bibnamefont {Maurer}}, \bibinfo
  {author} {\bibfnamefont {M.}~\bibnamefont {Baur}}, \bibinfo {author}
  {\bibfnamefont {R.}~\bibnamefont {Bianchetti}}, \bibinfo {author}
  {\bibfnamefont {J.~M.}\ \bibnamefont {Fink}}, \bibinfo {author}
  {\bibfnamefont {M.}~\bibnamefont {G\"oppl}}, \bibinfo {author} {\bibfnamefont
  {L.}~\bibnamefont {Steffen}},\ and\ \bibinfo {author} {\bibfnamefont
  {A.}~\bibnamefont {Wallraff}},\ }\bibfield  {title} {\bibinfo {title} {{Using
  sideband transitions for two-qubit operations in superconducting circuits}},\
  }\href@noop {} {\bibfield  {journal} {\bibinfo  {journal} {Phys. Rev. B}\
  }\textbf {\bibinfo {volume} {79}},\ \bibinfo {pages} {180511} (\bibinfo
  {year} {2009})}\BibitemShut {NoStop}%
\bibitem [{\citenamefont {Rigetti}\ and\ \citenamefont
  {Devoret}(2010)}]{Rigetti2010}%
  \BibitemOpen
  \bibfield  {author} {\bibinfo {author} {\bibfnamefont {C.}~\bibnamefont
  {Rigetti}}\ and\ \bibinfo {author} {\bibfnamefont {M.}~\bibnamefont
  {Devoret}},\ }\bibfield  {title} {\bibinfo {title} {Fully microwave-tunable
  universal gates in superconducting qubits with linear couplings and fixed
  transition frequencies},\ }\href@noop {} {\bibfield  {journal} {\bibinfo
  {journal} {Phys. Rev. B}\ }\textbf {\bibinfo {volume} {81}},\ \bibinfo
  {pages} {134507} (\bibinfo {year} {2010})}\BibitemShut {NoStop}%
\bibitem [{\citenamefont {Chow}\ \emph {et~al.}(2011)\citenamefont {Chow},
  \citenamefont {C\'orcoles}, \citenamefont {Gambetta}, \citenamefont
  {Rigetti}, \citenamefont {Johnson}, \citenamefont {Smolin}, \citenamefont
  {Rozen}, \citenamefont {Keefe}, \citenamefont {Rothwell}, \citenamefont
  {Ketchen},\ and\ \citenamefont {Steffen}}]{Chow2011}%
  \BibitemOpen
  \bibfield  {author} {\bibinfo {author} {\bibfnamefont {J.~M.}\ \bibnamefont
  {Chow}}, \bibinfo {author} {\bibfnamefont {A.~D.}\ \bibnamefont
  {C\'orcoles}}, \bibinfo {author} {\bibfnamefont {J.~M.}\ \bibnamefont
  {Gambetta}}, \bibinfo {author} {\bibfnamefont {C.}~\bibnamefont {Rigetti}},
  \bibinfo {author} {\bibfnamefont {B.~R.}\ \bibnamefont {Johnson}}, \bibinfo
  {author} {\bibfnamefont {J.~A.}\ \bibnamefont {Smolin}}, \bibinfo {author}
  {\bibfnamefont {J.~R.}\ \bibnamefont {Rozen}}, \bibinfo {author}
  {\bibfnamefont {G.~A.}\ \bibnamefont {Keefe}}, \bibinfo {author}
  {\bibfnamefont {M.~B.}\ \bibnamefont {Rothwell}}, \bibinfo {author}
  {\bibfnamefont {M.~B.}\ \bibnamefont {Ketchen}},\ and\ \bibinfo {author}
  {\bibfnamefont {M.}~\bibnamefont {Steffen}},\ }\bibfield  {title} {\bibinfo
  {title} {{Simple All-Microwave Entangling Gate for Fixed-Frequency
  Superconducting Qubits}},\ }\href@noop {} {\bibfield  {journal} {\bibinfo
  {journal} {Phys. Rev. Lett.}\ }\textbf {\bibinfo {volume} {107}},\ \bibinfo
  {pages} {080502} (\bibinfo {year} {2011})}\BibitemShut {NoStop}%
\bibitem [{\citenamefont {Beaudoin}\ \emph {et~al.}(2012)\citenamefont
  {Beaudoin}, \citenamefont {da~Silva}, \citenamefont {Dutton},\ and\
  \citenamefont {Blais}}]{Beaudoin2012}%
  \BibitemOpen
  \bibfield  {author} {\bibinfo {author} {\bibfnamefont {F.}~\bibnamefont
  {Beaudoin}}, \bibinfo {author} {\bibfnamefont {M.~P.}\ \bibnamefont
  {da~Silva}}, \bibinfo {author} {\bibfnamefont {Z.}~\bibnamefont {Dutton}},\
  and\ \bibinfo {author} {\bibfnamefont {A.}~\bibnamefont {Blais}},\ }\bibfield
   {title} {\bibinfo {title} {{First-order sidebands in circuit QED using qubit
  frequency modulation}},\ }\href@noop {} {\bibfield  {journal} {\bibinfo
  {journal} {Phys. Rev. A}\ }\textbf {\bibinfo {volume} {86}},\ \bibinfo
  {pages} {022305} (\bibinfo {year} {2012})}\BibitemShut {NoStop}%
\bibitem [{\citenamefont {McKay}\ \emph {et~al.}(2016)\citenamefont {McKay},
  \citenamefont {Filipp}, \citenamefont {Mezzacapo}, \citenamefont {Magesan},
  \citenamefont {Chow},\ and\ \citenamefont {Gambetta}}]{McKay2016}%
  \BibitemOpen
  \bibfield  {author} {\bibinfo {author} {\bibfnamefont {D.~C.}\ \bibnamefont
  {McKay}}, \bibinfo {author} {\bibfnamefont {S.}~\bibnamefont {Filipp}},
  \bibinfo {author} {\bibfnamefont {A.}~\bibnamefont {Mezzacapo}}, \bibinfo
  {author} {\bibfnamefont {E.}~\bibnamefont {Magesan}}, \bibinfo {author}
  {\bibfnamefont {J.~M.}\ \bibnamefont {Chow}},\ and\ \bibinfo {author}
  {\bibfnamefont {J.~M.}\ \bibnamefont {Gambetta}},\ }\bibfield  {title}
  {\bibinfo {title} {Universal gate for fixed-frequency qubits via a tunable
  bus},\ }\href@noop {} {\bibfield  {journal} {\bibinfo  {journal} {Phys. Rev.
  Appl.}\ }\textbf {\bibinfo {volume} {6}},\ \bibinfo {pages} {064007}
  (\bibinfo {year} {2016})}\BibitemShut {NoStop}%
\bibitem [{\citenamefont {Tosi}\ \emph {et~al.}(2018)\citenamefont {Tosi},
  \citenamefont {Mohiyaddin}, \citenamefont {Tenberg}, \citenamefont {Laucht},\
  and\ \citenamefont {Morello}}]{Tosi2018}%
  \BibitemOpen
  \bibfield  {author} {\bibinfo {author} {\bibfnamefont {G.}~\bibnamefont
  {Tosi}}, \bibinfo {author} {\bibfnamefont {F.~A.}\ \bibnamefont
  {Mohiyaddin}}, \bibinfo {author} {\bibfnamefont {S.}~\bibnamefont {Tenberg}},
  \bibinfo {author} {\bibfnamefont {A.}~\bibnamefont {Laucht}},\ and\ \bibinfo
  {author} {\bibfnamefont {A.}~\bibnamefont {Morello}},\ }\bibfield  {title}
  {\bibinfo {title} {Robust electric dipole transition at microwave frequencies
  for nuclear spin qubits in silicon},\ }\href@noop {} {\bibfield  {journal}
  {\bibinfo  {journal} {Phys. Rev. B}\ }\textbf {\bibinfo {volume} {98}},\
  \bibinfo {pages} {075313} (\bibinfo {year} {2018})}\BibitemShut {NoStop}%
\bibitem [{\citenamefont {Sigillito}\ \emph {et~al.}(2019)\citenamefont
  {Sigillito}, \citenamefont {Gullans}, \citenamefont {Edge}, \citenamefont
  {Borselli},\ and\ \citenamefont {Petta}}]{Sigillito2019npjQI}%
  \BibitemOpen
  \bibfield  {author} {\bibinfo {author} {\bibfnamefont {A.~J.}\ \bibnamefont
  {Sigillito}}, \bibinfo {author} {\bibfnamefont {M.~J.}\ \bibnamefont
  {Gullans}}, \bibinfo {author} {\bibfnamefont {L.~F.}\ \bibnamefont {Edge}},
  \bibinfo {author} {\bibfnamefont {M.}~\bibnamefont {Borselli}},\ and\
  \bibinfo {author} {\bibfnamefont {J.~R.}\ \bibnamefont {Petta}},\ }\bibfield
  {title} {\bibinfo {title} {Coherent transfer of quantum information in a
  silicon double quantum dot using resonant swap gates},\ }\href@noop {}
  {\bibfield  {journal} {\bibinfo  {journal} {npj Quantum Inf.}\ }\textbf
  {\bibinfo {volume} {5}},\ \bibinfo {pages} {110} (\bibinfo {year}
  {2019})}\BibitemShut {NoStop}%
\bibitem [{\citenamefont {Ruskov}\ and\ \citenamefont
  {Tahan}(2021)}]{Ruskov2021}%
  \BibitemOpen
  \bibfield  {author} {\bibinfo {author} {\bibfnamefont {R.}~\bibnamefont
  {Ruskov}}\ and\ \bibinfo {author} {\bibfnamefont {C.}~\bibnamefont {Tahan}},\
  }\bibfield  {title} {\bibinfo {title} {Modulated longitudinal gates on
  encoded spin qubits via curvature couplings to a superconducting cavity},\
  }\href@noop {} {\bibfield  {journal} {\bibinfo  {journal} {Phys. Rev. B}\
  }\textbf {\bibinfo {volume} {103}},\ \bibinfo {pages} {035301} (\bibinfo
  {year} {2021})}\BibitemShut {NoStop}%
\bibitem [{\citenamefont {Warren}\ \emph {et~al.}(2021)\citenamefont {Warren},
  \citenamefont {G\"ung\"ord\"u}, \citenamefont {Kestner}, \citenamefont
  {Barnes},\ and\ \citenamefont {Economou}}]{Warren2021}%
  \BibitemOpen
  \bibfield  {author} {\bibinfo {author} {\bibfnamefont {A.}~\bibnamefont
  {Warren}}, \bibinfo {author} {\bibfnamefont {U.}~\bibnamefont
  {G\"ung\"ord\"u}}, \bibinfo {author} {\bibfnamefont {J.~P.}\ \bibnamefont
  {Kestner}}, \bibinfo {author} {\bibfnamefont {E.}~\bibnamefont {Barnes}},\
  and\ \bibinfo {author} {\bibfnamefont {S.~E.}\ \bibnamefont {Economou}},\
  }\bibfield  {title} {\bibinfo {title} {Robust photon-mediated entangling
  gates between quantum dot spin qubits},\ }\href@noop {} {\bibfield  {journal}
  {\bibinfo  {journal} {Phys. Rev. B}\ }\textbf {\bibinfo {volume} {104}},\
  \bibinfo {pages} {115308} (\bibinfo {year} {2021})}\BibitemShut {NoStop}%
\bibitem [{\citenamefont {Hansen}\ \emph {et~al.}(2021)\citenamefont {Hansen},
  \citenamefont {Seedhouse}, \citenamefont {Saraiva}, \citenamefont {Laucht},
  \citenamefont {Dzurak},\ and\ \citenamefont {Yang}}]{Hansen2021}%
  \BibitemOpen
  \bibfield  {author} {\bibinfo {author} {\bibfnamefont {I.}~\bibnamefont
  {Hansen}}, \bibinfo {author} {\bibfnamefont {A.~E.}\ \bibnamefont
  {Seedhouse}}, \bibinfo {author} {\bibfnamefont {A.}~\bibnamefont {Saraiva}},
  \bibinfo {author} {\bibfnamefont {A.}~\bibnamefont {Laucht}}, \bibinfo
  {author} {\bibfnamefont {A.~S.}\ \bibnamefont {Dzurak}},\ and\ \bibinfo
  {author} {\bibfnamefont {C.~H.}\ \bibnamefont {Yang}},\ }\bibfield  {title}
  {\bibinfo {title} {Pulse engineering of a global field for robust and
  universal quantum computation},\ }\href@noop {} {\bibfield  {journal}
  {\bibinfo  {journal} {Phys. Rev. A}\ }\textbf {\bibinfo {volume} {104}},\
  \bibinfo {pages} {062415} (\bibinfo {year} {2021})}\BibitemShut {NoStop}%
\bibitem [{\citenamefont {McMillan}\ and\ \citenamefont
  {Burkard}(2022)}]{McMillan2022}%
  \BibitemOpen
  \bibfield  {author} {\bibinfo {author} {\bibfnamefont {S.~R.}\ \bibnamefont
  {McMillan}}\ and\ \bibinfo {author} {\bibfnamefont {G.}~\bibnamefont
  {Burkard}},\ }\bibfield  {title} {\bibinfo {title} {{Resonant single-shot
  CNOT in remote double quantum dot spin qubits}},\ }\href@noop {} {\bibfield
  {journal} {\bibinfo  {journal} {arXiv:2207.13588}\ } (\bibinfo {year}
  {2022})}\BibitemShut {NoStop}%
\bibitem [{\citenamefont {Mielke}\ and\ \citenamefont
  {Burkard}(2023)}]{Mielke2023}%
  \BibitemOpen
  \bibfield  {author} {\bibinfo {author} {\bibfnamefont {J.}~\bibnamefont
  {Mielke}}\ and\ \bibinfo {author} {\bibfnamefont {G.}~\bibnamefont
  {Burkard}},\ }\bibfield  {title} {\bibinfo {title} {Dispersive
  cavity-mediated quantum gate between driven dot-donor nuclear spins},\
  }\href@noop {} {\bibfield  {journal} {\bibinfo  {journal} {Phys. Rev. B}\
  }\textbf {\bibinfo {volume} {107}},\ \bibinfo {pages} {155302} (\bibinfo
  {year} {2023})}\BibitemShut {NoStop}%
\bibitem [{\citenamefont {Mollow}(1969)}]{Mollow1969}%
  \BibitemOpen
  \bibfield  {author} {\bibinfo {author} {\bibfnamefont {B.~R.}\ \bibnamefont
  {Mollow}},\ }\bibfield  {title} {\bibinfo {title} {{Power Spectrum of Light
  Scattered by Two-Level Systems}},\ }\href@noop {} {\bibfield  {journal}
  {\bibinfo  {journal} {Phys. Rev.}\ }\textbf {\bibinfo {volume} {188}},\
  \bibinfo {pages} {1969} (\bibinfo {year} {1969})}\BibitemShut {NoStop}%
\bibitem [{\citenamefont {Kim}\ \emph {et~al.}(2014)\citenamefont {Kim},
  \citenamefont {Shen}, \citenamefont {Roy-Choudhury}, \citenamefont
  {Solomon},\ and\ \citenamefont {Waks}}]{Kim2014PRL}%
  \BibitemOpen
  \bibfield  {author} {\bibinfo {author} {\bibfnamefont {H.}~\bibnamefont
  {Kim}}, \bibinfo {author} {\bibfnamefont {T.~C.}\ \bibnamefont {Shen}},
  \bibinfo {author} {\bibfnamefont {K.}~\bibnamefont {Roy-Choudhury}}, \bibinfo
  {author} {\bibfnamefont {G.~S.}\ \bibnamefont {Solomon}},\ and\ \bibinfo
  {author} {\bibfnamefont {E.}~\bibnamefont {Waks}},\ }\bibfield  {title}
  {\bibinfo {title} {{Resonant Interactions between a Mollow Triplet Sideband
  and a Strongly Coupled Cavity}},\ }\href@noop {} {\bibfield  {journal}
  {\bibinfo  {journal} {Phys. Rev. Lett.}\ }\textbf {\bibinfo {volume} {113}},\
  \bibinfo {pages} {027403} (\bibinfo {year} {2014})}\BibitemShut {NoStop}%
\bibitem [{\citenamefont {Benito}\ \emph {et~al.}(2017)\citenamefont {Benito},
  \citenamefont {Mi}, \citenamefont {Taylor}, \citenamefont {Petta},\ and\
  \citenamefont {Burkard}}]{Benito2017}%
  \BibitemOpen
  \bibfield  {author} {\bibinfo {author} {\bibfnamefont {M.}~\bibnamefont
  {Benito}}, \bibinfo {author} {\bibfnamefont {X.}~\bibnamefont {Mi}}, \bibinfo
  {author} {\bibfnamefont {J.~M.}\ \bibnamefont {Taylor}}, \bibinfo {author}
  {\bibfnamefont {J.~R.}\ \bibnamefont {Petta}},\ and\ \bibinfo {author}
  {\bibfnamefont {G.}~\bibnamefont {Burkard}},\ }\bibfield  {title} {\bibinfo
  {title} {Input-output theory for spin-photon coupling in si double quantum
  dots},\ }\href@noop {} {\bibfield  {journal} {\bibinfo  {journal} {Phys. Rev.
  B}\ }\textbf {\bibinfo {volume} {96}},\ \bibinfo {pages} {235434} (\bibinfo
  {year} {2017})}\BibitemShut {NoStop}%
\bibitem [{\citenamefont {Croot}\ \emph {et~al.}(2020)\citenamefont {Croot},
  \citenamefont {Mi}, \citenamefont {Putz}, \citenamefont {Benito},
  \citenamefont {Borjans}, \citenamefont {Burkard},\ and\ \citenamefont
  {Petta}}]{Croot2020}%
  \BibitemOpen
  \bibfield  {author} {\bibinfo {author} {\bibfnamefont {X.}~\bibnamefont
  {Croot}}, \bibinfo {author} {\bibfnamefont {X.}~\bibnamefont {Mi}}, \bibinfo
  {author} {\bibfnamefont {S.}~\bibnamefont {Putz}}, \bibinfo {author}
  {\bibfnamefont {M.}~\bibnamefont {Benito}}, \bibinfo {author} {\bibfnamefont
  {F.}~\bibnamefont {Borjans}}, \bibinfo {author} {\bibfnamefont
  {G.}~\bibnamefont {Burkard}},\ and\ \bibinfo {author} {\bibfnamefont {J.~R.}\
  \bibnamefont {Petta}},\ }\bibfield  {title} {\bibinfo {title} {Flopping-mode
  electric dipole spin resonance},\ }\href@noop {} {\bibfield  {journal}
  {\bibinfo  {journal} {Phys. Rev. Research}\ }\textbf {\bibinfo {volume}
  {2}},\ \bibinfo {pages} {012006} (\bibinfo {year} {2020})}\BibitemShut
  {NoStop}%
\bibitem [{\citenamefont {Medford}\ \emph
  {et~al.}(2013{\natexlab{a}})\citenamefont {Medford}, \citenamefont {Beil},
  \citenamefont {Taylor}, \citenamefont {Rashba}, \citenamefont {Lu},
  \citenamefont {Gossard},\ and\ \citenamefont {Marcus}}]{Medford2013}%
  \BibitemOpen
  \bibfield  {author} {\bibinfo {author} {\bibfnamefont {J.}~\bibnamefont
  {Medford}}, \bibinfo {author} {\bibfnamefont {J.}~\bibnamefont {Beil}},
  \bibinfo {author} {\bibfnamefont {J.~M.}\ \bibnamefont {Taylor}}, \bibinfo
  {author} {\bibfnamefont {E.~I.}\ \bibnamefont {Rashba}}, \bibinfo {author}
  {\bibfnamefont {H.}~\bibnamefont {Lu}}, \bibinfo {author} {\bibfnamefont
  {A.~C.}\ \bibnamefont {Gossard}},\ and\ \bibinfo {author} {\bibfnamefont
  {C.~M.}\ \bibnamefont {Marcus}},\ }\bibfield  {title} {\bibinfo {title}
  {{Quantum-Dot-Based Resonant Exchange Qubit}},\ }\href@noop {} {\bibfield
  {journal} {\bibinfo  {journal} {Phys. Rev. Lett.}\ }\textbf {\bibinfo
  {volume} {111}},\ \bibinfo {pages} {050501} (\bibinfo {year}
  {2013}{\natexlab{a}})}\BibitemShut {NoStop}%
\bibitem [{\citenamefont {Taylor}\ \emph {et~al.}(2013)\citenamefont {Taylor},
  \citenamefont {Srinivasa},\ and\ \citenamefont {Medford}}]{Taylor2013}%
  \BibitemOpen
  \bibfield  {author} {\bibinfo {author} {\bibfnamefont {J.~M.}\ \bibnamefont
  {Taylor}}, \bibinfo {author} {\bibfnamefont {V.}~\bibnamefont {Srinivasa}},\
  and\ \bibinfo {author} {\bibfnamefont {J.}~\bibnamefont {Medford}},\
  }\bibfield  {title} {\bibinfo {title} {{Electrically Protected Resonant
  Exchange Qubits in Triple Quantum Dots}},\ }\href@noop {} {\bibfield
  {journal} {\bibinfo  {journal} {Phys. Rev. Lett.}\ }\textbf {\bibinfo
  {volume} {111}},\ \bibinfo {pages} {050502} (\bibinfo {year}
  {2013})}\BibitemShut {NoStop}%
\bibitem [{\citenamefont {Srinivasa}\ \emph {et~al.}(2016)\citenamefont
  {Srinivasa}, \citenamefont {Taylor},\ and\ \citenamefont
  {Tahan}}]{Srinivasa2016}%
  \BibitemOpen
  \bibfield  {author} {\bibinfo {author} {\bibfnamefont {V.}~\bibnamefont
  {Srinivasa}}, \bibinfo {author} {\bibfnamefont {J.~M.}\ \bibnamefont
  {Taylor}},\ and\ \bibinfo {author} {\bibfnamefont {C.}~\bibnamefont
  {Tahan}},\ }\bibfield  {title} {\bibinfo {title} {Entangling distant resonant
  exchange qubits via circuit quantum electrodynamics},\ }\href@noop {}
  {\bibfield  {journal} {\bibinfo  {journal} {Phys. Rev. B}\ }\textbf {\bibinfo
  {volume} {94}},\ \bibinfo {pages} {205421} (\bibinfo {year}
  {2016})}\BibitemShut {NoStop}%
\bibitem [{\citenamefont {Abadillo-Uriel}\ \emph {et~al.}(2021)\citenamefont
  {Abadillo-Uriel}, \citenamefont {King}, \citenamefont {Coppersmith},\ and\
  \citenamefont {Friesen}}]{Abadillo-Uriel2021}%
  \BibitemOpen
  \bibfield  {author} {\bibinfo {author} {\bibfnamefont {J.~C.}\ \bibnamefont
  {Abadillo-Uriel}}, \bibinfo {author} {\bibfnamefont {C.}~\bibnamefont
  {King}}, \bibinfo {author} {\bibfnamefont {S.~N.}\ \bibnamefont
  {Coppersmith}},\ and\ \bibinfo {author} {\bibfnamefont {M.}~\bibnamefont
  {Friesen}},\ }\bibfield  {title} {\bibinfo {title} {Long-range
  two-hybrid-qubit gates mediated by a microwave cavity with red sidebands},\
  }\href@noop {} {\bibfield  {journal} {\bibinfo  {journal} {Phys. Rev. A}\
  }\textbf {\bibinfo {volume} {104}},\ \bibinfo {pages} {032612} (\bibinfo
  {year} {2021})}\BibitemShut {NoStop}%
\bibitem [{\citenamefont {Russ}\ and\ \citenamefont
  {Burkard}(2015{\natexlab{a}})}]{Russ2015b}%
  \BibitemOpen
  \bibfield  {author} {\bibinfo {author} {\bibfnamefont {M.}~\bibnamefont
  {Russ}}\ and\ \bibinfo {author} {\bibfnamefont {G.}~\bibnamefont {Burkard}},\
  }\bibfield  {title} {\bibinfo {title} {{Long distance coupling of resonant
  exchange qubits}},\ }\href@noop {} {\bibfield  {journal} {\bibinfo  {journal}
  {Phys. Rev. B}\ }\textbf {\bibinfo {volume} {92}},\ \bibinfo {pages} {205412}
  (\bibinfo {year} {2015}{\natexlab{a}})}\BibitemShut {NoStop}%
\bibitem [{\citenamefont {Levy}(2002)}]{Levy2002}%
  \BibitemOpen
  \bibfield  {author} {\bibinfo {author} {\bibfnamefont {J.}~\bibnamefont
  {Levy}},\ }\bibfield  {title} {\bibinfo {title} {{Universal Quantum
  Computation with Spin-1/2 Pairs and Heisenberg Exchange}},\ }\href@noop {}
  {\bibfield  {journal} {\bibinfo  {journal} {Phys. Rev. Lett.}\ }\textbf
  {\bibinfo {volume} {89}},\ \bibinfo {pages} {147902} (\bibinfo {year}
  {2002})}\BibitemShut {NoStop}%
\bibitem [{\citenamefont {Petta}\ \emph {et~al.}(2005)\citenamefont {Petta},
  \citenamefont {Johnson}, \citenamefont {Taylor}, \citenamefont {Laird},
  \citenamefont {Yacoby}, \citenamefont {Lukin}, \citenamefont {Marcus},
  \citenamefont {Hanson},\ and\ \citenamefont {Gossard}}]{Petta2005}%
  \BibitemOpen
  \bibfield  {author} {\bibinfo {author} {\bibfnamefont {J.~R.}\ \bibnamefont
  {Petta}}, \bibinfo {author} {\bibfnamefont {A.~C.}\ \bibnamefont {Johnson}},
  \bibinfo {author} {\bibfnamefont {J.~M.}\ \bibnamefont {Taylor}}, \bibinfo
  {author} {\bibfnamefont {E.~A.}\ \bibnamefont {Laird}}, \bibinfo {author}
  {\bibfnamefont {A.}~\bibnamefont {Yacoby}}, \bibinfo {author} {\bibfnamefont
  {M.~D.}\ \bibnamefont {Lukin}}, \bibinfo {author} {\bibfnamefont {C.~M.}\
  \bibnamefont {Marcus}}, \bibinfo {author} {\bibfnamefont {M.~P.}\
  \bibnamefont {Hanson}},\ and\ \bibinfo {author} {\bibfnamefont {A.~C.}\
  \bibnamefont {Gossard}},\ }\bibfield  {title} {\bibinfo {title} {{Coherent
  Manipulation of Coupled Electron Spins in Semiconductor Quantum Dots}},\
  }\href@noop {} {\bibfield  {journal} {\bibinfo  {journal} {Science}\ }\textbf
  {\bibinfo {volume} {309}},\ \bibinfo {pages} {2180} (\bibinfo {year}
  {2005})}\BibitemShut {NoStop}%
\bibitem [{\citenamefont {Burkard}\ and\ \citenamefont
  {Imamoglu}(2006)}]{Burkard2006}%
  \BibitemOpen
  \bibfield  {author} {\bibinfo {author} {\bibfnamefont {G.}~\bibnamefont
  {Burkard}}\ and\ \bibinfo {author} {\bibfnamefont {A.}~\bibnamefont
  {Imamoglu}},\ }\bibfield  {title} {\bibinfo {title} {{Ultra-long-distance
  interaction between spin qubits}},\ }\href@noop {} {\bibfield  {journal}
  {\bibinfo  {journal} {Phys. Rev. B}\ }\textbf {\bibinfo {volume} {74}},\
  \bibinfo {pages} {041307} (\bibinfo {year} {2006})}\BibitemShut {NoStop}%
\bibitem [{\citenamefont {Taylor}\ and\ \citenamefont
  {Lukin}(2006)}]{Taylor2006}%
  \BibitemOpen
  \bibfield  {author} {\bibinfo {author} {\bibfnamefont {J.~M.}\ \bibnamefont
  {Taylor}}\ and\ \bibinfo {author} {\bibfnamefont {M.~D.}\ \bibnamefont
  {Lukin}},\ }\bibfield  {title} {\bibinfo {title} {{{Cavity quantum
  electrodynamics with semiconductor double-dot molecules on a chip}}},\
  }\href@noop {} {\bibfield  {journal} {\bibinfo  {journal}
  {arXiv:cond-mat/0605144}\ } (\bibinfo {year} {2006})}\BibitemShut {NoStop}%
\bibitem [{\citenamefont {Taylor}\ \emph {et~al.}(2007)\citenamefont {Taylor},
  \citenamefont {Petta}, \citenamefont {Johnson}, \citenamefont {Yacoby},
  \citenamefont {Marcus},\ and\ \citenamefont {Lukin}}]{Taylor2007}%
  \BibitemOpen
  \bibfield  {author} {\bibinfo {author} {\bibfnamefont {J.~M.}\ \bibnamefont
  {Taylor}}, \bibinfo {author} {\bibfnamefont {J.~R.}\ \bibnamefont {Petta}},
  \bibinfo {author} {\bibfnamefont {A.~C.}\ \bibnamefont {Johnson}}, \bibinfo
  {author} {\bibfnamefont {A.}~\bibnamefont {Yacoby}}, \bibinfo {author}
  {\bibfnamefont {C.~M.}\ \bibnamefont {Marcus}},\ and\ \bibinfo {author}
  {\bibfnamefont {M.~D.}\ \bibnamefont {Lukin}},\ }\bibfield  {title} {\bibinfo
  {title} {{Relaxation, dephasing, and quantum control of electron spins in
  double quantum dots}},\ }\href@noop {} {\bibfield  {journal} {\bibinfo
  {journal} {Phys. Rev. B}\ }\textbf {\bibinfo {volume} {76}},\ \bibinfo
  {pages} {035315} (\bibinfo {year} {2007})}\BibitemShut {NoStop}%
\bibitem [{\citenamefont {B{\o}ttcher}\ \emph {et~al.}(2022)\citenamefont
  {B{\o}ttcher}, \citenamefont {Harvey}, \citenamefont {Fallahi}, \citenamefont
  {Gardner}, \citenamefont {Manfra}, \citenamefont {Vool}, \citenamefont
  {Bartlett},\ and\ \citenamefont {Yacoby}}]{Bottcher2022}%
  \BibitemOpen
  \bibfield  {author} {\bibinfo {author} {\bibfnamefont {C.~G.~L.}\
  \bibnamefont {B{\o}ttcher}}, \bibinfo {author} {\bibfnamefont {S.~P.}\
  \bibnamefont {Harvey}}, \bibinfo {author} {\bibfnamefont {S.}~\bibnamefont
  {Fallahi}}, \bibinfo {author} {\bibfnamefont {G.~C.}\ \bibnamefont
  {Gardner}}, \bibinfo {author} {\bibfnamefont {M.~J.}\ \bibnamefont {Manfra}},
  \bibinfo {author} {\bibfnamefont {U.}~\bibnamefont {Vool}}, \bibinfo {author}
  {\bibfnamefont {S.~D.}\ \bibnamefont {Bartlett}},\ and\ \bibinfo {author}
  {\bibfnamefont {A.}~\bibnamefont {Yacoby}},\ }\bibfield  {title} {\bibinfo
  {title} {Parametric longitudinal coupling between a high-impedance
  superconducting resonator and a semiconductor quantum dot singlet-triplet
  spin qubit},\ }\href@noop {} {\bibfield  {journal} {\bibinfo  {journal} {Nat.
  Commun.}\ }\textbf {\bibinfo {volume} {13}},\ \bibinfo {pages} {4773}
  (\bibinfo {year} {2022})}\BibitemShut {NoStop}%
\bibitem [{\citenamefont {Corrigan}\ \emph {et~al.}(2023)\citenamefont
  {Corrigan}, \citenamefont {Harpt}, \citenamefont {Holman}, \citenamefont
  {Ruskov}, \citenamefont {Marciniec}, \citenamefont {Rosenberg}, \citenamefont
  {Yost}, \citenamefont {Das}, \citenamefont {Oliver}, \citenamefont
  {McDermott}, \citenamefont {Tahan}, \citenamefont {Friesen},\ and\
  \citenamefont {Eriksson}}]{Corrigan2023arxiv}%
  \BibitemOpen
  \bibfield  {author} {\bibinfo {author} {\bibfnamefont {J.}~\bibnamefont
  {Corrigan}}, \bibinfo {author} {\bibfnamefont {B.}~\bibnamefont {Harpt}},
  \bibinfo {author} {\bibfnamefont {N.}~\bibnamefont {Holman}}, \bibinfo
  {author} {\bibfnamefont {R.}~\bibnamefont {Ruskov}}, \bibinfo {author}
  {\bibfnamefont {P.}~\bibnamefont {Marciniec}}, \bibinfo {author}
  {\bibfnamefont {D.}~\bibnamefont {Rosenberg}}, \bibinfo {author}
  {\bibfnamefont {D.}~\bibnamefont {Yost}}, \bibinfo {author} {\bibfnamefont
  {R.}~\bibnamefont {Das}}, \bibinfo {author} {\bibfnamefont {W.~D.}\
  \bibnamefont {Oliver}}, \bibinfo {author} {\bibfnamefont {R.}~\bibnamefont
  {McDermott}}, \bibinfo {author} {\bibfnamefont {C.}~\bibnamefont {Tahan}},
  \bibinfo {author} {\bibfnamefont {M.}~\bibnamefont {Friesen}},\ and\ \bibinfo
  {author} {\bibfnamefont {M.~A.}\ \bibnamefont {Eriksson}},\ }\bibfield
  {title} {\bibinfo {title} {{Longitudinal coupling between a Si/SiGe quantum
  dot and an off-chip TiN resonator}},\ }\href@noop {} {\bibfield  {journal}
  {\bibinfo  {journal} {arXiv:2212.02736}\ } (\bibinfo {year}
  {2023})}\BibitemShut {NoStop}%
\bibitem [{\citenamefont {Scappucci}\ \emph {et~al.}(2021)\citenamefont
  {Scappucci}, \citenamefont {Kloeffel}, \citenamefont {Zwanenburg},
  \citenamefont {Loss}, \citenamefont {Myronov}, \citenamefont {Zhang},
  \citenamefont {De~Franceschi}, \citenamefont {Katsaros},\ and\ \citenamefont
  {Veldhorst}}]{Scappucci2021}%
  \BibitemOpen
  \bibfield  {author} {\bibinfo {author} {\bibfnamefont {G.}~\bibnamefont
  {Scappucci}}, \bibinfo {author} {\bibfnamefont {C.}~\bibnamefont {Kloeffel}},
  \bibinfo {author} {\bibfnamefont {F.~A.}\ \bibnamefont {Zwanenburg}},
  \bibinfo {author} {\bibfnamefont {D.}~\bibnamefont {Loss}}, \bibinfo {author}
  {\bibfnamefont {M.}~\bibnamefont {Myronov}}, \bibinfo {author} {\bibfnamefont
  {J.-J.}\ \bibnamefont {Zhang}}, \bibinfo {author} {\bibfnamefont
  {S.}~\bibnamefont {De~Franceschi}}, \bibinfo {author} {\bibfnamefont
  {G.}~\bibnamefont {Katsaros}},\ and\ \bibinfo {author} {\bibfnamefont
  {M.}~\bibnamefont {Veldhorst}},\ }\bibfield  {title} {\bibinfo {title} {The
  germanium quantum information route},\ }\href@noop {} {\bibfield  {journal}
  {\bibinfo  {journal} {Nat. Rev. Mater.}\ }\textbf {\bibinfo {volume} {6}},\
  \bibinfo {pages} {926} (\bibinfo {year} {2021})}\BibitemShut {NoStop}%
\bibitem [{\citenamefont {Benito}\ \emph
  {et~al.}(2019{\natexlab{b}})\citenamefont {Benito}, \citenamefont {Croot},
  \citenamefont {Adelsberger}, \citenamefont {Putz}, \citenamefont {Mi},
  \citenamefont {Petta},\ and\ \citenamefont {Burkard}}]{Benito2019}%
  \BibitemOpen
  \bibfield  {author} {\bibinfo {author} {\bibfnamefont {M.}~\bibnamefont
  {Benito}}, \bibinfo {author} {\bibfnamefont {X.}~\bibnamefont {Croot}},
  \bibinfo {author} {\bibfnamefont {C.}~\bibnamefont {Adelsberger}}, \bibinfo
  {author} {\bibfnamefont {S.}~\bibnamefont {Putz}}, \bibinfo {author}
  {\bibfnamefont {X.}~\bibnamefont {Mi}}, \bibinfo {author} {\bibfnamefont
  {J.~R.}\ \bibnamefont {Petta}},\ and\ \bibinfo {author} {\bibfnamefont
  {G.}~\bibnamefont {Burkard}},\ }\bibfield  {title} {\bibinfo {title}
  {Electric-field control and noise protection of the flopping-mode spin
  qubit},\ }\href@noop {} {\bibfield  {journal} {\bibinfo  {journal} {Phys.
  Rev. B}\ }\textbf {\bibinfo {volume} {100}},\ \bibinfo {pages} {125430}
  (\bibinfo {year} {2019}{\natexlab{b}})}\BibitemShut {NoStop}%
\bibitem [{\citenamefont {Koch}\ \emph {et~al.}(2007)\citenamefont {Koch},
  \citenamefont {Yu}, \citenamefont {Gambetta}, \citenamefont {Houck},
  \citenamefont {Schuster}, \citenamefont {Majer}, \citenamefont {Blais},
  \citenamefont {Devoret}, \citenamefont {Girvin},\ and\ \citenamefont
  {Schoelkopf}}]{Koch2007}%
  \BibitemOpen
  \bibfield  {author} {\bibinfo {author} {\bibfnamefont {J.}~\bibnamefont
  {Koch}}, \bibinfo {author} {\bibfnamefont {T.~M.}\ \bibnamefont {Yu}},
  \bibinfo {author} {\bibfnamefont {J.}~\bibnamefont {Gambetta}}, \bibinfo
  {author} {\bibfnamefont {A.~A.}\ \bibnamefont {Houck}}, \bibinfo {author}
  {\bibfnamefont {D.~I.}\ \bibnamefont {Schuster}}, \bibinfo {author}
  {\bibfnamefont {J.}~\bibnamefont {Majer}}, \bibinfo {author} {\bibfnamefont
  {A.}~\bibnamefont {Blais}}, \bibinfo {author} {\bibfnamefont {M.~H.}\
  \bibnamefont {Devoret}}, \bibinfo {author} {\bibfnamefont {S.~M.}\
  \bibnamefont {Girvin}},\ and\ \bibinfo {author} {\bibfnamefont {R.~J.}\
  \bibnamefont {Schoelkopf}},\ }\bibfield  {title} {\bibinfo {title}
  {{Charge-insensitive qubit design derived from the Cooper pair box}},\
  }\href@noop {} {\bibfield  {journal} {\bibinfo  {journal} {Phys. Rev. A}\
  }\textbf {\bibinfo {volume} {76}},\ \bibinfo {pages} {042319} (\bibinfo
  {year} {2007})}\BibitemShut {NoStop}%
\bibitem [{\citenamefont {Houck}\ \emph {et~al.}(2009)\citenamefont {Houck},
  \citenamefont {Koch}, \citenamefont {Devoret}, \citenamefont {Girvin},\ and\
  \citenamefont {Schoelkopf}}]{Houck2009}%
  \BibitemOpen
  \bibfield  {author} {\bibinfo {author} {\bibfnamefont {A.~A.}\ \bibnamefont
  {Houck}}, \bibinfo {author} {\bibfnamefont {J.}~\bibnamefont {Koch}},
  \bibinfo {author} {\bibfnamefont {M.~H.}\ \bibnamefont {Devoret}}, \bibinfo
  {author} {\bibfnamefont {S.~M.}\ \bibnamefont {Girvin}},\ and\ \bibinfo
  {author} {\bibfnamefont {R.~J.}\ \bibnamefont {Schoelkopf}},\ }\bibfield
  {title} {\bibinfo {title} {{Life after charge noise: recent results with
  transmon qubits}},\ }\href@noop {} {\bibfield  {journal} {\bibinfo  {journal}
  {Quantum Inf. Process.}\ }\textbf {\bibinfo {volume} {8}},\ \bibinfo {pages}
  {105} (\bibinfo {year} {2009})}\BibitemShut {NoStop}%
\bibitem [{\citenamefont {Fei}\ \emph {et~al.}(2015)\citenamefont {Fei},
  \citenamefont {Hung}, \citenamefont {Koh}, \citenamefont {Shim},
  \citenamefont {Coppersmith}, \citenamefont {Hu},\ and\ \citenamefont
  {Friesen}}]{Fei2015}%
  \BibitemOpen
  \bibfield  {author} {\bibinfo {author} {\bibfnamefont {J.}~\bibnamefont
  {Fei}}, \bibinfo {author} {\bibfnamefont {J.-T.}\ \bibnamefont {Hung}},
  \bibinfo {author} {\bibfnamefont {T.~S.}\ \bibnamefont {Koh}}, \bibinfo
  {author} {\bibfnamefont {Y.-P.}\ \bibnamefont {Shim}}, \bibinfo {author}
  {\bibfnamefont {S.~N.}\ \bibnamefont {Coppersmith}}, \bibinfo {author}
  {\bibfnamefont {X.}~\bibnamefont {Hu}},\ and\ \bibinfo {author}
  {\bibfnamefont {M.}~\bibnamefont {Friesen}},\ }\bibfield  {title} {\bibinfo
  {title} {{Characterizing gate operations near the sweet spot of an
  exchange-only qubit}},\ }\href@noop {} {\bibfield  {journal} {\bibinfo
  {journal} {Phys. Rev. B}\ }\textbf {\bibinfo {volume} {91}},\ \bibinfo
  {pages} {205434} (\bibinfo {year} {2015})}\BibitemShut {NoStop}%
\bibitem [{\citenamefont {Russ}\ and\ \citenamefont
  {Burkard}(2015{\natexlab{b}})}]{Russ2015}%
  \BibitemOpen
  \bibfield  {author} {\bibinfo {author} {\bibfnamefont {M.}~\bibnamefont
  {Russ}}\ and\ \bibinfo {author} {\bibfnamefont {G.}~\bibnamefont {Burkard}},\
  }\bibfield  {title} {\bibinfo {title} {{Asymmetric resonant exchange qubit
  under the influence of electrical noise}},\ }\href@noop {} {\bibfield
  {journal} {\bibinfo  {journal} {Phys. Rev. B}\ }\textbf {\bibinfo {volume}
  {91}},\ \bibinfo {pages} {235411} (\bibinfo {year}
  {2015}{\natexlab{b}})}\BibitemShut {NoStop}%
\bibitem [{\citenamefont {Shim}\ and\ \citenamefont {Tahan}(2016)}]{Shim2016}%
  \BibitemOpen
  \bibfield  {author} {\bibinfo {author} {\bibfnamefont {Y.-P.}\ \bibnamefont
  {Shim}}\ and\ \bibinfo {author} {\bibfnamefont {C.}~\bibnamefont {Tahan}},\
  }\bibfield  {title} {\bibinfo {title} {{Charge-noise-insensitive gate
  operations for always-on, exchange-only qubits}},\ }\href@noop {} {\bibfield
  {journal} {\bibinfo  {journal} {Phys. Rev. B}\ }\textbf {\bibinfo {volume}
  {93}},\ \bibinfo {pages} {121410} (\bibinfo {year} {2016})}\BibitemShut
  {NoStop}%
\bibitem [{\citenamefont {Russ}\ \emph {et~al.}(2016)\citenamefont {Russ},
  \citenamefont {Ginzel},\ and\ \citenamefont {Burkard}}]{Russ2016}%
  \BibitemOpen
  \bibfield  {author} {\bibinfo {author} {\bibfnamefont {M.}~\bibnamefont
  {Russ}}, \bibinfo {author} {\bibfnamefont {F.}~\bibnamefont {Ginzel}},\ and\
  \bibinfo {author} {\bibfnamefont {G.}~\bibnamefont {Burkard}},\ }\bibfield
  {title} {\bibinfo {title} {Coupling of three-spin qubits to their electric
  environment},\ }\href@noop {} {\bibfield  {journal} {\bibinfo  {journal}
  {Phys. Rev. B}\ }\textbf {\bibinfo {volume} {94}},\ \bibinfo {pages} {165411}
  (\bibinfo {year} {2016})}\BibitemShut {NoStop}%
\bibitem [{\citenamefont {Magnus}(1954)}]{Magnus1954}%
  \BibitemOpen
  \bibfield  {author} {\bibinfo {author} {\bibfnamefont {W.}~\bibnamefont
  {Magnus}},\ }\bibfield  {title} {\bibinfo {title} {On the exponential
  solution of differential equations for a linear operator},\ }\href@noop {}
  {\bibfield  {journal} {\bibinfo  {journal} {Commun. Pure Appl. Math.}\
  }\textbf {\bibinfo {volume} {7}},\ \bibinfo {pages} {649} (\bibinfo {year}
  {1954})}\BibitemShut {NoStop}%
\bibitem [{\citenamefont {Childs}\ and\ \citenamefont
  {Chuang}(2000)}]{Childs2000}%
  \BibitemOpen
  \bibfield  {author} {\bibinfo {author} {\bibfnamefont {A.~M.}\ \bibnamefont
  {Childs}}\ and\ \bibinfo {author} {\bibfnamefont {I.~L.}\ \bibnamefont
  {Chuang}},\ }\bibfield  {title} {\bibinfo {title} {{Universal quantum
  computation with two-level trapped ions}},\ }\href@noop {} {\bibfield
  {journal} {\bibinfo  {journal} {Phys. Rev. A}\ }\textbf {\bibinfo {volume}
  {63}},\ \bibinfo {pages} {012306} (\bibinfo {year} {2000})}\BibitemShut
  {NoStop}%
\bibitem [{\citenamefont {Noh}\ \emph {et~al.}(2023)\citenamefont {Noh},
  \citenamefont {Xiao}, \citenamefont {Jin}, \citenamefont {Cicak},
  \citenamefont {Doucet}, \citenamefont {Aumentado}, \citenamefont {Govia},
  \citenamefont {Ranzani}, \citenamefont {Kamal},\ and\ \citenamefont
  {Simmonds}}]{Noh2023}%
  \BibitemOpen
  \bibfield  {author} {\bibinfo {author} {\bibfnamefont {T.}~\bibnamefont
  {Noh}}, \bibinfo {author} {\bibfnamefont {Z.}~\bibnamefont {Xiao}}, \bibinfo
  {author} {\bibfnamefont {X.~Y.}\ \bibnamefont {Jin}}, \bibinfo {author}
  {\bibfnamefont {K.}~\bibnamefont {Cicak}}, \bibinfo {author} {\bibfnamefont
  {E.}~\bibnamefont {Doucet}}, \bibinfo {author} {\bibfnamefont
  {J.}~\bibnamefont {Aumentado}}, \bibinfo {author} {\bibfnamefont {L.~C.~G.}\
  \bibnamefont {Govia}}, \bibinfo {author} {\bibfnamefont {L.}~\bibnamefont
  {Ranzani}}, \bibinfo {author} {\bibfnamefont {A.}~\bibnamefont {Kamal}},\
  and\ \bibinfo {author} {\bibfnamefont {R.~W.}\ \bibnamefont {Simmonds}},\
  }\bibfield  {title} {\bibinfo {title} {{Strong parametric dispersive shifts
  in a statically decoupled multi-qubit cavity QED system}},\ }\href@noop {}
  {\bibfield  {journal} {\bibinfo  {journal} {Nat. Phys.}\ } (\bibinfo {year}
  {2023})}\BibitemShut {NoStop}%
\bibitem [{\citenamefont {Schuch}\ and\ \citenamefont
  {Siewert}(2003)}]{Schuch2003}%
  \BibitemOpen
  \bibfield  {author} {\bibinfo {author} {\bibfnamefont {N.}~\bibnamefont
  {Schuch}}\ and\ \bibinfo {author} {\bibfnamefont {J.}~\bibnamefont
  {Siewert}},\ }\bibfield  {title} {\bibinfo {title} {{Natural two-qubit gate
  for quantum computation using the $\mathrm{XY}$ interaction}},\ }\href@noop
  {} {\bibfield  {journal} {\bibinfo  {journal} {Phys. Rev. A}\ }\textbf
  {\bibinfo {volume} {67}},\ \bibinfo {pages} {032301} (\bibinfo {year}
  {2003})}\BibitemShut {NoStop}%
\bibitem [{\citenamefont {McEwen}\ \emph {et~al.}(2023)\citenamefont {McEwen},
  \citenamefont {Bacon},\ and\ \citenamefont {Gidney}}]{McEwen2023arxiv}%
  \BibitemOpen
  \bibfield  {author} {\bibinfo {author} {\bibfnamefont {M.}~\bibnamefont
  {McEwen}}, \bibinfo {author} {\bibfnamefont {D.}~\bibnamefont {Bacon}},\ and\
  \bibinfo {author} {\bibfnamefont {C.}~\bibnamefont {Gidney}},\ }\bibfield
  {title} {\bibinfo {title} {Relaxing hardware requirements for surface code
  circuits using time-dynamics},\ }\href@noop {} {\bibfield  {journal}
  {\bibinfo  {journal} {arXiv:2302.02192}\ } (\bibinfo {year}
  {2023})}\BibitemShut {NoStop}%
\bibitem [{\citenamefont {Gambetta}\ \emph {et~al.}(2011)\citenamefont
  {Gambetta}, \citenamefont {Motzoi}, \citenamefont {Merkel},\ and\
  \citenamefont {Wilhelm}}]{Gambetta2011}%
  \BibitemOpen
  \bibfield  {author} {\bibinfo {author} {\bibfnamefont {J.~M.}\ \bibnamefont
  {Gambetta}}, \bibinfo {author} {\bibfnamefont {F.}~\bibnamefont {Motzoi}},
  \bibinfo {author} {\bibfnamefont {S.~T.}\ \bibnamefont {Merkel}},\ and\
  \bibinfo {author} {\bibfnamefont {F.~K.}\ \bibnamefont {Wilhelm}},\
  }\bibfield  {title} {\bibinfo {title} {Analytic control methods for
  high-fidelity unitary operations in a weakly nonlinear oscillator},\
  }\href@noop {} {\bibfield  {journal} {\bibinfo  {journal} {Phys. Rev. A}\
  }\textbf {\bibinfo {volume} {83}},\ \bibinfo {pages} {012308} (\bibinfo
  {year} {2011})}\BibitemShut {NoStop}%
\bibitem [{\citenamefont {Werninghaus}\ \emph {et~al.}(2021)\citenamefont
  {Werninghaus}, \citenamefont {Egger}, \citenamefont {Roy}, \citenamefont
  {Machnes}, \citenamefont {Wilhelm},\ and\ \citenamefont
  {Filipp}}]{Werninghaus2021}%
  \BibitemOpen
  \bibfield  {author} {\bibinfo {author} {\bibfnamefont {M.}~\bibnamefont
  {Werninghaus}}, \bibinfo {author} {\bibfnamefont {D.~J.}\ \bibnamefont
  {Egger}}, \bibinfo {author} {\bibfnamefont {F.}~\bibnamefont {Roy}}, \bibinfo
  {author} {\bibfnamefont {S.}~\bibnamefont {Machnes}}, \bibinfo {author}
  {\bibfnamefont {F.~K.}\ \bibnamefont {Wilhelm}},\ and\ \bibinfo {author}
  {\bibfnamefont {S.}~\bibnamefont {Filipp}},\ }\bibfield  {title} {\bibinfo
  {title} {Leakage reduction in fast superconducting qubit gates via optimal
  control},\ }\href@noop {} {\bibfield  {journal} {\bibinfo  {journal} {npj
  Quantum Inf.}\ }\textbf {\bibinfo {volume} {7}},\ \bibinfo {pages} {14}
  (\bibinfo {year} {2021})}\BibitemShut {NoStop}%
\bibitem [{\citenamefont {Babu}\ \emph {et~al.}(2021)\citenamefont {Babu},
  \citenamefont {Tuorila},\ and\ \citenamefont {Ala-Nissila}}]{Babu2021}%
  \BibitemOpen
  \bibfield  {author} {\bibinfo {author} {\bibfnamefont {A.~P.}\ \bibnamefont
  {Babu}}, \bibinfo {author} {\bibfnamefont {J.}~\bibnamefont {Tuorila}},\ and\
  \bibinfo {author} {\bibfnamefont {T.}~\bibnamefont {Ala-Nissila}},\
  }\bibfield  {title} {\bibinfo {title} {State leakage during fast decay and
  control of a superconducting transmon qubit},\ }\href@noop {} {\bibfield
  {journal} {\bibinfo  {journal} {npj Quantum Inf.}\ }\textbf {\bibinfo
  {volume} {7}},\ \bibinfo {pages} {30} (\bibinfo {year} {2021})}\BibitemShut
  {NoStop}%
\bibitem [{\citenamefont {Zhang}\ \emph {et~al.}(2023)\citenamefont {Zhang},
  \citenamefont {Zhu}, \citenamefont {Ong},\ and\ \citenamefont
  {Petta}}]{Zhang2023arxiv}%
  \BibitemOpen
  \bibfield  {author} {\bibinfo {author} {\bibfnamefont {X.}~\bibnamefont
  {Zhang}}, \bibinfo {author} {\bibfnamefont {Z.}~\bibnamefont {Zhu}}, \bibinfo
  {author} {\bibfnamefont {N.~P.}\ \bibnamefont {Ong}},\ and\ \bibinfo {author}
  {\bibfnamefont {J.~R.}\ \bibnamefont {Petta}},\ }\bibfield  {title} {\bibinfo
  {title} {Developing high-impedance superconducting resonators and on-chip
  filters for semiconductor quantum dot circuit quantum electrodynamics},\
  }\href@noop {} {\bibfield  {journal} {\bibinfo  {journal} {arXiv:2306.16499}\
  } (\bibinfo {year} {2023})}\BibitemShut {NoStop}%
\bibitem [{\citenamefont {Srinivasa}\ \emph {et~al.}(2013)\citenamefont
  {Srinivasa}, \citenamefont {Nowack}, \citenamefont {Shafiei}, \citenamefont
  {Vandersypen},\ and\ \citenamefont {Taylor}}]{Srinivasa2013}%
  \BibitemOpen
  \bibfield  {author} {\bibinfo {author} {\bibfnamefont {V.}~\bibnamefont
  {Srinivasa}}, \bibinfo {author} {\bibfnamefont {K.~C.}\ \bibnamefont
  {Nowack}}, \bibinfo {author} {\bibfnamefont {M.}~\bibnamefont {Shafiei}},
  \bibinfo {author} {\bibfnamefont {L.~M.~K.}\ \bibnamefont {Vandersypen}},\
  and\ \bibinfo {author} {\bibfnamefont {J.~M.}\ \bibnamefont {Taylor}},\
  }\bibfield  {title} {\bibinfo {title} {{Simultaneous Spin-Charge Relaxation
  in Double Quantum Dots}},\ }\href@noop {} {\bibfield  {journal} {\bibinfo
  {journal} {Phys. Rev. Lett.}\ }\textbf {\bibinfo {volume} {110}},\ \bibinfo
  {pages} {196803} (\bibinfo {year} {2013})}\BibitemShut {NoStop}%
\bibitem [{\citenamefont {Petersson}\ \emph {et~al.}(2010)\citenamefont
  {Petersson}, \citenamefont {Petta}, \citenamefont {Lu},\ and\ \citenamefont
  {Gossard}}]{Petersson2010}%
  \BibitemOpen
  \bibfield  {author} {\bibinfo {author} {\bibfnamefont {K.~D.}\ \bibnamefont
  {Petersson}}, \bibinfo {author} {\bibfnamefont {J.~R.}\ \bibnamefont
  {Petta}}, \bibinfo {author} {\bibfnamefont {H.}~\bibnamefont {Lu}},\ and\
  \bibinfo {author} {\bibfnamefont {A.~C.}\ \bibnamefont {Gossard}},\
  }\bibfield  {title} {\bibinfo {title} {{Quantum Coherence in a One-Electron
  Semiconductor Charge Qubit}},\ }\href@noop {} {\bibfield  {journal} {\bibinfo
   {journal} {Phys. Rev. Lett.}\ }\textbf {\bibinfo {volume} {105}},\ \bibinfo
  {pages} {246804} (\bibinfo {year} {2010})}\BibitemShut {NoStop}%
\bibitem [{\citenamefont {Yang}\ \emph {et~al.}(2013)\citenamefont {Yang},
  \citenamefont {Rossi}, \citenamefont {Ruskov}, \citenamefont {Lai},
  \citenamefont {Mohiyaddin}, \citenamefont {Lee}, \citenamefont {Tahan},
  \citenamefont {Klimeck}, \citenamefont {Morello},\ and\ \citenamefont
  {Dzurak}}]{Yang2013}%
  \BibitemOpen
  \bibfield  {author} {\bibinfo {author} {\bibfnamefont {C.~H.}\ \bibnamefont
  {Yang}}, \bibinfo {author} {\bibfnamefont {A.}~\bibnamefont {Rossi}},
  \bibinfo {author} {\bibfnamefont {R.}~\bibnamefont {Ruskov}}, \bibinfo
  {author} {\bibfnamefont {N.~S.}\ \bibnamefont {Lai}}, \bibinfo {author}
  {\bibfnamefont {F.~A.}\ \bibnamefont {Mohiyaddin}}, \bibinfo {author}
  {\bibfnamefont {S.}~\bibnamefont {Lee}}, \bibinfo {author} {\bibfnamefont
  {C.}~\bibnamefont {Tahan}}, \bibinfo {author} {\bibfnamefont
  {G.}~\bibnamefont {Klimeck}}, \bibinfo {author} {\bibfnamefont
  {A.}~\bibnamefont {Morello}},\ and\ \bibinfo {author} {\bibfnamefont {A.~S.}\
  \bibnamefont {Dzurak}},\ }\bibfield  {title} {\bibinfo {title} {{Spin-valley
  lifetimes in a silicon quantum dot with tunable valley splitting}},\
  }\href@noop {} {\bibfield  {journal} {\bibinfo  {journal} {Nat. Commun.}\
  }\textbf {\bibinfo {volume} {4}},\ \bibinfo {pages} {2069} (\bibinfo {year}
  {2013})}\BibitemShut {NoStop}%
\bibitem [{\citenamefont {Hollmann}\ \emph {et~al.}(2020)\citenamefont
  {Hollmann}, \citenamefont {Struck}, \citenamefont {Langrock}, \citenamefont
  {Schmidbauer}, \citenamefont {Schauer}, \citenamefont {Leonhardt},
  \citenamefont {Sawano}, \citenamefont {Riemann}, \citenamefont {Abrosimov},
  \citenamefont {Bougeard},\ and\ \citenamefont {Schreiber}}]{Hollmann2020}%
  \BibitemOpen
  \bibfield  {author} {\bibinfo {author} {\bibfnamefont {A.}~\bibnamefont
  {Hollmann}}, \bibinfo {author} {\bibfnamefont {T.}~\bibnamefont {Struck}},
  \bibinfo {author} {\bibfnamefont {V.}~\bibnamefont {Langrock}}, \bibinfo
  {author} {\bibfnamefont {A.}~\bibnamefont {Schmidbauer}}, \bibinfo {author}
  {\bibfnamefont {F.}~\bibnamefont {Schauer}}, \bibinfo {author} {\bibfnamefont
  {T.}~\bibnamefont {Leonhardt}}, \bibinfo {author} {\bibfnamefont
  {K.}~\bibnamefont {Sawano}}, \bibinfo {author} {\bibfnamefont
  {H.}~\bibnamefont {Riemann}}, \bibinfo {author} {\bibfnamefont {N.~V.}\
  \bibnamefont {Abrosimov}}, \bibinfo {author} {\bibfnamefont {D.}~\bibnamefont
  {Bougeard}},\ and\ \bibinfo {author} {\bibfnamefont {L.~R.}\ \bibnamefont
  {Schreiber}},\ }\bibfield  {title} {\bibinfo {title} {{Large, Tunable Valley
  Splitting and Single-Spin Relaxation Mechanisms in a
  $\mathrm{Si}$/${\mathrm{Si}}_{x}$${\mathrm{Ge}}_{1\ensuremath{-}x}$ Quantum
  Dot}},\ }\href@noop {} {\bibfield  {journal} {\bibinfo  {journal} {Phys. Rev.
  Applied}\ }\textbf {\bibinfo {volume} {13}},\ \bibinfo {pages} {034068}
  (\bibinfo {year} {2020})}\BibitemShut {NoStop}%
\bibitem [{\citenamefont {McJunkin}\ \emph {et~al.}(2022)\citenamefont
  {McJunkin}, \citenamefont {Harpt}, \citenamefont {Feng}, \citenamefont
  {Losert}, \citenamefont {Rahman}, \citenamefont {Dodson}, \citenamefont
  {Wolfe}, \citenamefont {Savage}, \citenamefont {Lagally}, \citenamefont
  {Coppersmith}, \citenamefont {Friesen}, \citenamefont {Joynt},\ and\
  \citenamefont {Eriksson}}]{McJunkin2022}%
  \BibitemOpen
  \bibfield  {author} {\bibinfo {author} {\bibfnamefont {T.}~\bibnamefont
  {McJunkin}}, \bibinfo {author} {\bibfnamefont {B.}~\bibnamefont {Harpt}},
  \bibinfo {author} {\bibfnamefont {Y.}~\bibnamefont {Feng}}, \bibinfo {author}
  {\bibfnamefont {M.~P.}\ \bibnamefont {Losert}}, \bibinfo {author}
  {\bibfnamefont {R.}~\bibnamefont {Rahman}}, \bibinfo {author} {\bibfnamefont
  {J.~P.}\ \bibnamefont {Dodson}}, \bibinfo {author} {\bibfnamefont {M.~A.}\
  \bibnamefont {Wolfe}}, \bibinfo {author} {\bibfnamefont {D.~E.}\ \bibnamefont
  {Savage}}, \bibinfo {author} {\bibfnamefont {M.~G.}\ \bibnamefont {Lagally}},
  \bibinfo {author} {\bibfnamefont {S.~N.}\ \bibnamefont {Coppersmith}},
  \bibinfo {author} {\bibfnamefont {M.}~\bibnamefont {Friesen}}, \bibinfo
  {author} {\bibfnamefont {R.}~\bibnamefont {Joynt}},\ and\ \bibinfo {author}
  {\bibfnamefont {M.~A.}\ \bibnamefont {Eriksson}},\ }\bibfield  {title}
  {\bibinfo {title} {{SiGe quantum wells with oscillating Ge concentrations for
  quantum dot qubits}},\ }\href@noop {} {\bibfield  {journal} {\bibinfo
  {journal} {Nat. Commun.}\ }\textbf {\bibinfo {volume} {13}},\ \bibinfo
  {pages} {7777} (\bibinfo {year} {2022})}\BibitemShut {NoStop}%
\bibitem [{\citenamefont {DiVincenzo}\ \emph {et~al.}(2000)\citenamefont
  {DiVincenzo}, \citenamefont {Bacon}, \citenamefont {Kempe}, \citenamefont
  {Burkard},\ and\ \citenamefont {Whaley}}]{DiVincenzo2000Nature}%
  \BibitemOpen
  \bibfield  {author} {\bibinfo {author} {\bibfnamefont {D.~P.}\ \bibnamefont
  {DiVincenzo}}, \bibinfo {author} {\bibfnamefont {D.}~\bibnamefont {Bacon}},
  \bibinfo {author} {\bibfnamefont {J.}~\bibnamefont {Kempe}}, \bibinfo
  {author} {\bibfnamefont {G.}~\bibnamefont {Burkard}},\ and\ \bibinfo {author}
  {\bibfnamefont {K.~B.}\ \bibnamefont {Whaley}},\ }\bibfield  {title}
  {\bibinfo {title} {{Universal quantum computation with the exchange
  interaction}},\ }\href@noop {} {\bibfield  {journal} {\bibinfo  {journal}
  {Nature}\ }\textbf {\bibinfo {volume} {408}},\ \bibinfo {pages} {339}
  (\bibinfo {year} {2000})}\BibitemShut {NoStop}%
\bibitem [{\citenamefont {Medford}\ \emph
  {et~al.}(2013{\natexlab{b}})\citenamefont {Medford}, \citenamefont {Beil},
  \citenamefont {Taylor}, \citenamefont {Bartlett}, \citenamefont {Doherty},
  \citenamefont {Rashba}, \citenamefont {DiVincenzo}, \citenamefont {Lu},
  \citenamefont {Gossard},\ and\ \citenamefont {Marcus}}]{Medford2013NNano}%
  \BibitemOpen
  \bibfield  {author} {\bibinfo {author} {\bibfnamefont {J.}~\bibnamefont
  {Medford}}, \bibinfo {author} {\bibfnamefont {J.}~\bibnamefont {Beil}},
  \bibinfo {author} {\bibfnamefont {J.~M.}\ \bibnamefont {Taylor}}, \bibinfo
  {author} {\bibfnamefont {S.~D.}\ \bibnamefont {Bartlett}}, \bibinfo {author}
  {\bibfnamefont {A.~C.}\ \bibnamefont {Doherty}}, \bibinfo {author}
  {\bibfnamefont {E.~I.}\ \bibnamefont {Rashba}}, \bibinfo {author}
  {\bibfnamefont {D.~P.}\ \bibnamefont {DiVincenzo}}, \bibinfo {author}
  {\bibfnamefont {H.}~\bibnamefont {Lu}}, \bibinfo {author} {\bibfnamefont
  {A.~C.}\ \bibnamefont {Gossard}},\ and\ \bibinfo {author} {\bibfnamefont
  {C.~M.}\ \bibnamefont {Marcus}},\ }\bibfield  {title} {\bibinfo {title}
  {{Self-consistent measurement and state tomography of an exchange-only spin
  qubit}},\ }\href@noop {} {\bibfield  {journal} {\bibinfo  {journal} {Nat.
  Nanotechnol.}\ }\textbf {\bibinfo {volume} {8}},\ \bibinfo {pages} {654}
  (\bibinfo {year} {2013}{\natexlab{b}})}\BibitemShut {NoStop}%
\bibitem [{\citenamefont {S\o{}rensen}\ and\ \citenamefont
  {M\o{}lmer}(1999)}]{Sorensen1999}%
  \BibitemOpen
  \bibfield  {author} {\bibinfo {author} {\bibfnamefont {A.}~\bibnamefont
  {S\o{}rensen}}\ and\ \bibinfo {author} {\bibfnamefont {K.}~\bibnamefont
  {M\o{}lmer}},\ }\bibfield  {title} {\bibinfo {title} {{Quantum Computation
  with Ions in Thermal Motion}},\ }\href@noop {} {\bibfield  {journal}
  {\bibinfo  {journal} {Phys. Rev. Lett.}\ }\textbf {\bibinfo {volume} {82}},\
  \bibinfo {pages} {1971} (\bibinfo {year} {1999})}\BibitemShut {NoStop}%
\end{thebibliography}%

\end{document}